\documentclass[11pt]{article}
\pdfoutput=1
\usepackage{jheppub}
\usepackage{amsmath}
\usepackage{bm}
\usepackage{slashed}
\usepackage{graphicx}

\newcommand{\tr}{\mathrm{tr}\,}
\newcommand{\Tr}{\mathrm{Tr}\,}
\newlength{\dummysp}
\newcommand{\T}{\mathbb{T}}

\newcommand{\beq}{\begin{eqnarray}}
\newcommand{\eeq}{\end{eqnarray}}

\newcommand{\gappeq}{\mathrel{\rlap {\raise.5ex\hbox{$>$}}
{\lower.5ex\hbox{$\sim$}}}}
\newcommand{\lappeq}{\mathrel{\rlap{\raise.5ex\hbox{$<$}}
{\lower.5ex\hbox{$\sim$}}}}

\newcommand{\ben}{\begin{enumerate}}
\newcommand{\een}{\end{enumerate}}

\newcommand{\bit}{\begin{itemize}}
\newcommand{\eit}{\end{itemize}}

\def\[{\left [}
\def\]{\right ]}
\def\({\left (}
\def\){\right )}
\def\R{{\mathbb R}}
\def\S{{\mathbb S}}
\def\Z{{\mathbb Z}}

\def\cS{{\cal{S}}}
\def\cW{{\cal{W}}}

 \usepackage{dsfont}
 \newcommand{\identity}{\mathds{1}}
 
\newcommand{\basispl}{
   \put(-.5,-.5){\line(1,0){1}}
   \put(.5,-.5){\line(0,1){1}}
   \put(.5,.5){\line(-1,0){1}}
   \put(-.5,.5){\line(0,-1){1}}}
\newcommand{\basisar}{
   \put(0,-.5){\vector(1,0){0}}
   \put(.5,0){\vector(0,1){0}}
   \put(0,.5){\vector(-1,0){0}}
   \put(-.5,0){\vector(0,-1){0}}}
\newcommand{\plaq}[2]{\setlength{\unitlength}{.5cm}\raisebox{-.2cm}{
   \begin{picture}(1.2,1.2)(-.6,-.6)
   \basispl\basisar
   \put(-.5,-.5){\circle*{.2}}
   \put(-.55,-.55){\makebox(0,0)[tr]{\footnotesize $n$}}
   \put(-.55,0){\makebox(0,0)[r]{\footnotesize $#2$}}
   \put(0,-.55){\makebox(0,0)[t]{\footnotesize $#1$}}
   \end{picture}}}
\newcommand{\stapup}{\setlength{\unitlength}{.5cm}\raisebox{-.2cm}{
   \begin{picture}(1.2,1.2)(-.6,-.6)
   \put(.5,-.5){\line(0,1){1}}
   \put(.5,.5){\line(-1,0){1}}
   \put(-.5,.5){\line(0,-1){1}}
   \put(.5,0){\vector(0,-1){0}}
   \put(0,.5){\vector(1,0){0}}
   \put(-.5,0){\vector(0,1){0}}
   \put(-.5,-.5){\circle*{.2}}
   \put(-.55,-.55){\makebox(0,0)[tr]{\footnotesize $n$}}
   \put(-.55,0){\makebox(0,0)[r]{\footnotesize $\nu$}}
   \put(0,.55){\makebox(0,0)[b]{\footnotesize $\mu$}}
   \end{picture}}}
\newcommand{\stapdw}{\setlength{\unitlength}{.5cm}\raisebox{-.2cm}{
   \begin{picture}(1.2,1.2)(-.6,-.6)
   \put(.5,-.5){\line(0,1){1}}
   \put(.5,-.5){\line(-1,0){1}}
   \put(-.5,.5){\line(0,-1){1}}
   \put(.5,0){\vector(0,1){0}}
   \put(0,-.5){\vector(1,0){0}}
   \put(-.5,0){\vector(0,-1){0}}
   \put(-.5,.5){\circle*{.2}}
   \put(-.55,.75){\makebox(0,0)[tr]{\footnotesize $n$}}
   \put(-.55,0){\makebox(0,0)[r]{\footnotesize $\nu$}}
   \put(0,-.55){\makebox(0,0)[t]{\footnotesize $\mu$}}
   \end{picture}}}
\newcommand{\twoplaq}{\setlength{\unitlength}{1cm}\raisebox{-.5cm}{
   \begin{picture}(1.2,1.2)(-.6,-.6)
   \basispl
   \put(-.5,-.5){\circle*{.1}}
   \put(-.5,.5){\circle*{.1}}
   \put(.5,-.5){\circle*{.1}}
   \put(.5,.5){\circle*{.1}}
   \put(0,-.5){\circle*{.1}}
   \put(0,.5){\circle*{.1}}
   \put(.5,0){\circle*{.1}}
   \put(-.5,0){\circle*{.1}}
   \put(-.25,-.5){\vector(1,0){0}}
   \put(.25,-.5){\vector(1,0){0}}
   \put(.5,-.25){\vector(0,1){0}}
   \put(.5,.25){\vector(0,1){0}}
   \put(-.25,.5){\vector(-1,0){0}}
   \put(.25,.5){\vector(-1,0){0}}
   \put(-.5,-.25){\vector(0,-1){0}}
   \put(-.5,.25){\vector(0,-1){0}}
   \put(-.55,-.55){\makebox(0,0)[tr]{\footnotesize $n$}}
   \put(-.55,0){\makebox(0,0)[r]{\footnotesize $\nu$}}
   \put(0,-.55){\makebox(0,0)[t]{\footnotesize $\mu$}}
   \end{picture}}}

\title{On the moduli space of multi-fractional instantons on the twisted $\T^4$ }

 \author[a]{Mohamed M. Anber,}\author[b]{Andrew A. Cox,} \author[b]{Erich Poppitz} 
  
\affiliation[a]{Centre for Particle Theory, Department of Mathematical Sciences, Durham University, South Road, Durham DH1 3LE, UK}
\affiliation[b]{Department of Physics,   University of Toronto, 60 St George St., 
Toronto, ON M5S 1A7, Canada}
\emailAdd{mohamed.anber@durham.ac.uk}\emailAdd{aacox@physics.utoronto.ca}\emailAdd{poppitz@physics.utoronto.ca}

\abstract{
{\flushleft{The}}  moduli space of self-dual $SU(N)$ Yang-Mills  instantons on $\T^4$ of  topological charge  $Q = r/N$,  $1 $$\leq$$ r $$\leq$$ N-1$, is of current interest, yet is not fully understood. In this paper, starting from 't Hooft's constant field strength ($F$) instantons, the only known exact solutions on $\T^4$, we explore the moduli space via analytical and lattice tools. These solutions are characterized  by two positive integers $k, \ell$,  $k$$+$$\ell$$=$$N$, and are  self-dual for $\T^4$ sides $L_\mu$ tuned to $k L_1 L_2$$ =$$ r \ell L_3 L_4$.  For gcd$(k,r)$$=$$r$, we show, analytically and numerically (for $N$$=$$3$) that the constant-$F$ solutions are the only self-dual solutions on the tuned $\T^4$, with $4r$ holonomy moduli. In contrast,  when gcd$(k,r)$$ \ne$$ r$, we argue that the self-dual constant-$F$ solutions  acquire,  in addition to the $4$gcd$(k,r)$ holonomies, $4 r$$-$$4\text{gcd}(k,r)$ extra moduli, whose turning on makes the  field strength nonabelian and non-constant. Thus, for gcd($k,r) \ne r$,   't Hooft's constant-$F$ solutions are a measure-zero subset of the moduli space on the tuned $\T^4$, a fact explaining a puzzle encountered in \cite{Anber:2023sjn}.   We also  show that,  for $r$$=$$k$$=$$2$, $N$$=$$3$,  the agreement between the  approximate  analytic  solutions on the slightly detuned $\T^4$ and the $Q=2/3$ self-dual configurations obtained by minimizing the lattice action is remarkable.
}

\begin{document}

\maketitle

\flushbottom
  
\section{Introduction, summary, and outlook}

\subsection{Motivation}

The recently renewed interest in studying the role that instantons with fractional topological charge  \cite{tHooft:1979rtg,tHooft:1981sps,tHooft:1981nnx,vanBaal:1982ag} play in gauge dynamics is motivated by several developments. The newest one is  the discovery of generalized anomalies  involving $1$-form (center) symmetry \cite{Gaiotto:2014kfa,Gaiotto:2017yup}. Here, the introduction of fractional charge backgrounds  is used to reveal various   't Hooft anomalies involving center symmetry (these lead to, for example, exact spectral degeneracies at finite volume \cite{Cox:2021vsa}). Another development is the  somewhat older observation \cite{RTN:1993ilw,Gonzalez-Arroyo:1995ynx}  that objects with fractional topological charge are responsible for nonperturbative gauge theory  phenomena such as confinement and chiral symmetry breaking. This has been shown in many cases  using reliable semiclassics, sometimes  combined with insight from lattice studies \cite{Unsal:2007jx,Unsal:2008ch,Tanizaki:2022ngt,Tanizaki:2022plm}. The topic is reviewed in \cite{Poppitz:2021cxe,Gonzalez-Arroyo:2023kqv} and is still under intense scrutiny, including on the lattice, see \cite{Mickley:2023exg,Soler:2025vwc,Mickley:2025uqd}.

Instantons of fractional topological charge $Q={r \over N}$ in $SU(N)$ gauge theories on $\T^4$ were discovered by 't Hooft \cite{tHooft:1979rtg,tHooft:1981sps}, who found explicit solutions with commuting constant field strength \cite{tHooft:1981nnx} but nonabelian transition functions, as reviewed in  \cite{Gonzalez-Arroyo:1997ugn}. Further studies were slowed by the lack of explicit solutions for more general---nonabelian, with space-time dependent field strength---instantons. These had to be studied using numerical techniques \cite{GarciaPerez:1992fj,Gonzalez-Arroyo:1998hjb}. 

An important advance in the study of fractional instantons was the introduction \cite{GarciaPerez:2000aiw} of an analytic technique for finding self-dual, or minimum action,  fractional instantons via an expansion in a small parameter $\Delta$, the asymmetry parameter of $\T^4$. This ``$\Delta$-expansion'' technique,  further developed in \cite{Gonzalez-Arroyo:2004fmv,Gonzalez-Arroyo:2019wpu} and \cite{Anber:2023sjn},  shall be important in our discussions below.

The  (multi-) fractional instantons\footnote{We call instantons with $Q=r/N$ and $r>1$ ``multi-fractional.'' In an analytic  study using the $\Delta$-expansion they  resemble an ``instanton liquid,'' comprised  of $r$ strongly overlapping identical constituents \cite{Anber:2023sjn}.}  found in \cite{GarciaPerez:2000aiw,Gonzalez-Arroyo:2019wpu,Anber:2023sjn} were  used by two of the authors of this paper in the semiclassical  calculation of the (higher-order) gaugino condensates in super-Yang-Mills theory on a small twisted-$\T^4$ \cite{Anber:2022qsz,Anber:2024mco}. After understanding many subtle points, see \cite{Anber:2024mco}, we found complete agreement with the weak-coupling semiclassical $\R^4$ and $\R^3 \times \S^1$ results for the gaugino condensates (see \cite{Shifman:1999kf,Shifman:1999mv} and the review \cite{Dorey:2002ik}) and the recent independent large-$N$ lattice determination \cite{Bonanno:2024bqg}. We stress that our completion of a calculation first attempted 40 years ago \cite{Cohen:1983fd} was  only possible  after the recent understanding of both the relevant generalized anomalies in the Hamiltonian framework  \cite{Cox:2021vsa} and  the somewhat intricate multi-fractional instanton moduli space  \cite{Anber:2023sjn,Anber:2024mco}.

Another  recent observation involving fractional instantons is  that seemingly different objects of fractional topological charge, such as center vortices and monopole-instantons, shown to be responsible for semiclassical confinement in  different geometries with small\footnote{Of size $L$ such that  $L \Lambda \ll 1$, where $\Lambda$ is the gauge theory strong coupling scale. In addition to $L$ being small, depending on the matter content of the theory, one requires a double-trace deformation potential or a 't Hooft flux to ensure semiclassical calculability.} compact spaces, $\R \times \T^3$, $\R^3 \times \S^1$, or $\R^2 \times \T^2$  \cite{RTN:1993ilw,Gonzalez-Arroyo:1995ynx,Unsal:2007jx,Unsal:2008ch,Tanizaki:2022ngt,Tanizaki:2022plm}, are  smoothly related to each other by taking different limits of twists and $\T^4$ periods \cite{GarciaPerez:1999hs,Unsal:2020yeh,Poppitz:2022rxv,Hayashi:2024yjc,Guvendik:2024umd, Wandler:2024hsq,Hayashi:2024psa}. 
Using Monte-Carlo simulations, center vortices and monopole-instantons  have long ago been argued to be relevant for confinement, even outside of the semiclassical regime, as described in e.g. the monograph \cite{Greensite:2011zz}. The novelty here is their embedding in a rich framework of intertwined semiclassically-calculable regimes. Apart from the obvious advantage of being analytic, the semiclassical approach allows the   study of a range of gauge theories, notably ones with fermions in different representations \cite{Anber:2017pak,Anber:2019nfu,Anber:2021lzb,Hayashi:2023wwi,Hayashi:2024qkm,Hayashi:2024gxv},  which are difficult and prohibitively expensive  to access via  lattice simulations.

{\flushleft{In}} summary, we believe that the  developments mentioned above clearly  motivate the need for better understanding of fractional instantons and their role in gauge dynamics.

\subsection{Overview and summary of results}
   
   \subsubsection{Overview of necessary  background}
   
This paper is driven by our desire to resolve a  puzzle encountered when studying the $\Delta$-expansion of multi-fractional instantons  \cite{Anber:2023sjn}. To describe it even briefly,  we need to introduce a minimum amount of background, with complete detail  found in the body of the paper and the Appendices.

The starting point of the $\Delta$-expansion are 't Hooft's constant commuting field strength fractional instantons in $SU(N)$ gauge theory on $\T^4$, with  twisted boundary conditions parameterized by an antisymmetric twist tensor $n_{\mu\nu}$ ($\mu,\nu = 1,2,3,4$), comprised of (mod $N$) integers. Thus, we begin with a lightning review of

{\flushleft{\bf{Constant-$F$ backgrounds with $Q=r/N$:}}} For definiteness, we take a $\T^4$ of periods $L_\mu$ and 't Hooft twisted boundary conditions with only nonzero $n_{12} = -r$ and $n_{34}=1$. Then,  the fractional part of the topological charge is $Q=r/N$, taking for now $1 \le r < N$. The constant-$F$ fractional instantons   \cite{tHooft:1981nnx} are constructed by taking two arbitrary positive integers $k$ and $\ell$ such that $k + \ell = N$. 
Then,   one defines the $U(1)_\omega \in SU(N)$ generator
 \begin{equation}
\label{omega}
\omega = 2 \pi \; {\rm diag} (\underbrace{\ell, \ell, ..., \ell}_{k \; {\rm times}},  \underbrace{- k, -k, ...., -k}_{\ell \; {\rm times}})
\end{equation}
and without further ado\footnote{For brevity, here we skip details such as the gauge choice for the transition functions implementing the twists and simply assure the reader that (\ref{solution1}) obeys all appropriate boundary conditions on the twisted $\T^4$, as shown in great detail in section \ref{sec:tunedgeneral}. }  writes the solution as
\begin{eqnarray}\label{solution1}
\bar A_1 =0, \; \bar A_2 &=& - \omega \;{r x_1 \over N k L_1 L_2}, \bar A_3 = 0, \; \bar A_4 = - \omega \;{x_3 \over N \ell L_3 L_4},\\  \; \implies \;
\bar F_{12} &=& - \omega \;{r  \over N k L_1 L_2}, \; \bar F_{34} =  - \omega\; {1 \over N \ell L_3 L_4}. \nonumber
\end{eqnarray}
Clearly, the background (\ref{solution1})  has constant field strength, with only nonzero $\bar F_{12}$ and $\bar F_{34}$. That the topological charge is $Q=r/N$ can be easily verified explicitly. Furthermore,  (\ref{solution1}) 
is self dual (and thus has minimum action ${r \over N} {8 \pi^2  \over  g^2}$) whenever $\bar F_{12} = \bar F_{34}$. From (\ref{solution1}) we see that this only occurs  for $\Delta = 0$, with $\Delta$ defined  for future use as\footnote{The self-dual constant-$F$ solutions have a simple realization in string theory. They are  $T$-dual to two stacks of $k$ and $\ell$ $D2$-branes, wrapped at an  angle (determined by $r$)  on the $\T^4$, with the BPS condition being precisely  $\Delta(r,k,\ell)=0$, see \cite{Guralnik:1997sy,Hashimoto:1997gm}.}
\begin{equation}\label{deltadef}
\Delta(r, k, \ell) = {r \ell L_3 L_4 - k L_1 L_2 \over \sqrt{L_1 L_2 L_3 L_4}}\,,
\end{equation}
and we consider only the case $\Delta>0$.\footnote{Because the results of \cite{Anber:2023sjn} assumed $\Delta >0$. These can be trivially extended to $\Delta<0$, but modifications of some formulae there are required.} 
The fluctuation spectrum around some of these constant-$F$ solutions was studied long ago \cite{vanBaal:1984ar}, albeit not in an exhaustive manner. As we see below, there are features, even for the self-dual constant-$F$ solutions, that were not understood until now, prompted by our study of the multi-fractional instantons in \cite{Anber:2023sjn}. 

{\flushleft{\bf{The $\Delta$-expansion:}}}  As shown in    \cite{GarciaPerez:2000aiw}, one can analytically construct self dual, nonabelian- and non-constant-$F$ solutions with  $Q=r/N$ on a $\T^4$ whose periods are detuned away from $\Delta =0$, using a series expansion in $\sqrt\Delta$. 

Schematically, one writes the deformed-$\T^4$ (i.e. the one with small nonzero $\Delta$)  background  as the constant-$F$ background $\bar A$ of  (\ref{solution1}) plus general fluctuations $\cS, \cW$ covering the entire $SU(N)$ (omitting the spacetime index $\mu$ for brevity):
\begin{equation}\label{solution2}
A = \bar A + \omega \cS^{ \omega } +  \left(\begin{array}{cc} \cS^{ k } & \cW^{k \times \ell} \cr (\cW^{k \times \ell})^\dagger  & \cS^{ \ell } \end{array} \right)~,
\end{equation}
where $\cS^{ k, \ell }$ are in $SU(k)$ and $SU(\ell)$ respectively and $\cS^{ \omega }$ is in $U(1)_\omega$, recall (\ref{omega}).
One then imposes 
 self duality  on the field strength of $A$, along with  the background gauge condition $D_\mu(\bar A) A_\mu = 0$. 
 
 Solving the  nonlinear  self-duality condition, which is explicitly given later in the paper,\footnote{The  nonlinear self-duality conditions are similar to eqns.~(\ref{main equations we need to solve k eq r}) with $\lambda =1$, but with an extra constant term $\sim \Delta$ on the r.h.s. of the first two equations (which can be found in \cite{Anber:2023sjn}).} is done consistently via an iterative procedure, using a power series in $\Delta$. One expands $\cS$ (denoting either $\omega, k, \ell$ component) and $\cW$ in powers of $\Delta^p$ and $\Delta^{p + {1\over 2}}$, $p =0,1,2,...$, respectively, 
 \begin{equation}\label{delta1}
 \cS \sim \cS^{(0)} + \Delta \cS^{(1)} +...,~~\cW \sim \sqrt{\Delta}\;( \cW^{(0)} + \Delta \cW^{(1)} +...),
 \end{equation} showing only the few lowest-order terms. We now state the salient features of the $\Delta$-expansion: \begin{enumerate}
 \item The $\cS^{(0) \; \omega, k, \ell}$ terms are constant pieces  omitted from (\ref{solution1}). These represent the only allowed holonomies, the ones commuting with the transition functions. The fact that, as shown in   \cite{Anber:2023sjn}, there are only $4 {\rm gcd}(k,r)$ such nonvanishing holonomies is important in what follows. These holonomies become moduli of the nonabelian solution. 
 \item One can show that the nonlinear self-duality equations for $A$ of (\ref{solution2}) can be solved iteratively. One begins with $\cW^{(0)}$, which is determined by $\bar A$ and $\cS^{(0)}$. Then, one finds  that $\cS^{(1)}$ is determined by $\bar A, \cS^{(0)}$ and $\cW^{(0)}$. This iterative procedure continues to any desired order in $\Delta$, but higher orders quickly become analytically cumbersome to treat.\footnote{For a study of an analogous expansion for self-dual vortices in two dimensions, where high orders can be computed more easily, see \cite{Gonzalez-Arroyo:2004fmv}.}
 
 The solution is nonabelian and the field strength is non-constant on the torus. For $r>1$, gcd$(k,r)=r$, and to leading nontrivial order in $\Delta$, it exhibits instanton-liquid like appearance of strongly overlapping $r$ lumps  \cite{Anber:2023sjn}.
 \item The $\Delta$-expansion has been confronted with the ``exact'' fractional instanton, the one obtained by minimizing the lattice action with twists. This was done already in  \cite{GarciaPerez:2000aiw} for $r=1, N=2$ (here only $k=\ell=1$ are possible). The agreement with the numerics, even upon keeping only the leading $\cW^{(0)}$ term, is remarkable, for values of $\Delta$ as large as $0.02-0.09$. 
 
 Further in this paper, in section~\ref{sec:testdelta}, we similarly favourably compare the small-$\Delta$ expansion for $r=2$ multi-fractional instanton for $ N=3, k=2$ with the ``exact'' numerical solution on a  detuned $\T^4$ with small $\Delta(2,2,1)$. We find excellent agreement of the analytic and numeric fractional instantons for   $\Delta = 0.129$ and $0.236$.
 
 \item Our study of the multi-fractional instanton contribution to the gaugino condensate utilized the analytic solution for the $Q=r/N$ multi-fractional instanton with $k=r$ via the $\Delta$-expansion  \cite{Anber:2023sjn}.  As shown in \cite{Anber:2023sjn},   gauge invariant densities evaluated in the background of the $\Delta$-expansion solutions are composed of $r$ ``lumps.'' More precisely, to leading order in the $\Delta$-expansion the gauge invariants are given by a sum of $r$ identical functions that only differ by their position on the $\T^4$, accounting thus for the total number, $4r$, of  moduli. Generally, however, these $r$ objects strongly overlap and their gauge invariant densities do not  resemble those expected of a dilute gas; for an illustration, see the plots  for $r=2$, $N=3$, $\Delta=0.129$, on Figure~\ref{fig:Delta_fitSmall} below.\footnote{However, if one considers the regime of large-$\Delta$, e.g. $L_1 L_2 \gg L_3 L_4$  (not accessible analytically) resembling a $\T^2_{large} \times \T^2_{small}$ geometry, one numerically finds $r$ well-separated objects, of size determined by $L_3 L_4$,  localized in the large $L_1 L_2$ torus (this geometry was used, e.g., recently in   \cite{Soler:2025vwc}).}  
 
In  \cite{Anber:2024mco}, we determined the shape and volume of the moduli space and 
  showed that the integral over the $r$-lump moduli space reproduces the order-$r$  gaugino condensate, $\langle (\lambda\lambda)^r \rangle$ (with  gcd$(r,N)=1$). This was calculated  on  $\R^4$ in \cite{Dorey:2002ik}, via the---significantly more complicated, in our view---ADHM construction. 
 \end{enumerate}
 Finally, we can state 

{\flushleft{\bf{The puzzle of}}}  \cite{Anber:2023sjn}: In our studies of multi-fractional instantons (with $1 < r < N$), we encountered a problem with the $\Delta$-expansion around  constant-$F$ solutions with gcd$(k,r) \ne r$. We found that the $\Delta$-expansion, already at the leading nontrivial order,  leads to new moduli and a seemingly noncompact moduli space. This is the feature we set out to better understand here. We shall do this using a combination of analytic and numerical tools.

To describe the resolution, we first turn  to the number of constant holonomies $\cS_0$, which  is $4 {\rm gcd}(k,r)$. It is well known that  a $Q=r/N$ self-dual fractional instanton should have $4 r$ moduli, as follows from the index theorem \cite{Schwarz:1977az,Weinberg:1979ma,Dorey:2002ik}. The number of constant holonomies $\cS_0$ thus equals $4 r$ only when gcd$(r,k)=r$, in agreement with the index theorem. Those with $r=k$ (with $r>1$) and $r=1$ (any $k$), are indeed the cases where we found that the $\Delta$-expansion produces a unique\footnote{\label{foot:unique}Here, ``unique'' is meant in  the sense that each constant-$F$ solution (for given values of its $4r$ moduli) gives rise to a unique non-constant-$F$ solution on the detuned torus, as a series expansion in $\Delta$. We have not proven that the solutions obtained within the $\Delta$-expansion are the only fractional instantons for the given values of $L_\mu$. However,  the comparison   of the numerical  multi-fractional instantons with the ones obtained from the analytic $\Delta$-expansion, see section~\ref{sec:testdelta}, as well as the correct  determination of the higher-order gaugino condensates \cite{Anber:2024mco}, provide evidence that these are, indeed, the only saddles.   } solution, with moduli determined by the $4 {\rm gcd}(k,r)= 4 r$ constant holonomies. The integration over the latter determine the condensate $\langle (\lambda\lambda)^r \rangle$.

\subsubsection{Summary of results and outline of the paper} 

{\bf \flushleft{T}he first focus of this paper} is the study of the moduli space of self-dual backgrounds of charge $Q=r/N$ for different choices of $k$, on a tuned $\T^4$, with $\Delta(r,k,\ell)=0$, or $r \ell L_3 L_4 = k L_1 L_2$. For these values of $L_\mu$, the constant-$F$ solution (\ref{solution1}) is self-dual. Here, we find that:
\begin{enumerate}
 \item {\bf gcd$\mathbf{(k,r)=r}$:} For $k$ such that gcd$(k,r)=r$,   there is a $4 r$ dimensional moduli space parameterized by the constant holonomies $\cS_0$. All self-dual backgrounds have the same constant $F$, and the constant holonomies comprise all moduli of the $\Delta(r,k,\ell)=0$ self-dual solution.
 
 First, we show this analytically in section \ref{sec:tunedgcd=r}. We consider   general self-duality preserving fluctuations (orthogonal to small gauge transformations) around the constant-$F$ self-dual solution, parameterized exactly like in (\ref{solution2}), but now on the tuned $\T^4$. We expand  the self-duality condition in a formal series in the nonlinearity. We call this the $\lambda$-expansion, where $\lambda$ is a parameter introduced to count the order of the nonlinear terms (as advocated in \cite{Schwarz:1977az,Weinberg:1979ma,Taubes:1982qem}). The expansion used to solve the self-duality condition of eqn.~(\ref{main equations we need to solve k eq r}) for (\ref{solution2}) is now, in contrast with (\ref{delta1}):
  \begin{equation}\label{lambda1}
 \cS \sim \cS_0 + \lambda \cS_1 + \lambda^2 \cS_2 + ...,~~\cW \sim \cW_0+  \lambda \cW_1+ \lambda^2 \cW_2 +...,
 \end{equation}
Proceeding with iteratively solving the self-duality condition in powers of $\lambda$, we find that 
when gcd$(k,r)=r$,  only the constant holonomies $\cS_0$ are nonzero. All other terms, $\cW_0, \cS_1, \cW_1, \cS_2, \cW_2,...$, are set to zero, order by order in the $\lambda$-expansion. 

Second, in section \ref{sec:numericgcd=r}, we confirm this analytic result using the ``exact'' (multi-) fractional instantons for $N=3$, obtained by minimizing the lattice action for  $k=r=1$, on a $\T^4$ with $\Delta(1,1,2)=0$, or $2 L_3 L_4 = L_1 L_2$ (see Figure~\ref{fig:r=k=1action}),  as well as   $k=r=2$, on a $\T^4$  with $\Delta(2,2,1)=0$, or $L_3 L_4 = L_1 L_2)$ (see  Figure~\ref{fig:r=k=2action}).
In each case, we find that 
when the $\T^4$ is tuned so that $\Delta(r,r,3-r)=0$, the lattice minimization only produces constant-$F$ solutions. As a check of the numerical simulations, we also study the moduli space of these constant-$F$ solutions by computing the winding Wilson loops of the numerical solutions. We find precise agreement with the winding Wilson loops, calculated for the analytic solutions in \cite{Anber:2024mco}. The relevant expressions for $SU(3)$ that we plot are presented in Appendix~\ref{appx:wilson}. See Figures~\ref{fig:r=k=1wilson} and \ref{fig:r=k=2wilson}. 

Thus, we conclude that for gcd$(k,r)=r$ and $\Delta(r,k,\ell)=0$ (or $r \ell L_3 L_4 = k L_1 L_2$), all charge $r/N$ multi-fractional self-dual instantons are constant-$F$ ones, with moduli given by the $4 r$ constant holonomies. Upon detuning the $\T^4$, a small-$\Delta$ expansion around such backgrounds produces a unique (in the sense of footnote~\ref{foot:unique}) nonabelian solution, determined by the constant-$F$ abelian background (\ref{solution1}) and the $\cS_0$ holonomies. 

\item {\bf gcd$\mathbf{(k,r)\ne r}$:} Next, we turn to a study of the space of self-dual solutions of charge $r/N$ and $k$ such that gcd$(k,r) \ne r$,  on a tuned $\T^4$   with $\Delta(r,k,\ell)=0$, or $r \ell L_3 L_4 = k L_1 L_2$. While the constant-$F$  self-dual background (\ref{solution1}) still exists, perturbing it by self-duality preserving perturbations, we find a behaviour quite different  from that  for gcd$(k,r)=r$ studied above. 

First, in section \ref{sec:analyticgcdnotr}, we use the leading-order analytic $\lambda$-expansion to study self-duality preserving perturbations around the self-dual constant-$F$ solution. We show that, in contrast to the gcd$(k,r)=r$ case, this expansion now produces solutions with nonabelian parts which are space-time dependent, already at the first nontrivial order. There are additional moduli which appear at leading order in $\lambda$ and, within the leading-order $\lambda$-expansion, the moduli space appears noncompact.  The total number of moduli,  the $4 {\rm gcd}(k,r)$ constant holonomies, plus the  moduli appearing in the $\lambda$-expansion add up to $4 r$, consistent with the index theorem. However,  the   moduli space  is expected to be compact on the $\T^4$, and we do not yet know how to determine its global structure using the leading-order expansion  in nonlinearity. We leave this important problem for future work. 

Second, in section \ref{sec:numericgcdnotr},
this analytic finding is further supported by studying the multi-fractional instanton for gauge group $SU(3)$ on the lattice, for $r=2, k=1$,  using a $\T^4$ with $\Delta(2,1,2)=0$ (or $4 L_3 L_4 = L_1 L_2$). In contrast to the $r=2,k=2$, $\Delta(2,2,1)=0$ (or $L_1 L_2 = L_3 L_4$) case, the numerical minimization found no  constant-$F$ solutions---see Figures~\ref{fig:SU3lattice1} and \ref{fig:SU3lattice2} for two different sizes of $\T^4$. By the analytic argument of section~\ref{sec:analyticgcdnotr}, the constant-$F$ self-dual instantons are a set of measure zero. As numerics generates only a finite number of configurations, the fact that the numerical algorithm   found no constant-$F$ backgrounds is consistent with our analytic result. As Figure~\ref{fig:SU3lattice2}  shows, among the $214$ self-dual configurations ($Q=2/3$, $k=1$) on the larger $(32,8,8,8)$ lattice, we found several with small spatial variation of $F$. For these configurations, we fitted the analytical prediction of the leading-order $\lambda$ expansion of section \ref{sec:analyticgcdnotr} for the gauge invariant $\tr F_{13}^2$ with the numerics. We found reasonable agreement for small values of the new ``non-compact'' moduli, making this rather preliminary study self-consistent. See section~\ref{sec:lambdafit}. 

We further corroborate our finding by evaluating the winding Wilson loops in these non-constant-$F$ configurations. In contrast to the gcd$(k,r)=r$ constant-$F$ solutions, here we have no analytic results for the Wilson loops. However, we  show that, when summed over all numerically generated self-dual configurations, these average to zero (consistent with the finite size of the sample, see Figures~\ref{fig:r=2-k=1wilson} and \ref{fig:avg_Wilson}). In  \cite{Anber:2023sjn,Anber:2024mco}, we showed  that the integral of the winding Wilson loop over the fractional instanton moduli space should vanish (this vanishing follows from the Hamiltonian interpretation of the twisted partition function).  The vanishing of the winding Wilson loops averaged over all numerically generated configurations (shown on Figures~\ref{fig:r=2-k=1wilson} and \ref{fig:avg_Wilson})  provides evidence  that the entire moduli space of the $r=2, k=1$ fractional instanton is covered by our numerical procedure. 

Our results indicate that a small self-duality preserving fluctuation around the $\Delta(r,k,\ell)=0$ $(r \ell L_3 L_4 = L_1 L_2)$ self-dual constant-$F$ fractional instanton with gcd$(k,r)\ne r$ leads to a non-constant-$F$ solution. Thus, the constant-$F$ self-dual fractional instanton is a measure zero set in the moduli space of charge $r/N$ self-dual multi-fractional instantons on $\T^4$ with $\Delta(r,k,\ell)=0$ and gcd$(k,r) \ne r$.

In conclusion, we take the analysis here to suggest  that the reason behind the failure of the $\Delta$-expansion for gcd$(k,r)\ne r$ observed in \cite{Anber:2023sjn} is that the $\Delta=0$ background one expands around is generically not known. Except for a measure zero set on the moduli space, it is not a constant-$F$ one. However, as we discuss further below, it is desirable to better understand the relation between the $\Delta$- and $\lambda$-expansions.

\item {\bf $\mathbf{SU(2)}$ with $\mathbf{Q=r/2, r>1}$:} Finally, in section~\ref{appx:su2}, we point out that a similar issue (to the one described above for gcd$(k,r)\ne r$) exists for all $SU(2)$ instantons of charge $Q=r/2$, for any integer $r > 1$, thus including also all integer-charged self-dual instantons\footnote{It is well known that without any twists, there is no $Q=1$ self-dual instanton on the $\T^4$, as follows from the Nahm transform \cite{Schenk:1986xe,Braam:1988qk}.} on the twisted $\T^4$.

While a constant-$F$ solution exists for $\Delta(r,1,1)=0$  or $r L_3 L_4 = L_1 L_2$, it only allows for $4$ constant holonomies, or translational moduli. 
The other moduli correspond to deformations that produce   non-constant field strength. 
We show this by considering general fluctuations around the constant-$F$ solution and studying the nonlinear self-duality equation on the tuned $\T^4$. Via an analytic argument that goes beyond the $\lambda$-expansion  of section~\ref{sec:analyticgcdnotr}, we  argue in section~\ref{appx:su2} that additional $4r -4$ moduli appear and that their turning on makes the solution non-constant. We are, however, unable to show that the moduli space becomes compact (this is expected on $\T^4$, see  also the remarks of \cite{BraamTodorov} on the $r=2$ solution)  leaving this as an interesting area of future study.

\end{enumerate}

{\bf \flushleft{The second} focus of this paper} is our use of the $SU(3)$ numerical simulations  to provide a comparison of the ``exact" solution for the $2$-lump multi-fractional instanton of charge $Q=2/3$ with the analytic solution obtained via the $\Delta$-expansion on a detuned $\T^4$. This is presented in section~\ref{sec:testdelta}. We show, for two values of $\Delta<1$, $\Delta = 0.236$ on Figure~\ref{fig:Delta_fit}, and $\Delta=0.129$ on Figure~\ref{fig:Delta_fitSmall}, that the comparison, obtained by a fit of the moduli of the analytical solution (found using the $\Delta$-expansion) to the numerical self-dual configurations is rather impressive.

\subsection{Outlook}

The moduli space of fractional instantons on $\mathbb{T}^4$ is a subject of mathematical interest, which also has significant implications for understanding the semiclassical expansion in field theory. While this paper marks a step toward understanding its structure, many directions remain open for better comprehension.

\begin{enumerate}

\item In this paper, we conducted a comprehensive analytical and numerical study of the case gcd$(k,r) = r$, on the tuned-$\T^4$ with $\Delta(r,k,\ell) = 0$ ($r \ell L_3 L_4 = k L_1 L_2$), demonstrating that an abelian constant-$F$ solution remains unchanged under self-duality-preserving perturbations. Additionally, we found that the analytical $\Delta$-expansion solution aligns with numerical results when $\gcd(k,r) = r$ for small enough values of $\Delta$, suggesting that perturbations on top of the analytical $\Delta$-expansion solution preserve it. Yet, it is desirable to find an analytical mathematical statement to confirm this finding. In particular,   a future research direction would be the exploration of the interplay between the $\lambda$- and $\Delta$-expansions.

\item  Our findings for the case gcd$(k,r) \neq r$ and $\Delta(r,k,\ell) = 0$ highlight the need for further investigation. A key open question concerns the number of bosonic and fermionic zero modes in this background (which is only understood numerically). Even though one expects from the index theorem that this number is $4r$ and $2r$ for the bosonic and adjoint fermion zero modes, a numerical confirmation is desired. The fermionic zero modes, in particular, can be explored through numerical lattice studies by analyzing the spectrum of adjoint fermions in the fractional instanton background, following approaches such as those in \cite{GarciaPerez:1997fq,Berruto:2000fx,GarciaPerez:2007ne} or the more recent \cite{Soler:2023qzi}.

\item Our numerical studies of the cases $r$$=$$k$$=$$2$ and $r$$=$$2$, $k$$=$$1$ with $\Delta(r,k,\ell)$$=$$0$  $(r \ell L_3 L_4 = k L_1 L_2)$, all within the gauge group $SU(3)$, showed drastically different qualitative behavior between the two cases with the same topological charge $2/3$. Contrasting these results with the $\Delta$-expansion gives the following puzzle.  In principle, the analytical solution that corresponds to the numerical study with $r$$=$$2$, $k$$=$$1$ and $\Delta(r,k,\ell)$$=$$\Delta(2,1,2)$$=$$0$  can be obtained starting from the constant-$F$ analytical solution with $r$$=$$k$$=$$2$ (recall that this analytical solution has $8$ compact moduli identified as translations and holonomies) taking  $\Delta(r,k,\ell)$$=$$\Delta(2,2,1)$$=$$-3$.  We know from the numerical studies that the analytical solution with $r$$=$$k$$=$$2$ and  $\Delta(r,k,\ell)$$=$$\Delta(2,2,1)$$=$$-3$ has a drastically different qualitative behavior compared to the analytical solution  $r$$=$$2$, $k$$=$$2$ and $\Delta(r,k,\ell)$$=$$\Delta(2,2,1) = 0$. This means that as we consider large values of $\Delta$, an intriguing mathematical structure should account for this dramatic shift in behavior. 

\item From the more formal mathematical side, a further study of  the  moduli space of the $Q=r/2$, $r \ge 2$, instantons in $SU(2)$, the topic of  section~\ref{appx:su2}, is of great interest.  We considered the full set of nonlinear equations obeyed by the self-duality preserving fluctuations around the constant-$F$ solution on the tuned-$\T^4$, but were only able to show that these imply the correct dimensionality of the moduli space. Showing that the  nonlinear equations for $\cS$, eqns.~(\ref{sequation}, \ref{wsolution},\ref{sconstr1}) of section ~\ref{appx:su2},  lead to compactification of the moduli space  and determine its global structure (for $r=2$ studied in \cite{BraamTodorov}) remains a challenge. In addition, it would be of interest to generalize the analytic arguments of  section~\ref{appx:su2} to general $N, r, k$ with gcd$(r,k)\ne r$.\end{enumerate}

 \section{General $Q={r \over N}$ self-dual instantons on the tuned $\T^4$: $r \ell L_3 L_4 = k L_1 L_2$}
  \label{sec:tunedgeneral}
  
  This section begins by spelling out the details of the $SU(N)$ boundary conditions on the twisted four-torus $\mathbb T^4$ and by presenting the constant-$F$ solution,  already given in (\ref{solution1}), and the allowed constant holonomies. 
  
   We take the torus to have periods of length $L_\mu$, $\mu=1,2,3,4$, where  $\mu,\nu$ runs over the spacetime dimensions. The gauge fields $A_\mu$ are Hermitian traceless $N\times N$ matrices and taken to obey the boundary conditions (BCS)
\begin{eqnarray}
A_\nu(x+L_\mu \hat e_\mu)=\Omega_\mu(x) A_\nu(x) \Omega_\mu^{-1}(x)-i \Omega_\mu(x) \partial_\nu \Omega_\mu^{-1}(x)\,,
\label{conditions on gauge field}
\end{eqnarray}
upon traversing $\mathbb T^4$ in each direction. The transition functions $\Omega_\mu$ are $N\times N$ unitary matrices, and  $\hat e_\nu$ are unit vectors in the $x_\nu$ direction. The subscript $\mu$ in $\Omega_\mu$ indicates that the function $\Omega_\mu$ does not depend on the coordinate $x_\mu$. Then, the compatibility of (\ref{conditions on gauge field}) at the corners of the $x_\mu$-$x_\nu$ plane of $\mathbb T^4$ gives the cocylce condition
\begin{eqnarray}\label{cocycle}
\Omega_\mu (x + \hat{e}_\nu L_\nu) \; \Omega_\nu (x) = e^{i {2 \pi n_{\mu\nu} \over N}} \Omega_\nu (x+ \hat{e}_\mu L_\mu) \; \Omega_\mu (x)~.
\end{eqnarray}
In order to write the transition functions giving rise to topological charge $Q= r/N$, we introduce the $k \times k$ matrices
 $P_k$ and $Q_k$, the shift and clock matrices:
\begin{eqnarray}\label{pandq}
P_k=\gamma_k\left[\begin{array}{cccc} 0& 1&0&...\\ 0&0&1&...\\... \\ ...&  &0&1 \\1&0&...&0\end{array}\right]\,,\quad Q_k=\gamma_k\; \mbox{diag}\left[1, e^{\frac{i 2\pi}{k}}, e^{2\frac{i 2\pi}{k}},...\right]\,,
\end{eqnarray}
which satisfy the relation $P_kQ_k=e^{i\frac{2\pi}{k}}Q_kP_k$. The factor $\gamma_k\equiv e^{\frac{i\pi (1-k)}{k}}$ ensures that $\det Q_k=1$ and $\det P_k=1$ (the matrices $P_\ell$ and $Q_\ell$ used below are defined similarly). 

To write the $SU(N)$ transition functions, we use a $k \times \ell$ block matrix notation, where $k + \ell = N$. 
The  transition functions  now read
\begin{eqnarray}
\nonumber
\Omega_1&=&P_k^{-r}\otimes I_\ell e^{i \omega \frac{r x_2}{Nk L_2}} = \left[\begin{array}{cc}P_k^{-r}e^{i2\pi \ell r \frac{x_2}{Nk L_2}}&0\\0& e^{-i 2\pi r\frac{x_2}{NL_2}}I_\ell\end{array}\right],\\
\nonumber
\Omega_2&=&Q_k\otimes I_\ell = \left[\begin{array}{cc}Q_k&0\\0& I_\ell\end{array}\right],\\
\nonumber
\Omega_3&=&I_k\otimes P_\ell e^{i \omega \frac{x_4}{N\ell L_4}} = \left[\begin{array}{cc} e^{i2\pi  \frac{x_4}{N L_4}} I_k&0\\0& e^{-i 2\pi k\frac{x_4}{N \ell L_4}}P_\ell\end{array}\right],\\
\Omega_4&=&I_k\otimes Q_\ell = \left[\begin{array}{cc}I_k&0\\0& Q_\ell\end{array}\right].
\label{transition1}
\end{eqnarray}
where we recall that  $\omega$ is given by (\ref{omega}), $P_{k \; (\ell)}$ and $Q_{k \; (\ell)}$ in (\ref{pandq}), and $I_k$ ($I_\ell$) denote $k\times k$ ($\ell\times \ell$) unit matrices. 

The transition functions (\ref{transition1}) can be explicitly seen to obey the cocycle condition (\ref{cocycle}) with  only nonzero
 $n_{12}=-r$ and $n_{34} = 1$; hence, the topological charge of the gauge field obeying (\ref{conditions on gauge field}) is $Q={r \over N}$. That the constant-$F$ background of (\ref{solution1}) or (\ref{solution11}) obeys (\ref{conditions on gauge field}) can also be seen explicitly. 
 
 Here, we reproduce the background already given in (\ref{solution1}): 
\begin{eqnarray}\label{solution11}
\bar A_1 =0, \; \bar A_2 &=& - \omega \;{r x_1 \over N k L_1 L_2}, \bar A_3 = 0, \; \bar A_4 = - \omega \;{x_3 \over N \ell L_3 L_4},\\  \;  
\bar F_{12} &=& - \omega \;{r  \over N k L_1 L_2}, \; \bar F_{34} =  - \omega\; {1 \over N \ell L_3 L_4}. \nonumber
\end{eqnarray}
As discussed in the Introduction, the solution (\ref{solution11}) does not include any   allowed holonomies. These are constant contributions to $\bar A_\mu$, which, by (\ref{conditions on gauge field}), have to commute with the transition functions (\ref{transition1}). 
We parameterize the constant contributions to $\bar A_\mu$ of (\ref{solution11}) by $\phi_\mu$, which we include by replacing (\ref{solution11}) as indicated below:
\begin{equation}\label{solution12}
\bar A_\mu \rightarrow \bar A_\mu(\phi_\mu)~.
\end{equation}
To  write (\ref{solution12}) explicitly, it is convenient to switch to an index notation where $C', D',... = 1,..., k$ and $C,D,... = 1,... \ell$. 
 Then, recalling the expression (\ref{omega}) for the $U(1)$-generator $\omega$, we write the $k\times k$ components of the background (\ref{solution11}), with the $\phi_\mu$ moduli included:\footnote{To avoid confusion, we warn the reader that there can be different parameterizations of the moduli. The one used here  is from  \cite{Anber:2024uwl}  and differs from  the earlier \cite{Anber:2023sjn, Anber:2024mco}. The relation between the different parameterizations is given in eqn.~(\ref{phis}) and is important whenever translating moduli-dependent quantities between  different papers.}
\begin{eqnarray}
\label{Ukbackground}
\bar A_{1 \; C'D'}(\phi) &=& \delta_{C'D'} \; 2 \pi (  - \ell \; \phi_{1 \; C'}) \nonumber \\
\bar  A_{2 \; C'D'}(\phi) &=&\delta_{C'D'} \; 2 \pi ( -  \ell \phi_{2 \; C'}  - {\ell r    \over N k L_1 L_2} x_1)\\
\bar A_{3 \; C'D'}(\phi) &=&  \delta_{C'D'} \; 2 \pi(  - \ell \;\phi_{3 \; C'}) \nonumber\\
\bar A_{4 \; C'D'}(\phi) &=& \delta_{C'D'} \; 2 \pi ( -  \ell \;  \phi_{4 \; C'}    -{  1    \over N  L_3 L_4} x_3),~ C',D'=1,...,k~. \nonumber
\end{eqnarray}
Similarly, the $\ell \times\ell$ components are
\begin{eqnarray}
\label{Ulbackground}
\bar A_{1 \; CD}(\phi) &=&   \delta_{CD} \; 2 \pi( k \;  \tilde\phi_{1}) \nonumber \\ 
\bar A_{2 \; CD}(\phi) &=& \delta_{CD} \; 2 \pi ( k \;  \tilde\phi_{2} + {r \over N   L_1 L_2} x_1 )\\
\bar A_{3 \; CD}(\phi) &=&  \delta_{CD} \;  2 \pi( k\;  \tilde\phi_{3}) \nonumber\\
\bar A_{4 \; CD}(\phi) &=&   \delta_{CD}\; 2 \pi ( k \;   \tilde\phi_{4} + {k \over  N  \ell L_3 L_4} x_3),~C,D=1,...\ell.\nonumber
\end{eqnarray}
Here, we defined
\begin{equation}
\tilde\phi_{\mu}  \equiv  {1 \over k} \sum\limits_{C'=1}^k \phi_{\mu \; C'}~,
\end{equation} 
which ensures $SU(N)$ traceleness of (\ref{Ukbackground}, \ref{Ulbackground}). 
We note that in order to commute with the transition functions, the moduli $\phi_{\mu \; C'}$, $C'=1,...,k$, are not all independent  \cite{Anber:2023sjn}, but 
\begin{eqnarray} \label{SUNmoduli}
\phi_{\mu \; C'} &=& \phi_{\mu \; C'-r (\text{mod} \; k)} \equiv \phi_{\mu\; [C'-r]_k},  
\end{eqnarray}
where we introduced the notation 
\begin{equation}
[x]_k \equiv x (\text{mod} k),
\end{equation} and we note that $[k]_k$ can be taken to be either $0$ or $k$, depending on the whether the index $C'$ is taken to range from $0$ to $k-1$ or from $1$ to $k$.
That the $\phi_{\mu \; C'}$ obeying (\ref{SUNmoduli}) are the most general holonomies allowed by the transition functions was shown in \cite{Anber:2023sjn} (the readers can verify this themselves by considering the explicit form of $\Omega_\mu$ (\ref{transition1}) and demanding that the constant pieces commute with them). That, for each $\mu$, there are gcd$(k,r)$ independent $\phi_{\mu \; C'}$ follows from the fact that there are gcd$(k,r)$ different orbits of the  integers $C'=1,...k$, obtained by identifying $C'$ with $[C'-r]_k$, as per (\ref{SUNmoduli}).

In conclusion, as already stated, there are $4 \times$gcd$(k,r)$ independent $SU(N)$ holonomies $\phi_{\mu C'}$. For gcd$(k,r)=r$, this is exactly the number of bosonic moduli required by the index theorem for a charge-$r/N$ self-dual $SU(N)$ instanton.

Next, as described in the Introduction, we introduce a general fluctuation around the constant-$F$ background, already given in (\ref{solution2}), which we reproduce here with the index $\mu$ restored:\footnote{We recall, as discussed in the introduction, that the constant holonomies $\phi_{\mu \; C'}$ appearing in (\ref{Ukbackground}, \ref{Ulbackground}) can be absorbed into $\cS_\mu^{\omega, k, \ell}$.}
\begin{equation}\label{solution21}
A_\mu = \bar A_\mu + \omega \cS_\mu^{ \omega } +  \left(\begin{array}{cc} \cS_\mu^{ k } & \cW_\mu^{k \times \ell} \cr (\cW_\mu^{k \times \ell})^\dagger  & \cS_\mu^{ \ell } \end{array} \right).
\end{equation}

Note that $A_\mu$ has to obey the BCS (\ref{conditions on gauge field}), which in term determine the BCS on $\cS$ and $\cW$. Since they are important to our considerations, we list them explicitly, using the $k \times \ell$ index notation introduced in writing (\ref{Ukbackground}, \ref{Ulbackground}):
\begin{enumerate}
\item
 For $\cS_{\mu \; C'B'}$, with $C', B'$$=$$1,...,k$, combining $\cS_\mu^k$ and the first $k$ (diagonal) entries of $\omega \cS_\mu^\omega$ into a single $U(k)$ matrix using the index notation, they are
\begin{eqnarray}
\nonumber
\cS_{\mu \; C' B'}(x+L_1\hat e_1)&=&\cS_{\mu \; [C'-r]_k \; [B'-r]_k}(x)\,,\\
\nonumber
\cS_{\mu \;C' B'}(x+L_2\hat e_2)&=& e^{i2\pi\frac{C'-B'}{k}}\cS_{\mu \; C' B'}(x)\,,\\
\nonumber
\cS_{\mu \;C' B'}(x+L_3\hat e_3)&=&\cS_{\mu \; C' B'}(x)\,,\\
\cS_{\mu \;C' B'}(x+L_4\hat e_4)&=&\cS_{\mu \; C' B'}(x)\,. 
\label{BCS SC'B'}
\end{eqnarray}
\item
Similarly, $\cS_{\mu \; CB}$, with $C, B$$=$$1,...,\ell$, combines $\cS_\mu^\ell$ and the last $\ell$ (diagonal) entries of $\omega \cS_\mu^\omega$  into a single $U(\ell)$ matrix using the index notation, and obeys
\begin{eqnarray}
\nonumber
\cS_{\mu \; CB}(x+L_1\hat e_1)&=&\cS_{\mu \;C B}(x)\,,\\
\nonumber
\cS_{\mu \;CB}(x+L_2\hat e_2)&=&\cS_{\mu \;CB}(x)\,,\\
\nonumber
\cS_{\mu \;CB}(x+L_3\hat e_3)&=&\cS_{\mu \;[C+1]_\ell \; [B+1]_\ell}(x)\,,\\
\cS_{\mu \;CB}(x+L_4\hat e_4)&=& e^{i2\pi\frac{C-B}{\ell}}\; \cS_{\mu \;CB}(x)\,.\label{BCS SCB}
\end{eqnarray}
\item
Finally, for $\cW^{k \times \ell}_{\mu \; C'B}$ we have:
\begin{eqnarray}
\nonumber
\cW_{\mu \;C' B}(x+L_1\hat e_1)&=&\gamma_k^{-r}e^{i2\pi \frac{rx_2}{kL_2}} \;\cW_{\mu \;[C'-r]_k\;B}(x)\,,\\
\nonumber
\cW_{\mu \;C' B}(x+L_2\hat e_2)&=&\gamma_k e^{i2\pi\frac{(C'-1)}{k}} \;\cW_{\mu \;C'  B}(x)\,,\\
\nonumber
\cW_{\mu \;C' B}(x+L_3\hat e_3)&=&\gamma_\ell^{-1}e^{i2\pi \frac{x_4}{\ell L_4}}\; \cW_{\mu \;C' [B+1]_\ell}(x)\,,\\
\cW_{\mu \;C' B}(x+L_4\hat e_4)&=&\gamma_\ell^{-1} e^{-i2\pi\frac{(B-1)}{\ell}}\; \cW_{\mu \;C' B}(x)\,.
\label{BCS W}
\end{eqnarray}
Clearly, $(\cW^{k\times \ell}_{\mu})^\dagger_{C B'}$ obeys the conditions following from complex conjugating (\ref{BCS W}). 
\end{enumerate}
Our goal now is to study the space of self-dual configurations with $Q={r/N}$ in the neighborhood of the constant-$F$ solution of (\ref{solution11}),  parameterized by the general fluctuations of (\ref{solution21}). To this end, we impose the self-duality condition on  (\ref{solution21})  as well as  the background gauge condition, ensuring that the fluctuations are orthogonal to small gauge transformations:
\begin{equation} \label{gaugecondition}
D(\bar A)_\mu A_\mu = 0,
\end{equation} where $D(\bar A)$ is the adjoint-representation covariant derivative in the $\bar A_\mu$ background.
The self-duality condition is equivalent (see e.g. \cite{Dorey:2002ik}) to imposing the constraint on the field strength of (\ref{solution21}): \begin{eqnarray}
\bar \sigma_{\mu\nu}F_{\mu\nu}=0\,.\label{selfduality1}
\end{eqnarray}
where\footnote{ Here, $\sigma_\mu \equiv(i\vec\sigma,1)$, $\bar\sigma_\mu \equiv(-i\vec\sigma,1)$, $\vec \sigma$ are the Pauli matrices which determine the $\mu={1,2,3}$ components of the four-vectors $\sigma_\mu, \bar\sigma_\mu$.} 
  $\bar\sigma_{\mu\nu}=\frac{1}{2}(\bar\sigma_\mu\sigma_\nu-\bar\sigma_\nu\sigma_\mu)$.
  To write the self-duality condition, we use quaternionic notation: for every  four-vector ${\cal V}_\mu$, we define the quaternions  ${\cal V}\equiv \sigma_{\mu}{\cal V}_\mu$ and $\bar{\cal V}\equiv \bar\sigma_\mu{\cal V}_\mu$. Then,
  we compute the $SU(N)$ field strength of (\ref{solution21}), use the gauge condition (\ref{gaugecondition}), and obtain the self-duality conditions for each of the $k \times k$, $\ell \times \ell$ and $k \times \ell$ blocks of the $SU(N)$ matrix in (\ref{solution21}) as follows:\footnote{The definition of quaternions was already given in the paragraph after (\ref{selfduality1}).  We warn the reader, temporarily not denoting explicitly that these are $k \times \ell$ matrices, to keep in mind the difference between the quaternions, ${\cal W} \equiv {\cal W}_\mu \sigma_\mu$, $\bar{\cal W}= \bar\sigma_\mu \cW_\mu$, and the four-vector ${\cal W}_\mu$ and, furthermore, note that  ${\cal W}^\dagger = \sigma_\mu \cW^\dagger_\mu$ and $\bar{\cal W}^\dagger = \bar\sigma_\mu \cW^\dagger_\mu$. Here and below, the terms that have sums over $\mu$ should be multiplied by unit quaternion   $\sigma_4$, which we have omitted for brevity. }
  \begin{eqnarray}
\nonumber
&&2\pi\ell\bar\partial {\cal S}^\omega I_k+\bar\partial {\cal S}^k+\lambda\left\{-i\bar{\cal S}^k{\cal S}^k+i {\cal S}_\mu^k{\cal S}_\mu^k+i \bar {\cal W}^{k\times \ell}{\cal W}^{\dagger\ell\times k}-i{\cal W}_\mu^{k\times \ell}{\cal W}_\mu^{\dagger\ell\times k}\right\}=0\,,\\
\nonumber
&&-2\pi k\bar\partial {\cal S}^\omega I_\ell +\bar\partial {\cal S}^\ell+\lambda\{-i\bar{\cal S}^\ell{\cal S}^\ell+i {\cal S}_\mu^\ell{\cal S}_\mu^\ell+i \bar {\cal W}^{\dagger \ell\times k}{\cal W}^{k\times \ell}-i{\cal W}_\mu^{\dagger \ell\times k}{\cal W}_\mu^{  k\times \ell}\}=0\,,\\
\nonumber
&&\bar {D} {\cal W}^{k\times \ell}+\lambda\{i\bar{\cal S}^{k}{\cal W}^{k\times \ell}-i {\cal S}_\mu^k{\cal W}_\mu^{k\times \ell}+i\bar {\cal W}^{k\times \ell}{\cal S}^\ell-i{\cal W}_\mu^{k\times\ell} {\cal S}_\mu^\ell\\
&&+i 2\pi N\left(\bar {\cal S}^\omega{\cal W}^{k\times \ell}-{\cal S}_\mu^\omega {\cal W}_\mu^{k\times \ell}\right)\}=0\,,
\label{main equations we need to solve k eq r}  
\end{eqnarray}
Above, we introduced a parameter $\lambda$ multiplying the nonlinear terms, since our goal is to solve these equations using an expansion around the linearized equations (the ones with $\lambda=0$).   The background covariant derivative acting on $\cal{W}$ above is, explicitly, 
\begin{equation}\label{wderivative}
D_\mu {\cal{W}}_\nu^{k \times \ell} =(\partial_\mu + i 2 \pi N \bar A_\mu^\omega) {\cal{W}}_\nu^{k \times \ell},\end{equation}  to be contracted with the appropriate quaternions. Here $\bar A_\mu^\omega$  is the component of $\bar A_\mu$ (\ref{solution11}) along $\omega$, i.e. $\bar A_\mu = \omega \bar A_\mu^\omega$. 

Eqn.~(\ref{main equations we need to solve k eq r}) is obtained by setting $\Delta=0$ in eqn.~(4.15) in \cite{Anber:2023sjn}, where more details of the straightforward but somewhat tedious derivation can be found. On the tuned $\T^4$, the meaning of eqn.~(\ref{main equations we need to solve k eq r}) is that its solutions, subject to the BCS (\ref{BCS SC'B'}, \ref{BCS SCB}, \ref{BCS W}) span the manifold of self-dual configurations of charge $Q=r/N$ on the tuned $\T^4$. 
Below, we shall study the manifold of such self-dual configurations using a
  perturbation series in the nonlinearity $\lambda$. 
  
  To this end, we assume that a perturbative series of ${\cal S}$ and ${\cal W}$ exists such that
\begin{eqnarray}
\nonumber
{\cal W}^{k\times \ell}&=& \sum_{a=0}^\infty \lambda^a {\cal W}^{(a)k\times \ell}\, ,\\
{\cal S}&=&  \sum_{a=0}^\infty \lambda^a{\cal S}^{(a)}\,, \label{expansion1 r eq k}
\end{eqnarray}
where ${\cal S}$ accounts for ${\cal S}^\omega$, ${\cal S}^k$, and ${\cal S}^\ell$.   Expanding to ${\cal O}(\lambda^0)$, we find that the linearized self-dual fluctuations around the constant instanton obey the equations:\footnote{Eqns.~(\ref{orderlambdazero})  are obeyed by linearized self-dual fluctuations  around the instanton, subject to the background gauge condition (\ref{gaugecondition}). These   equations are related to the adjoint-fermion Dirac equation in the instanton background, see \cite{Anber:2023sjn}. Recall that the dimension of the moduli space of self-dual solutions is related to (twice) the index of the adjoint Dirac operator, see   \cite{Schwarz:1977az,Bernard:1977nr,Brown:1977bj,Weinberg:1979ma}  and the reviews \cite{Vandoren:2008xg,Dorey:2002ik}. }
\begin{eqnarray}\label{orderlambdazero}
{\cal O}(\lambda^0): ~\bar D{\cal W}^{(0) k\times\ell}=0\,, \quad 2\pi\ell\bar\partial {\cal S}^{(0)\omega} I_k+\bar\partial {\cal S}^{(0)k}=0\,,\quad -2\pi k\bar\partial {\cal S}^{(0)\omega} I_\ell +\bar\partial {\cal S}^{(0)\ell}=0\,, \nonumber \\
\end{eqnarray}
where the solutions must be endowed with the appropriate BCS from  (\ref{BCS SC'B'}, \ref{BCS SCB}, \ref{BCS W}). These leading-order equations were solved in \cite{Anber:2023sjn} and, in what follows, we shall make use of the solutions given there, and reproduced in Appendix \ref{appx:phip}. We now consider separately the cases gcd$(k,r)=r$ and gcd$(k,r) \ne r$.

\section{Moduli space of $Q$=${r \over N}$  instantons on the tuned $\T^4$ with  gcd$(k,r)=r$}

\label{sec:tunedgcd=r}
  
  We begin with the tuned $\T^4$, with $\Delta(k,r,\ell)=0$ (\ref{deltadef}) and gcd$(k,r)=r$. Here, we shall show that the equations defining the moduli space of $Q=r/N$ self-dual instantons (\ref{main equations we need to solve k eq r}) only admit constant-$F$ solutions. In section   \ref{sec:analyticgcd=r}, we    show this analytically, to all orders in the $\lambda$-expansion we set up above. Then, in section \ref{sec:numericgcd=r}, we  give numerical evidence, for $N=3$, $k=r=2$ and $k=r=1$ that, indeed, only constant field strength fractional instanton solutions exist. We also show that the numerical minimization of the action, starting from random configurations, covers the entire moduli space of the constant-$F$ fractional instanton solutions  known from analytic studies.
  
  \subsection{Analytic study of the moduli space for gcd$(k,r)=r$}
  
  \label{sec:analyticgcd=r}

{\flushleft{\bf {The ${\cal{O}}(\lambda^0)$ equations: }}}

 The general solution  of  the ${\cal{O}}(\lambda^0)$ equation in (\ref{expansion1 r eq k}) for ${\cal W}^{(0) k\times\ell}$  was found in \cite{Anber:2023sjn}. It is given  in Appendix \ref{appx:phip} and expressed  in terms of the  functions $\Phi^{(p)}$ defined there. Noting that for gcd$(k,r)=r$  only $p=0$ is relevant, we find:
\begin{eqnarray}\nonumber
{\cal W}^{(0) k\times\ell}_{2 C'C}(x)&=&i{\cal W}_{1 C'C}^{(0)k\times \ell}(x)=V^{-1/4}{\cal C}_2^{C'}\Phi^{(0)}_{C'C}(x)=: {W}_{2 \; C'C}\,,\\
 {\cal W}^{(0) k\times\ell}_{4 C'C}(x)&=&i{\cal W}_{3 C'C}^{(0)k\times \ell}(x)=V^{-1/4}{\cal C}_4^{C'}\Phi^{(0)}_{C'C}(x)=: {W}_{4 \; C'C}\,,
\label{Wsolution11} \end{eqnarray}
where  ${\cal{C}}_{2 (4)}^{C'}$ are yet-to-be-determined complex coefficients, not fixed to the leading order $\lambda^{0}$.
The boundary conditions for ${\cal S}_{CB}$ of eqn.~(\ref{BCS SCB}) imply that ${\cal S}^{(0)\ell}=0$, since $\cS_{CB}$ can only be proportional to the unit matrix. Clearly, then, $\cS_\mu^\omega$ as well as $\cS_\mu^k$ are constant, and, in addition, $\cS_\mu^k$ can only be diagonal due to the boundary conditions (\ref{BCS SC'B'}). Then, the second equation in (\ref{orderlambdazero}) implies that  the solutions  of ${\cal S}^{(0)k}_\mu$ and ${\cal S}^{(0)\omega}_\mu$ can be combined and are given by
\begin{eqnarray} \label{szero}
\left({\cal S}^{(0)k}_\mu+2\pi N{\cal S}^{(0)\omega}_\mu\right)_{C'D'}=\phi_{\mu C'}\delta_{C'D'}\,.
\end{eqnarray}

{\flushleft{\bf The next, ${\cal O}(\lambda)$, equation for $\cS$: }}

The ${\cal O}(\lambda)$ equation for $\cS$ in (\ref{main equations we need to solve k eq r}),   is\footnote{Here, for example $(W_{2} W^*_{4})_{C'B'} =\sum\limits_{C=1}^\ell  W_{2 \; C' C}  W^*_{4 \; B' C}$. We also note that no new constraints follow from the equation for $\cS^{(1) \ell}$, the second equation in (\ref{main equations we need to solve k eq r}).}
\begin{eqnarray}\label{equationforSk k eq r}
&&\left(2\pi \ell \bar\partial {\cal S}^{(1)\omega}I_k+\bar\partial {\cal S}^{(1)k}-i\bar {\cal S}^{(0) k} {\cal S}^{(0) k}+i {\cal S}^{(0) k}_\mu  {\cal S}^{(0) k}_\mu\right)_{C'B'}= \\\nonumber
&&i \left(\begin{array}{cc} - 2 \;(W_{2} W^*_2 -W_{4 } W^*_{4})_{C'B'} &
 4 \;(W_{2} W^*_{4})_{C'B'}  \cr
 4 \;(W_{4} W^*_{2})_{C'B'}   
  & + 2\; (W_{2} W^*_{2 } -W_{4 } W^*_{4})_{C'B'}   \end{array}\right)\,,~C',B'=1,..,k\,. \nonumber
\end{eqnarray}
This is a set of $k \times k$ quaternion equations and their consistency can be used to determine ${\cal C}_2^{C'}$ and ${\cal C}_4^{C'}$ as follows. We simply integrate the diagonal components of the $k\times k$ matrix (taking $B'=C'$) of the quaternion equation on $\mathbb T^4$. Noticing that $\bar {\cal S}^{(0) k} {\cal S}^{(0) k}= {\cal S}^{(0) k}_\mu  {\cal S}^{(0) k}_\mu$, and that the diagonal components of the matrix ${\cal S}_\mu^{(k)}$, for $r=k$,\footnote{\label{footnote:ktwor}For gcd$(k,r)=r$ but with $k \ne r$, e.g. for $k=2r$, the argument has to be modified, since $S_{\mu \; C'C'}$ are now not all periodic, as per the $x_1$ BC in (\ref{BCS SC'B'}). This and the general gcd$(k,r)$ case are treated in section 5 in \cite{Anber:2023sjn} (replacing the r.h.s. of equations (5.3) there with zero, to account for the fact that $\Delta=0$, leads to the identical conclusion that $\cW_\mu^{(0) \; k \times \ell}=0$ as well).}
 obey periodic BCS as per (\ref{BCS SC'B'}), we readily find:
\begin{eqnarray}\label{the most impt cond}
0&=& \int_{\mathbb T^4}\left(2\pi \ell \bar\partial {\cal S}^{(1)\omega}I_k+\bar\partial {\cal S}^{(1)k}\right)_{C'C'}\\
&=&\left( \begin{array}{cc} -2(|{\cal C}_2^{C'}|^2-|{\cal C}_4^{C'}|^2)& 4{\cal C}_2^{C'}{\cal C}_4^{*C'}\\ 4{\cal C}_2^{C'}{\cal C}_4^{*C'} &  2(|{\cal C}_2^{C'}|^2-|{\cal C}_4^{C'}|^2)  \end{array}\right) V^{-{1\over 2}}\int_{\mathbb T^4} \sum\limits_{C=1}^\ell \Phi^{(0)}_{C'C}(x) \Phi^{(0) *}_{C'C}(x)\,, \; \; C'=1,...k. \nonumber
\end{eqnarray}
Noting that by (\ref{property1}) that the integral (of $\Phi \Phi^*$) is nonzero, this fixes ${\cal C}_2^{C'}={\cal C}_4^{C'}=0$ for all $C'$. Thus, we have ${\cal W}_\mu^{(0)k\times \ell}=0$. 
Then, the remainder of the ${\cal{O}}(\lambda)$ equation for $\cS^{(1)}$ in (\ref{equationforSk k eq r}) gives
\begin{eqnarray} \label{sone}
2\pi \ell \bar\partial {\cal S}^{(1)\omega}I_k+\bar\partial {\cal S}^{(1)k}=0\,.
\end{eqnarray}
The solution of this equation, again, is a constant, which can be set to zero without loss of generality since any constant solution can be incorporated into ${\cal S}^{(0)k}$, as per (\ref{szero}).

Having solved  (\ref{main equations we need to solve k eq r}) to ${\cal{O}}(\lambda^0)$ for $\cW$ and to ${\cal{O}}(\lambda)$ for $\cS$, we have determined that the solutions  have the form: 
\begin{eqnarray}\nonumber
{\cal W}_\mu^{k\times \ell}(x)&=&0+\lambda {\cal W}_\mu^{(1) k\times \ell}(x)+\lambda^2 {\cal W}_\mu^{(2) k\times \ell}(x)\\
{\cal S}_\mu(x)&=&{\cal S}_\mu^{(0)}+0+\lambda^2 {\cal S}_\mu^{(2)}(x)\,, \label{expansion1}
\end{eqnarray}
with ${\cal S}_\mu^{(0)}$ from (\ref{szero}).

 {\flushleft{\bf Higher orders, ${\cal{O}}(\lambda^a)$, $a \ge 1$:}}
 
 We now continue and 
substitute (\ref{expansion1}) in   (\ref{main equations we need to solve k eq r}) giving  the equations to ${\cal{O}}(\lambda)$ for $\cW$ and to ${\cal{O}}(\lambda^2)$ for $\cS$ as follows:
\begin{eqnarray}\label{stwo}
\bar {\hat D} {\cal W}^{(1)k\times \ell}=0\,, \quad 2\pi\ell\bar\partial {\cal S}^{(2)\omega} I_k+\bar\partial {\cal S}^{(2)k}=0\,.
\end{eqnarray}
The solution of the second equation, to  ${\cal{O}}(\lambda^2)$ for $\cS^{(2)}$,  is a constant, which, again, can be absorbed in ${\cal S}^{(0)}$. Most importantly, we have found that, 
  due to (\ref{expansion1}), the r.h.s. of the ${\cal{O}}(\lambda)$ equation for $\cW$ vanishes. 
  Thus, the equation for $\cW^{(1) \; k \times \ell}$ has a solution  identical to the one for ${\cal W}^{(0)\; k\times\ell}$ found in (\ref{Wsolution11}), with new arbitrary constants ${\cal C}_{2,4}^{C'}$. Now, one can invoke the  ${\cal O}(\lambda^3)$  equation for $\cS$  in (\ref{main equations we need to solve k eq r}) to find that both of these constants must be set to zero. Then, one finds that $\cS^{(3)}$ obeys the same equation as $\cS^{(2)}$ (\ref{stwo}) (and $\cS^{(1)}$ (\ref{sone})), giving only the constant solution readily absorbed into (\ref{szero}).

{\flushleft{W}}e assure the reader that this inductive procedure continues to all orders in the nonlinearity $\lambda$. One finds that at each order in $\lambda$, the equation for $\cW^{(a)}$ has vanishing r.h.s., determined upon solving the lower-order equations, which set $\cW^{(0)}= \cW^{(1)}= ... =\cW^{(a-1)}=0$ (this is sufficient for the vanishing of the r.h.s. of the $\bar D {\cW^{(a) \; k \times \ell}}$ equation in (\ref{main equations we need to solve k eq r})). After this, the equation for $\cS^{(a+1)}$ sets the lower-order $\cW^{(a) \; k \times \ell} = 0$ and gives only the constant $\cS^{(a+1)}$ solution, absorbable into eqn.~(\ref{szero}) (explicitly presenting this straightforward inductive argument, however, is increasingly messy and we skip it for brevity).

{\flushleft{In}} conclusion,  induction shows that the compact $4$gcd$(k, r) = 4 r$ moduli $\phi_\mu^{C'}$ obeying (\ref{SUNmoduli}) are the only self-dual  perturbations  on the top of the abelian solution. We now show that this conclusion is corroborated by the ``exact" numerical solution.
                \begin{figure}[h] 
   \centering
   \includegraphics[width=6in]{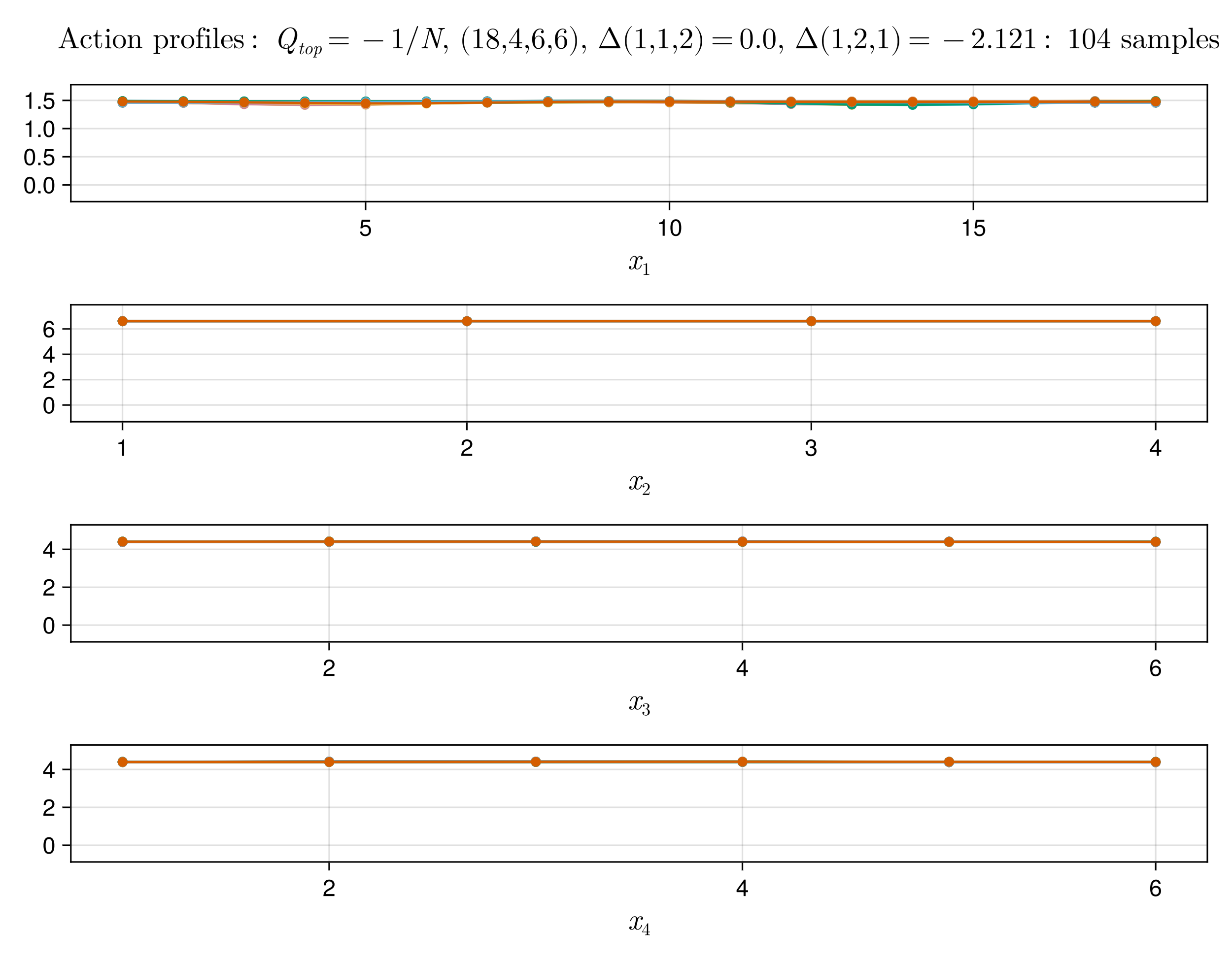} 
   \caption{Action density of the $|Q|=1/3$ self-dual solution with $r=k=1$. On the four plots, we show the action density, in each case  integrated over all but one $\T^4$ coordinate. The lattice size is $L_1, L_2, L_3, L_4 = 18, 4, 6, 6$, corresponding to a tuned $\T^4$ with $\Delta(1,1,2)=0$. Thus, multiplying the integrated action density by the appropriate $L_\mu$ in each case yields   $8 \pi^2 \times {1 \over 3} \simeq 26.3$ (this number can be verified simply by a look at the above plot). We generate $104$  minimum-action configurations starting from random initial conditions. We display the action density of all the $104$ configurations plotted on top of each other. The action densities of all these configurations coincide within numerical error. }
  \label{fig:r=k=1action}
\end{figure}

  \subsection{Numerical study of the moduli space for gcd$(k,r)=r$}
  
\label{sec:numericgcd=r}

  The fractional instantons can be also constructed on the lattice, by minimizing the lattice action with appropriate twists $n_{12}$ and $n_{34}$, by inserting two intersecting center vortices in the lattice action (thus twisting the boundary conditions on the lattice). Here, we generalize the recent $SU(2)$ study of \cite{Wandler:2024hsq} to the case of $SU(3)$. The lattice setup and the methods used are described in Appendix \ref{appx:lattice}.
  
      \begin{figure}[h] 
   \centering
   \includegraphics[width=5in]{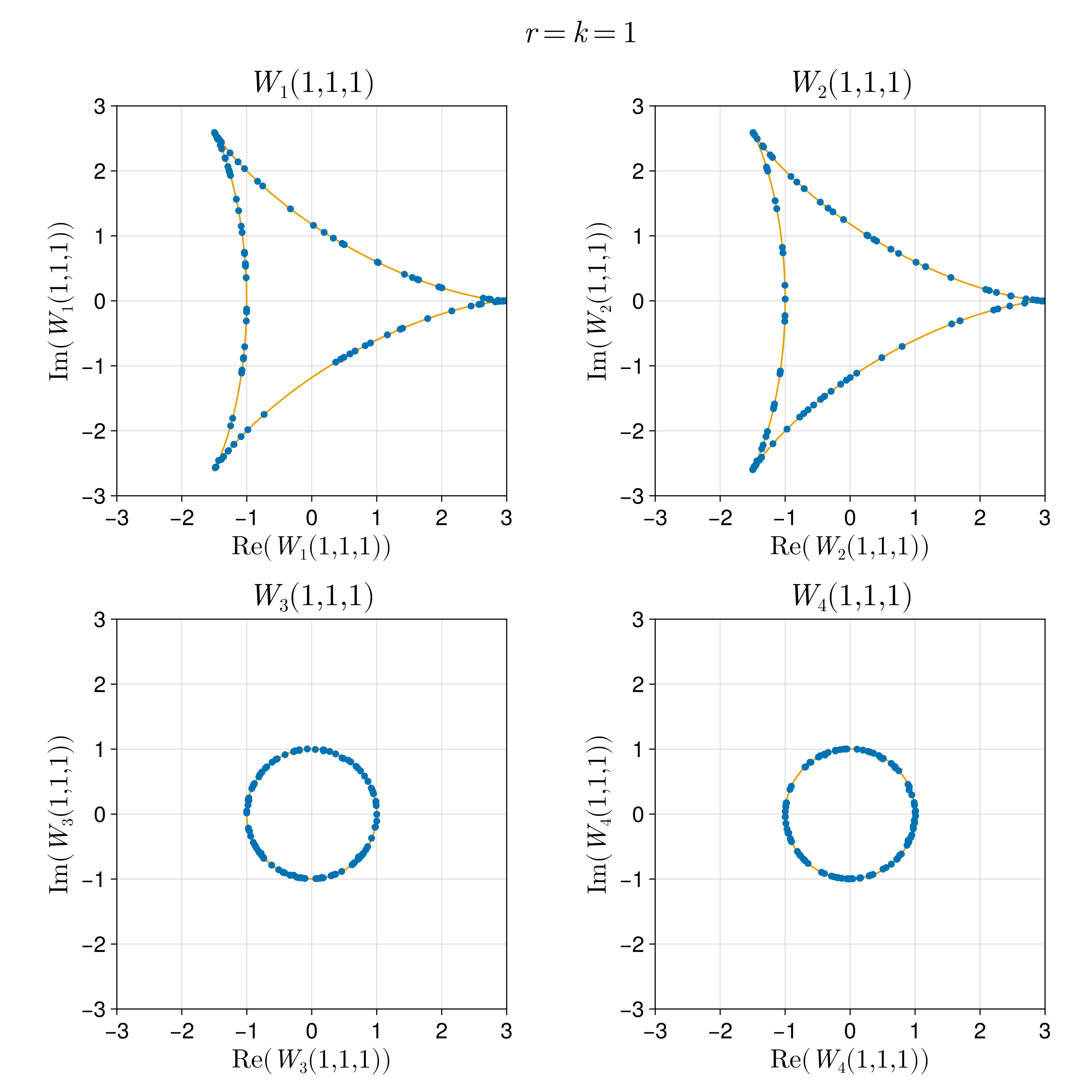} 
   \caption{The imaginary vs. the real parts of the fundamental representation Wilson loops  winding once in $x_\mu=1,2,3,4$, evaluated at fixed values of the coordinates, for the $|Q|=1/3$ self-dual solution on the tuned-$\T^4$ with $r=k=1$ (whose action density is shown on Figure \ref{fig:r=k=1action}). The solid lines show the analytical curves of eqn.~(\ref{wilsoneqns1}) that the real and imaginary part of the winding Wilson loop $W_\mu$ trace  as the translational modulus $\phi_\mu^{[C']_1}$  is varied. The $104$ configurations generated randomly, each denoted by a dot, appear to cover the moduli space of the constant-$F$ solutions.}
   \label{fig:r=k=1wilson}
\end{figure}

\begin{figure}[h] 
   \centering
   \includegraphics[width=5.3in]{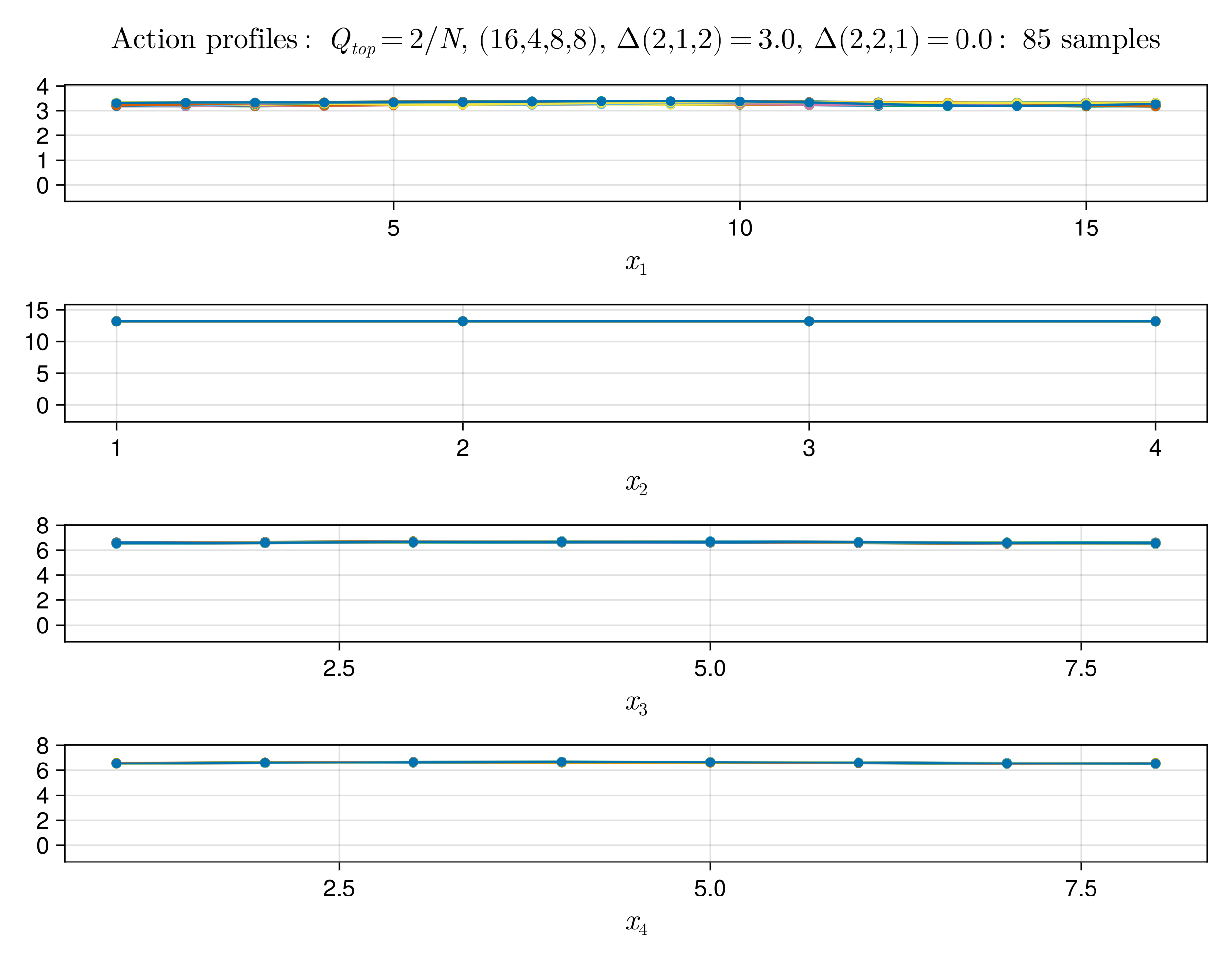} 
   \caption{Action density of the $Q=2/3$ self-dual solution with $r=k=2$. The lattice size is $L_1, L_2, L_3, L_4 = 16, 4, 8, 8$, corresponding to a tuned $\T^4$ with $\Delta(2,2,1)=0$, eqn.~(\ref{deltadef}). In each case, we show the action density integrated over all coordinates but the one shown. We show the action densities for all $85$ configurations generated, plotted on top of each other. }
   \label{fig:r=k=2action}
\end{figure}

  In this section, we present our results for the lattice minimum action configurations   with both $Q=-{1 \over 3}$ and $Q={2\over 3}$ for the tuned $\T^4$.\footnote{To avoid confusion, we note that our $SU(3)$ lattice simulations use the twists $n_{12}=1$ and $n_{34}=1$, corresponding to topological charge $Q=-{1\over 3} \; (\text{mod} \; 1)$. Thus the minimum action lattice configurations have topological charges either $-1/3$ or $2/3$; see Appendix~\ref{appx:lattice}.} Thus, we take $ 2 L_3 L_4 = L_1 L_2$ for $Q = -{1\over 3}$, such that  $\Delta(1,1,2)=0$, by (\ref{deltadef}). Likewise, we take $L_3 L_4 = L_1 L_2$ for $Q={2 \over 3}$, such that $\Delta(2,2,1)=0$. In both cases, we find that, starting the cooling algorithm from arbitrary configurations, the minimum action configurations always correspond to a constant-$F$ solution.  Two representative cases (two particular choices of $\T^4$ sides $L_\mu$, appropriately tuned) are shown on Figure~\ref{fig:r=k=1action}  and Figure~\ref{fig:r=k=2action}, for $Q=-{1\over 3}$ and $Q={2\over 3}$, respectively.

In order to further corroborate our findings, we also study the moduli space of these constant-$F$ solutions and show that the  lattice minimum action configurations  generated from random initial conditions cover the entire moduli space (subject to the limitation that we only generate a finite number of configurations).

We begin with $k=r=1$, where there are only $4$ moduli, the holonomies $\phi_\mu^{C'=1}$. As for any solution with $Q=1/N$, these moduli correspond to translations of the solution on $\T^4$. The action density is independent of the moduli, but the winding Wilson loops depend on them. A Wilson loop winding in $x_\mu$ only depends on $\phi_\mu^1$, as is clear from the explicit form of the constant-$F$ solutions (\ref{Ukbackground}, \ref{Ulbackground}) and the relation (\ref{SUNmoduli}). The Wilson loops, as functions of $\phi_\mu^1$ (for   fixed $x_\mu$)  were studied in detail in \cite{Anber:2024mco} and were an important tool in figuring out the shape and size of the moduli space. As the modulus $\phi_\mu^1$, is varied, the Wilson loop in the $\mu$-direction traces out the solid curves shown in the $\Re W_\mu$-$\Im W_\mu$ plane on Figure~\ref{fig:r=k=1wilson}. The continuous lines are obtained from the analytic solution (\ref{Ukbackground}, \ref{Ulbackground}) and their  equations in each  $\Re W_\mu$-$\Im W_\mu$ plane are shown in eqn.~(\ref{wilsoneqns1}) in Appendix \ref{appx:wilson}.
From the figure, it is clear that the lattice minimum-action configurations (recall that there are $104$ of them) cover the entire moduli space.\footnote{To avoid confusion, we stress that the vanishing of the expectation values of winding Wilson loops is a consequence of center symmetry. However, in the semiclassical approximation, any given instanton configuration, with some fixed values of the moduli, leads to a nonzero expectation value of the winding loops. Analytically, only after integration over all moduli, i.e.~over the moduli space, does center symmetry get restored. The fact that, when summed over our randomly generated numerical instanton configurations, the Wilson loops vanish provides evidence that our numerical procedure covers the entire moduli space.}

\begin{figure}[h] 
    \centering
     \includegraphics[width=5in]{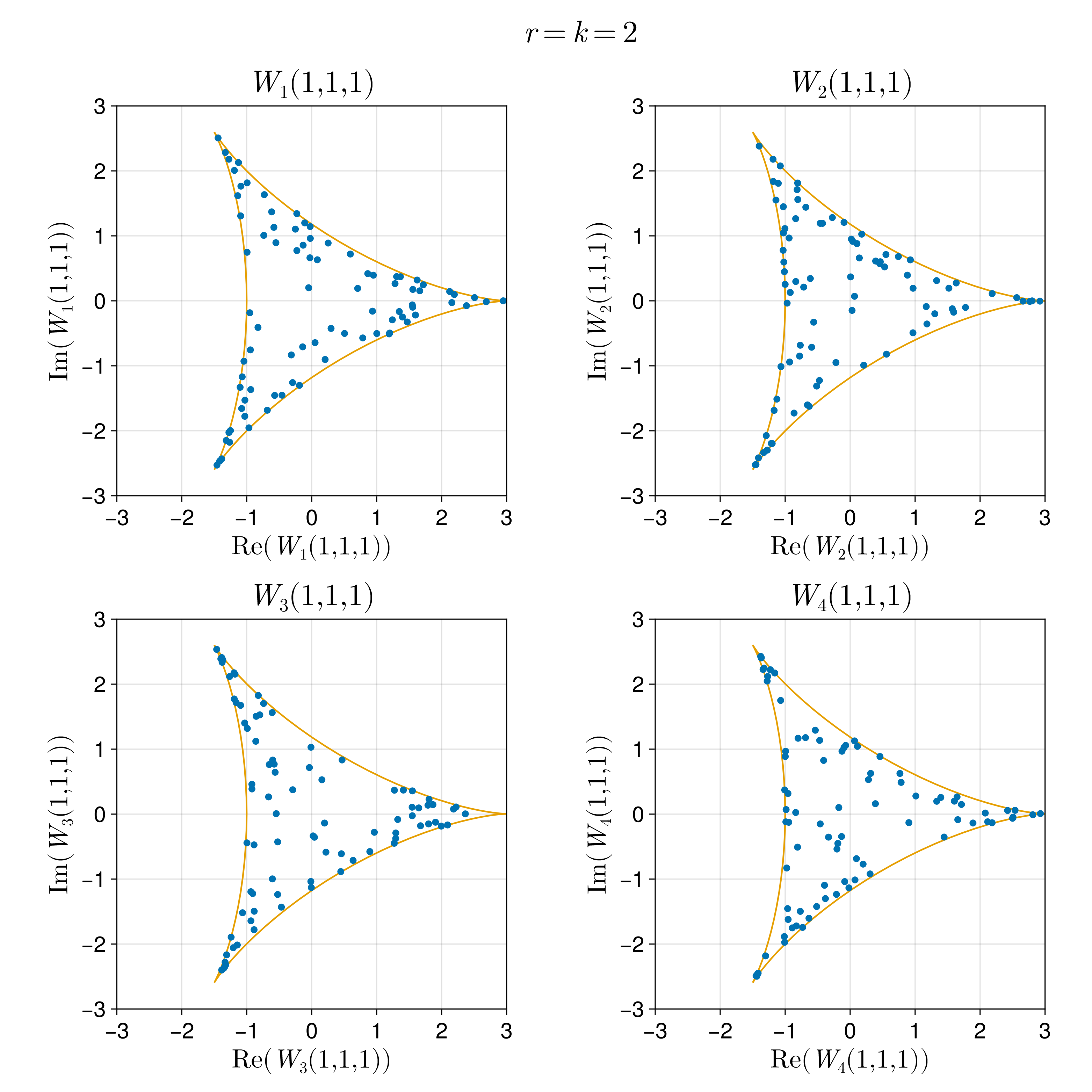} 
    \caption{The imaginary vs. the real parts of the fundamental representation Wilson loops  winding once in $x_\mu=1,2,3,4$, evaluated at fixed values of the coordinates, for the $Q=2/3$ self-dual solution with $\Delta(2,2,1)=0$, eqn.~(\ref{deltadef}), with $r=k=2$ (with action density  of Figure \ref{fig:r=k=2action}). The  analytical results of eqn.~(\ref{wilsonr=k=2}) as the $8$ moduli are varied fill the inside of the smooth curves shown (the boundary of the region corresponds to varying the translational moduli only). As in Figure~\ref{fig:r=k=1wilson}, the $85$ configurations appear to cover the moduli space of the constant-$F$ solutions.}
   \label{fig:r=k=2wilson}
 \end{figure}

For the multi-fractional instanton with $r=k=2$, there are now $8$ moduli, $\phi_\mu^{1}$ and $\phi_\mu^2$. Four linear combinations of them can be interpreted as translations of the center of mass of the $r=2$ ``composite instanton" and four can be interpreted as relative positions of the $r=2$ lumps,\footnote{Strictly,  the $2$-lump picture applies only on the detuned torus, as there are no lumps for $\Delta=0$.} as per \cite{Anber:2024mco}. Here again, as (\ref{Ukbackground}, \ref{Ulbackground}) and (\ref{SUNmoduli}) show, each winding Wilson loop $W_\mu$ only depends on $2$ moduli, $\phi_\mu^{C'=1}$ and $\phi_\mu^{C'=2}$. As these moduli are varied, the analytic expressions for $W_\mu$ from (\ref{wilsonr=k=2}) show that the  solid curves on Figure~\ref{fig:r=k=2wilson} represent the boundary of the moduli space (traced by varying the center of mass coordinate only, at vanishing relative position) with the inside now filled by varying the ``relative position'' modulus.  
Again, we find that the lattice minimum-action configurations appear to fill the moduli space.

{\bf{ \flushleft{Conclusion}} of section \ref{sec:tunedgcd=r}:} We have presented ample analytic and numerical evidence that for gcd$(r,k)=r$, the constant-$F$ solutions of 
't Hooft on the tuned $\T^4$ with $\Delta(r,k,\ell)=0$, eqn.~(\ref{deltadef}), are the only fractional instantons with $Q={r \over N}$. The gcd$(r,k)=r$ case is singled out by the fact that the number of constant holonomies commuting with the transition functions equals $4 r$ and  equals the dimensionality of the moduli space of $Q={r \over N}$ fractional instantons determined by the index theorem. No new undetermined moduli appear in either the $\lambda$-expansion on the tuned-$\T^4$, as we showed above in section \ref{sec:analyticgcd=r}, or in the $\Delta$-expansion on the detuned $\T^4$ as we showed in \cite{Anber:2023sjn}. 
This is in stark contrast with the constant-$F$ solutions on the tuned-$\T^4$ with gcd$(k,r)\ne r$, which we study next.

  \section{Moduli space of $Q$=${r \over N}$   instantons on the tuned $\T^4$ with gcd$(k,r) \ne r$}
  
  \label{sec:tunedgcdnotr}
  
   We  now turn to the tuned $\T^4$ with $\Delta(k,r,\ell)=0$, eqn.~(\ref{deltadef}), but gcd$(k,r) \ne r$. Here, in contrast with section~\ref{sec:tunedgcd=r}, we argue that among the solutions of the equations  (\ref{main equations we need to solve k eq r})  defining the moduli space of $Q=r/N$ self-dual instantons, the constant-$F$ solutions (\ref{Ukbackground}, \ref{Ulbackground}) are a set of measure zero. 
   
   In section   \ref{sec:analyticgcdnotr}, we    show this analytically, to leading order in the $\lambda$-expansion we set up in the beginning of section \ref{sec:tunedgeneral}. Then, in section \ref{sec:numericgcdnotr}, we  give numerical evidence, for $N=3$, $r=2$, $k=1$  that, starting from a random configuration,  constant field strength fractional instanton solutions never appear in our numerical minimization of the action for a tuned-$\T^4$ with $\Delta(2,1,1)=0$.\footnote{Numerically, we are only able to find constant-$F$ backgrounds by starting the minimization algorithm with a lattice configuration determined by discretizing the analytic constant-$F$ background (\ref{Ukbackground}, \ref{Ulbackground}).}
   
  That our findings make sense follows from the fact that with gcd$(k,r)\ne r$, the number of constant holonomies of the constant-$F$ solution is only $4$ gcd$(k,r)$, which is strictly smaller than $4 r$, the number required by the index theorem. As we show, by studying the  leading nontrivial order of the $\lambda$-expansion, we find that precisely the missing number of $4 r - 4$ gcd$(k,r)$ new moduli appear. Unfortunately, the $\lambda$-expansion is (in contrast to the all-orders 
  inductive argument of section \ref{sec:analyticgcd=r}) difficult to pursue beyond the leading order; hence determining the global structure of the moduli space in this case remains beyond our scope here. 
      
  Further below, in section \ref{appx:su2} we also give a general argument for the case of $Q=r/2$ instantons for $SU(2)$ 
   on the tuned $\T^4$, for any natural $r>1$, determining the number of moduli.
   
  \subsection{Leading-order analytic study of the moduli space for gcd$(k,r)\ne r$}
  \label{sec:analyticgcdnotr}  
  
  Our discussion follows the steps we used in section~\ref{sec:analyticgcd=r}. We begin with  (\ref{orderlambdazero}) and consider
  
 {\flushleft{\bf The ${\cal{O}}(\lambda^0)$ equations}}:
 
 The  equations for  $\cS$ in (\ref{orderlambdazero}) clearly imply that $\cS^{(0) \; \omega, k, \ell}$ are all constant and the boundary conditions determine $\cS^{(0) \ell}=0$, while the rest is given in terms of the $4$gcd$(k,r)$ allowed holonomies $\phi_{\mu \; C'}$ obeying (\ref{SUNmoduli}), exactly as in eqn.~(\ref{szero}).  
 
On the other hand, the solution of the equation for $\cW^{(0)\; k \times \ell}$ in (\ref{orderlambdazero}),   derived in \cite{Anber:2023sjn},  involves a larger number, $r/{\rm gcd}(k,r)$, of functions $\Phi^{(p)}$ (\ref{form of Phi}) and, as per
eqn.~(\ref{expressions of W2 and W4 with holonomies}), reads:
\begin{eqnarray}
\nonumber
\left({\cal W}_2^{(0)k\times \ell}\right)_{C'C}=V^{-1/4} \sum_{p=0}^{\scriptsize \frac{r}{\mbox{gcd}(k,r)}-1}{\cal C}^{[C'+pk]_r}_2\Phi^{(p)}_{C'C}(x,\hat\phi) =: {W}_{2 \; C'C}\,,\\
\left({\cal W}_4^{(0) k\times \ell}\right)_{C'C}=V^{-1/4} \sum_{p=0}^{\scriptsize \frac{r}{\mbox{gcd}(k,r)}-1}{\cal C}^{[C'+pk]_r}_4\Phi^{(p)}_{C'C}(x,\hat\phi)  =:{ W}_{4 \; C'C}\,,
\label{expressions of W2 and W4 gcdnotr}
\end{eqnarray}
As in section \ref{sec:analyticgcd=r}, we still have that, as per (\ref{assertion1}), $
{\cal W}_4^{(0)k\times \ell}=i{\cal W}_3^{(0)k\times \ell}$, ${\cal W}_2^{(0)k\times \ell}=i{\cal W}_1^{(0)k\times \ell}$.
The number of undetermined complex constants ${\cal{C}}_{2,4}$ is still $2 r$, equal to the number of values $[C' + pk]_r$  can take for $C'$ taking $k$ integer values and  $p$ taking $r/{\rm gcd}(k,r)$ ones.
We now continue to

{\flushleft{\bf The next, ${\cal O}(\lambda)$, equation for $\cS$: }}

This is the same as (\ref{equationforSk k eq r}), which we reproduce for completeness
\begin{eqnarray}\label{equationforSk k not r}
&&\left(2\pi \ell \bar\partial {\cal S}^{(1)\omega}I_k+\bar\partial {\cal S}^{(1)k}-i\bar {\cal S}^{(0) k} {\cal S}^{(0) k}+i {\cal S}^{(0) k}_\mu  {\cal S}^{(0) k}_\mu\right)_{C'B'}= \\\nonumber
&&i \left(\begin{array}{cc} - 2 \;(W_{2} W^*_2 -W_{4 } W^*_{4})_{C'B'} &
 4 \;(W_{2} W^*_{4})_{C'B'}  \cr
 4 \;(W_{4} W^*_{2})_{C'B'}   
  & + 2\; (W_{2} W^*_{2 } -W_{4 } W^*_{4})_{C'B'}   \end{array}\right)\,,~C',B'=1,..,k\,, \nonumber
\end{eqnarray}
however, we now note that we have to use ${W}_{2, 4 \; C'C}$ from (\ref{expressions of W2 and W4 gcdnotr}). To boot, the analysis is more complicated than the one for $k=r$ from section~\ref{sec:analyticgcd=r}, for the same reason mentioned in footnote \ref{footnote:ktwor}. 

Thus, to continue, we  now borrow the results from Section 5 of \cite{Anber:2023sjn}. 
What is done there is easy to describe (but the details are a bit messy): one  plugs the solutions (\ref{expressions of W2 and W4 gcdnotr}) into the r.h.s. of eqn.~(\ref{equationforSk k not r}) and demands that it be orthogonal to the zero modes of the adjoint of the $\bar\partial$ operator appearing on the l.h.s.~of (\ref{equationforSk k not r}).\footnote{\label{foot1}Consider the equation $\bar\partial X = F$. Take the inner product of both sides with $\zeta$, a zero mode of the adjoint of $\bar\partial$,  and, integrating by parts on the l.h.s.,   find that the inner product $(\zeta, F)=0$.}  One  thus 
 finds that the consistency conditions imposed by (\ref{equationforSk k not r}) on the $2 r$ complex coefficients ${\cal{C}}_{2,4}^{[C'+ pk]_r}$, denoted below by ${\cal{C}}_{2,4}^{A}$, $A=1,...r$, are as follows:\footnote{Eqn.~(\ref{constraintssubbed}) is eqn.~(5.3) in \cite{Anber:2023sjn} with the r.h.s. set to zero to account for the fact that $\Delta=0$.}
\begin{eqnarray}\label{constraintssubbed}
\sum\limits_{A_j \in S_j} {\cal C}_2^{A_j} \;{\cal C}_2^{* \; A_j} - {\cal C}_4^{A_j} \;{\cal C}_4^{* \; A_j} &=&0  \;\; - \text{there are gcd$(k,r)$ sets of indices $S_j$, see (\ref{sjdef})}, \nonumber 
 \\
\sum\limits_{A_j \in S_j} {\cal C}_2^{ A_j}\; {\cal C}_4^{* \; A_j} &=&0~,
\end{eqnarray}
where $S_j$ are  ${\rm gcd}(k,r)$ sets of $r \over {\rm gcd}(k,r)$ integers taking values in $\in \{0,...,r-1\}$. These are defined by
\begin{equation}\label{sjdef}
S_j = \bigg\{ [[j+nr]_k+ pk]_r, \text{for} \; n = 0,...\frac{k}{{\rm gcd}(k,r)}-1,  \text{and} \; p = 0,...,{r\over {\rm gcd}(k,r)}-1 \bigg\}, \end{equation} and repeated entries in $S_j$ are identified so that each set has  $r \over {\rm gcd}(k,r)$ elements. The union of all sets $S_j$ is the set $ \{0,...,r-1\}$.   

The first point to make is that there are extra moduli appearing at this level of the expansion, due to the fact that the equations (\ref{constraintssubbed}) do not fix all moduli. 
To see this, let us  count the number of moduli for general $k$ and $r > 1$, taking into account the constraints (\ref{constraintssubbed}), modding by the constant gauge transforms leaving the boundary conditions and gauge condition  invariant. First, there are $4$gcd$(k,r)$ constant holonomies $\phi_\mu^{C'}$, as per (\ref{SUNmoduli}). 
Then, there are $2 r$ real components of ${\cal C}_2^A$ and $2 r$ real components of ${\cal C}_4^A$. Thus the total number of real moduli is $4 r + 4 {\rm gcd}(k,r)$. These are subject to the constraints of eqn.~(\ref{constraintssubbed}): the gcd$(k,r)$ real constraints on the first line and $2$gcd$(k,r)$ real constraints on the second line.
Thus, it would appear that the number of moduli minus the number of constraints is $4 r + {\rm gcd}(k,r)$. 
We notice, however, that  the gauge conditions (\ref{gaugecondition}) are invariant under   constant gauge transformations  in the gcd$(k,r)$ Cartan directions, the ones along the directions of the allowed holonomies and commuting with the transition functions. Thus, the total number of bosonic moduli---the holonomies $\phi_\mu^{[C']_r}$ and the remaining components of ${\cal C}_2^A, {\cal C}_4^A$---for $k \ne r > 1$ is  $4 r$, as required by the index theorem for a selfdual solution. 

Two remarks are now due. 

First, 
as we already mentioned, the above argument, using the results of ref.~\cite{Anber:2023sjn}, is quite general, as it holds for any $N$ and any $r$ with gcd$(k,r)\ne r$. However, it is not trivially  extended to higher orders in $\lambda$.\footnote{The difficulty, compared to other studies of the ``$\lambda$-expansion,'' in the physics (e.g. \cite{Schwarz:1977az,Weinberg:1979ma}) or mathematics (e.g. \cite{Taubes:1982qem}) literature, stems from the non-invertibility of the $\bar\partial \partial$ operator.} Thus, in section \ref{appx:su2}, we give an argument for $SU(2)$ which---at least in principle---is exact, but is only valid for $N=2$ and any $r > 1$.\footnote{We believe that it should be possible to extend the argument of section~\ref{appx:su2}  to any $N$ and $r$, but leave this for the future.} 

Second, we test, further below in section~\ref{sec:lambdafit},  the leading-order $\lambda$-expansion (which is only   valid for small values of the coefficients ${\cal{C}}_{2,4}^{A}$, $A=1,...r$ appearing in (\ref{expressions of W2 and W4 gcdnotr}) and obeying (\ref{constraintssubbed})) for the almost-constant $F$ configurations we find numerically (in the section below) for $N=3$, $r=2$, $k=1$.

    \subsection{Results for numerical fractional instantons  with gcd$(k,r)\ne r$}
    
\label{sec:numericgcdnotr}

\begin{figure}[h]
   \centering
   \includegraphics[width=5.1in]{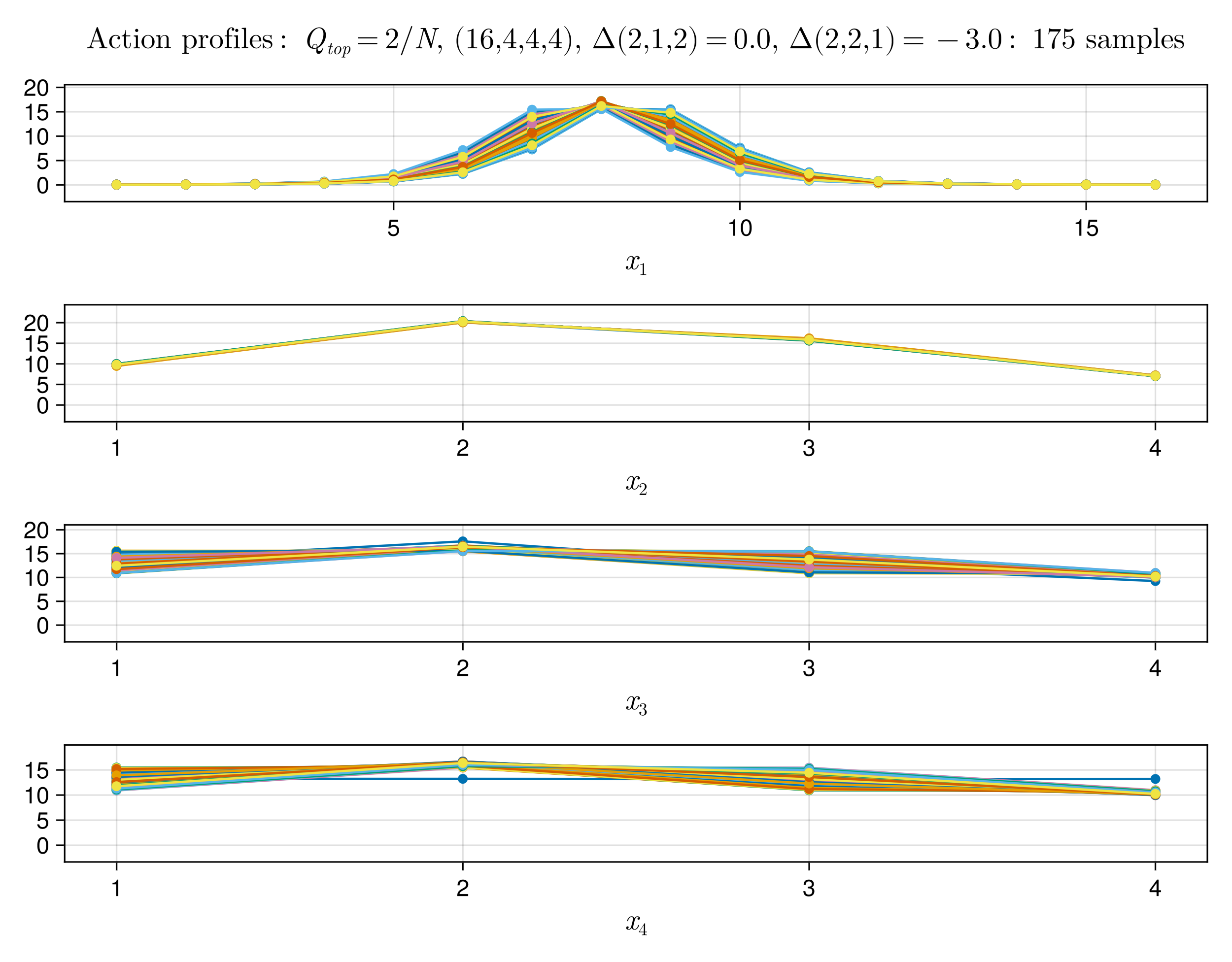} 
   \caption{The action density, integrated over all but one direction of the torus, for $SU(3)$ with $r=2, k=1$, on the tuned torus with $\Delta(2,1,2)=0$ and sides $(16, 4, 4,4)$. All $175$ self dual configurations generated starting from a random one are found to have non-constant action density.}
   \label{fig:SU3lattice1}
\end{figure}

\begin{figure}[h]
   \centering
   \includegraphics[width=5.1in]{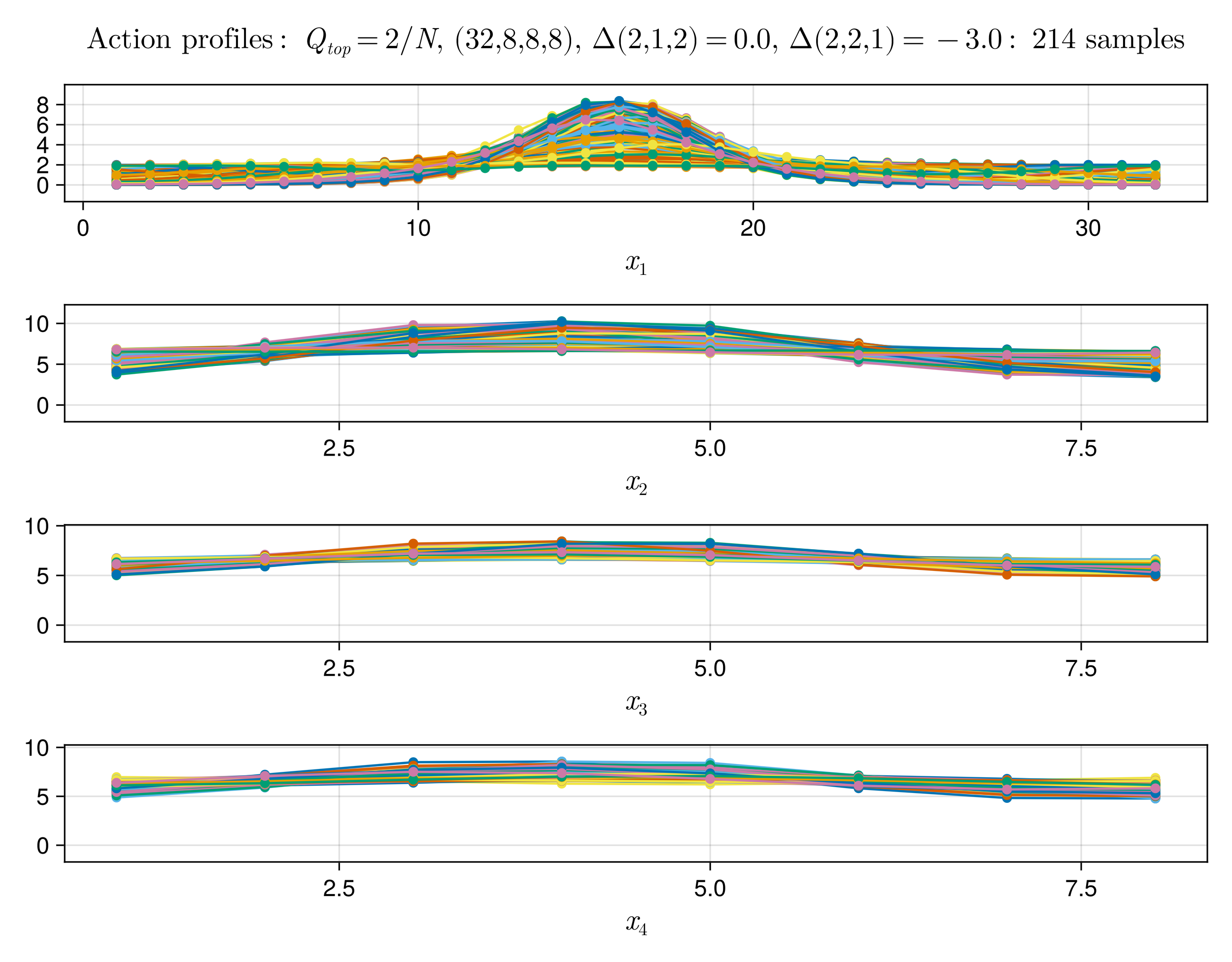} 
   \caption{The action density, integrated over all but one direction of the torus, for $SU(3)$ with $r=2, k=1$, on the tuned torus with $\Delta(2,1,2)=0$ and sides $(32, 8, 8,8)$. All of the $214$ configurations are found to have non-constant action density. Only a few appear that are almost constant. We use these to test the $\lambda$-expansion, see section~\ref{sec:lambdafit}. }
   \label{fig:SU3lattice2}
\end{figure}

Here we present a result of our numerical study of fractional instantons for $SU(3)$ with $Q={2 \over 3}$ with $r=2$, $k=1$. On Figures~\ref{fig:SU3lattice1} and \ref{fig:SU3lattice2} we present the numerical results for the action density profiles for fractional instantons of charge $2/3$ and gcd$(k,r)\ne r$, on the tuned $\T^4$, for two different lattices with $\Delta(2,1,2)=0$, i.e. with  $4 L_3 L_4 = L_1 L_2$ (recalling eqn.~(\ref{deltadef})). The lattice on Figure~\ref{fig:SU3lattice2} is twice as large (than the one on Figure~\ref{fig:SU3lattice1}) in all directions. Starting from a random configuration, as described in Appendix \ref{appx:lattice}, in each case we find that the action profile is non-constant, consistent with the argument that the constant-$F$ instantons with gcd$(k,r)\ne r$ are a measure-zero set. 

\begin{figure}[h]
 \centering
   \includegraphics[width=5.1in]{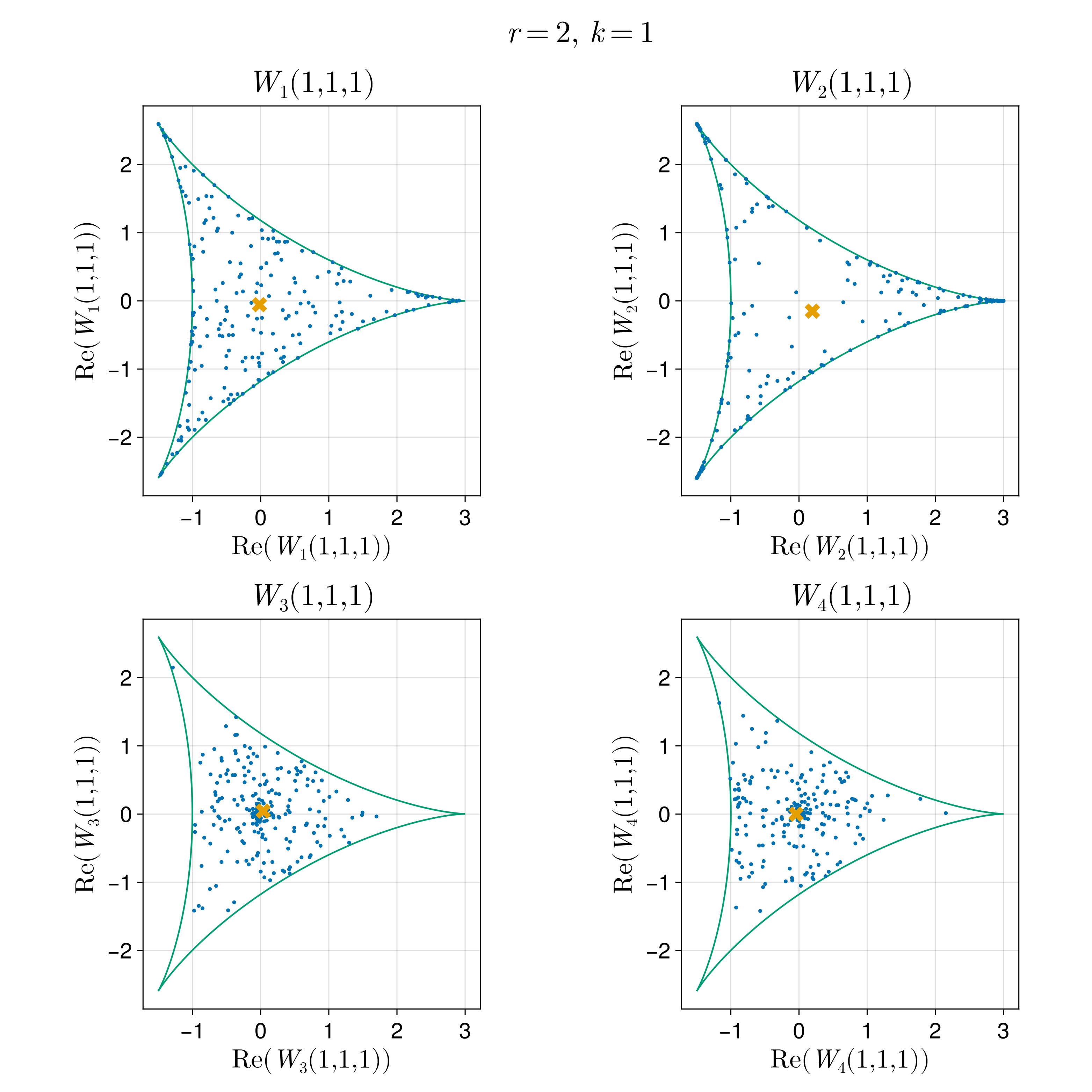} 
   \caption{Wilson loops evaluated at the origin of the lattice for 214 configurations of Figure~\ref{fig:SU3lattice2} on a lattice of size (32,8,8,8) corresponding to $\Delta(2,1,2)=0$. Data points are plotted in blue, the averages are plotted as orange crosses, and the solid boundary is the same as that plotted in Figure \ref{fig:r=k=2wilson}, corresponding to turning on only the translational moduli of the constant-$F$ solution.}
   \label{fig:r=2-k=1wilson}
\end{figure}

As a further evidence that the non-constant-$F$ lattice configurations we generated with $r=2, k=1$ and $\Delta(2,1,2)=0$ span the entire moduli space, we show two plots with results regarding the Wilson loop expectation value in these configurations. As shown in \cite{Anber:2024mco} from considerations of the Hamiltonian interpretation of the $\T^4$ partition function with twists,  the  value of the Wilson loop winding in any direction should vanish  when integrated over the moduli space of the multi-fractional instantons. We test this on Figure~\ref{fig:r=2-k=1wilson} by showing the values of the real and imaginary parts of the Wilson loops (taken at the origin) for all 214 configurations generated. The numerical sum of the Wilson loops over all configurations is shown by a cross, a value consistent with zero in each of the four directions (as required by the Hamiltonian interpretation of the twisted partition function \cite{Anber:2024mco}).

\begin{figure}[h]
   \centering
   \includegraphics[width=5.1in]{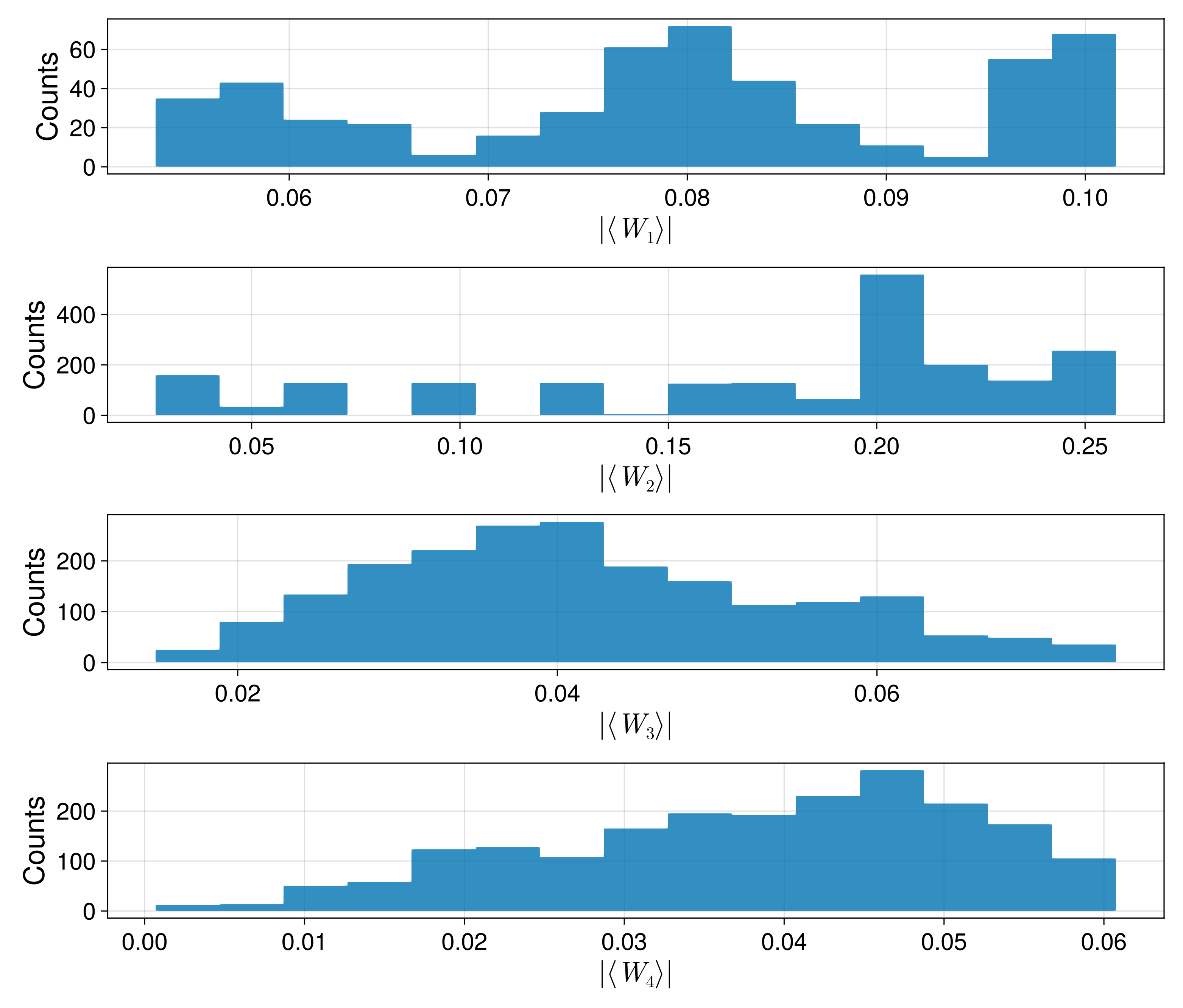} 
   \caption{Histogram of averaged Wilson loops over all configurations for $214$ configurations on a lattice of size $(32,8,8,8)$ corresponding to $\Delta(2,1,2)=0$. For each configuration, the Wilson loop winding around each direction is computed at every point on the lattice. Those Wilson loops are then averaged over all configurations to give an average value of each Wilson loop at every point on the lattice. We then make histograms of those averages to show how they are distributed across the lattice. In other words, each count in the histogram corresponds to a different lattice point.}
  \label{fig:avg_Wilson}
\end{figure}

A more systematic study of the average value of the Wilson loops calculated at every point on the lattice is depicted in Figure~\ref{fig:avg_Wilson}, where we display the histogram of the averaged Wilson loops over all configurations, for $214$ configurations on a lattice of size $(32,8,8,8)$, corresponding to $\Delta(2,1,2)=0$. In particular, we display the histogram of the absolute value of the averaged Wilson loops, $|\langle W_\mu (x)\rangle|=\left\lvert \sum_{\scriptsize\mbox{config}}\frac{W_\mu(x)}{{\cal N}}\right\rvert$, where ${\cal N}=214$, the total number of configurations. The histogram is consistent with $|\langle W_\mu(x)\rangle |\approx 0$.  

\subsection{The $\lambda$-expansion vs. numerics, for $N=3$, $r=2$, $k=1$}
 \label{sec:lambdafit} 

Here, we study the $SU(3)$, $r=2$, $k=1$ leading-order solution,  eqn.~(\ref{expressions of W2 and W4 gcdnotr}), of the $\lambda$-expansion around the constant-$F$ solution. This is the case we studied numerically in section~\ref{sec:numericgcdnotr} above.  We shall select a few configurations obtained on the lattice that have field strengths that are almost constant (as seen in Figure~\ref{fig:SU3lattice2}) and compare their gauge invariant densities with the analytic expressions obtained in the leading-order $\lambda$-expansion.
As we already discussed, the best we can do in the framework of the $\lambda$ expansion is to stop at this order. Higher orders are increasingly complicated due to the fact that (as opposed to the gcd$(k,r)=r$ case), $\cW_\mu^{(0)}$ is not set to zero and a nonvanishing r.h.s. of all equations appear at higher orders. 

Here, since $p=0,1$ and $C'=1, C=1,2$, the solution (\ref{expressions of W2 and W4 gcdnotr}) takes the form, remembering that $\cW_{1,3}$ are determined from the $\cW_{2,4}$ components given below:
 \begin{eqnarray}
\left({\cal W}_{2 \; (4)}^{(0)k\times \ell}\right)_{1 C}= V^{-{1\over 4}} \sum_{p=0}^{1}{\cal C}^{[1+p]_2}_{2 \; (4)}\Phi^{(p)}_{1  C}(x,\hat\phi) =   V^{-{1\over 4}}\left(  {\cal C}^{1}_{2 \; (4)} \Phi^{(0)}_{1 C} + {\cal C}^{0}_{2 \; (4)} \Phi^{(1)}_{1 C}\right), 
\label{W2SU3}
\end{eqnarray}
while there is only  a single set $S_0=\{0,1\}$, see (\ref{sjdef}), in this simple example. Thus, from (\ref{constraintssubbed}), we have that
  \begin{eqnarray}\label{su3constraints}
  |{\cal C}_2^0|^2 + |{\cal C}_2^1|^2 =  |{\cal C}_4^0|^2 + |{\cal C}_4^1|^2 \; \text{and} \; {\cal C}_2^0 \; ({\cal C}_4^0)^*+ {\cal C}_2^1 \; ({\cal C}_4^1)^* = 0.
  \end{eqnarray}
 These are $3$ real conditions on the $8$ real coefficients in ${\cal{C}}_{1, 2}^{0,1}$. There is also an unphysical phase corresponding to a $U(1)_\omega$, a constant gauge transformation in the $\omega$ direction, eqn.~(\ref{omega}). Thus, together with the $4$ translational moduli $\hat\phi_\mu^{C'=1}\equiv \hat\phi_\mu$, the parts of ${\cal{C}}_{1, 2}^{0,1}$ not fixed by (\ref{su3constraints}), make up the eight moduli of the $Q=2/3$ instanton.
  Locally, we  choose a  parameterization of the solutions of (\ref{su3constraints}) as follows (noting that the $U(1)_\omega$ acts as a common phase on all ${\cal{C}}_{2,4}^{0,1}$):
 \begin{eqnarray}\label{parametrization of moduli}\nonumber
{\cal C}_2^1&=&e^{i\chi_1}\eta \cos\theta\,,\quad
{\cal C}_2^0=e^{-i\chi_2}\eta \sin\theta\,,\\
{\cal C}_4^1&=&-ie^{i\chi_2}\eta \sin\theta\,,\quad
{\cal C}_4^0=ie^{-i\chi_1}\eta \cos\theta\,,
\end{eqnarray}
with a single real noncompact modulus $\eta$ and compact moduli $\theta, \chi_1, \chi_2$.  We note, however, that the solution is only trustable for small values of $\eta$, since it is obtained by neglecting the nonlinear terms in the self-duality condition (\ref{selfduality1}, \ref{main equations we need to solve k eq r}). We have, using our index notation, remembering that we have $k=1$, $\ell=2$, the order $\lambda^0$ solution has the form:
\begin{eqnarray}
A_\mu \simeq \left(\begin{array}{ccc} \bar{A}_{\mu \; 1' 1'}(x, \phi_\mu^1)& \cW_{\mu \; 1'1}^{(0)}  &\cW_{\mu \; 1'2}^{(0)}\cr (\cW_{\mu\;1' 1}^{(0)})^* &\bar{A}_{\mu \; 11}(x, \phi_\mu^1) &0 \cr (\cW_{\mu\;1' 2}^{(0)})^*&0& \bar{A}_{\mu \; 22}(x, \phi_\mu^1)  
\end{array}\right)~,\label{a1}
\end{eqnarray}
where  $\bar{A}_{\mu \; 1'1'}$ is from\footnote{Where now $C'=1$ only and the primes $1'$ remind us that this is the $k\times k$ part.}  (\ref{Ukbackground}) and $\bar{A}_{\mu D D}$, $D=1,2$ is from (\ref{Ulbackground}), with the substitutions $N=2, r=2, k=1,\ell=2$ understood. We also recall that the $0$-th order $\cS^0$ contributions are the constant moduli which are absorbed in $\bar A$. The substitution (\ref{parametrization of moduli}) is understood to be made in the expressions for $\cW_\mu$ read off from eqn.~(\ref{W2SU3}) and can be trusted only for small values of the noncompact modulus $\eta$.
  
Using (\ref{a1}), with (\ref{W2SU3}) and (\ref{parametrization of  moduli}), as well as  the expressions from \cite{Anber:2023sjn}, the computation of $\left[F_{13}F_{13} \right]$ is consistent\footnote{This is similar to section~\ref{sec:testdelta}, where we compare the $\Delta$-expansion to numerics. The point is that, as in the $\Delta$-expansion,  gauge invariants of the form $\tr F_{12}^2$, with two indices in a plane of nonzero twist receive contributions to leading order also from components $\cS^{(1)}$, for which a formal expression can be written, but which are difficult to compute explicitly. This is discussed in Appendix C of \cite{Anber:2023sjn}. } to ${\cal O}(\lambda^0)$, and its integral over the $2$-$4$ plane of the torus is given by
\begin{eqnarray}\label{integrated F13 eta}\nonumber
&&f_{\scriptsize\mbox{theory}}(x_1,x_3;\eta,\theta,\bar x_1,\bar x_3)=\int_{24}\mbox{tr}\left[F_{13}F_{13} \right]=\\\nonumber
&&2V^{-1/2}L_2L_4\eta^2\left[ \sin^2\theta |{\cal G}_1^{(0)}|^2+\cos^2\theta |{\cal G}_3^{(0)}|^2+\cos^2\theta |{\cal G}_1^{(1)}|^2+\sin^2\theta |{\cal G}_3^{(1)}|^2 \right]_{x_2=x_4=\hat\phi_1=\hat\phi_3=0}\,.\\
\end{eqnarray}
Here $|{\cal G}_1^{(p)}|^2 \equiv  \sum\limits_{C=1}^2 {\cal G}^{(p)}_{1 C} ({\cal G}^{(p)}_{1 C})^*$ (and likewise for $|{\cal G}_3^{(p)}|^2$) and the functions ${\cal G}^{(p)}_{1 C}$ and ${\cal G}^{(p)}_{3 C}$ are defined in eqns.~(\ref{the G1 function}, \ref{the G3 function}), see Appendix~\ref{appx:phip}. We note that integrating over the $2$-$4$ plane eliminates $4$ out of the $8$ moduli, leaving $\eta$, $\theta$, $\hat \phi_2$, and $\hat\phi_4$ in (\ref{integrated F13 eta}). We also define $\bar x_1\equiv \frac{L_1L_2}{4\pi}\hat \phi_2$ and $\bar x_3\equiv \frac{L_3L_4}{\pi}\hat \phi_4$.
\begin{figure}[h]
   \centering
   \includegraphics[width=6.1in]{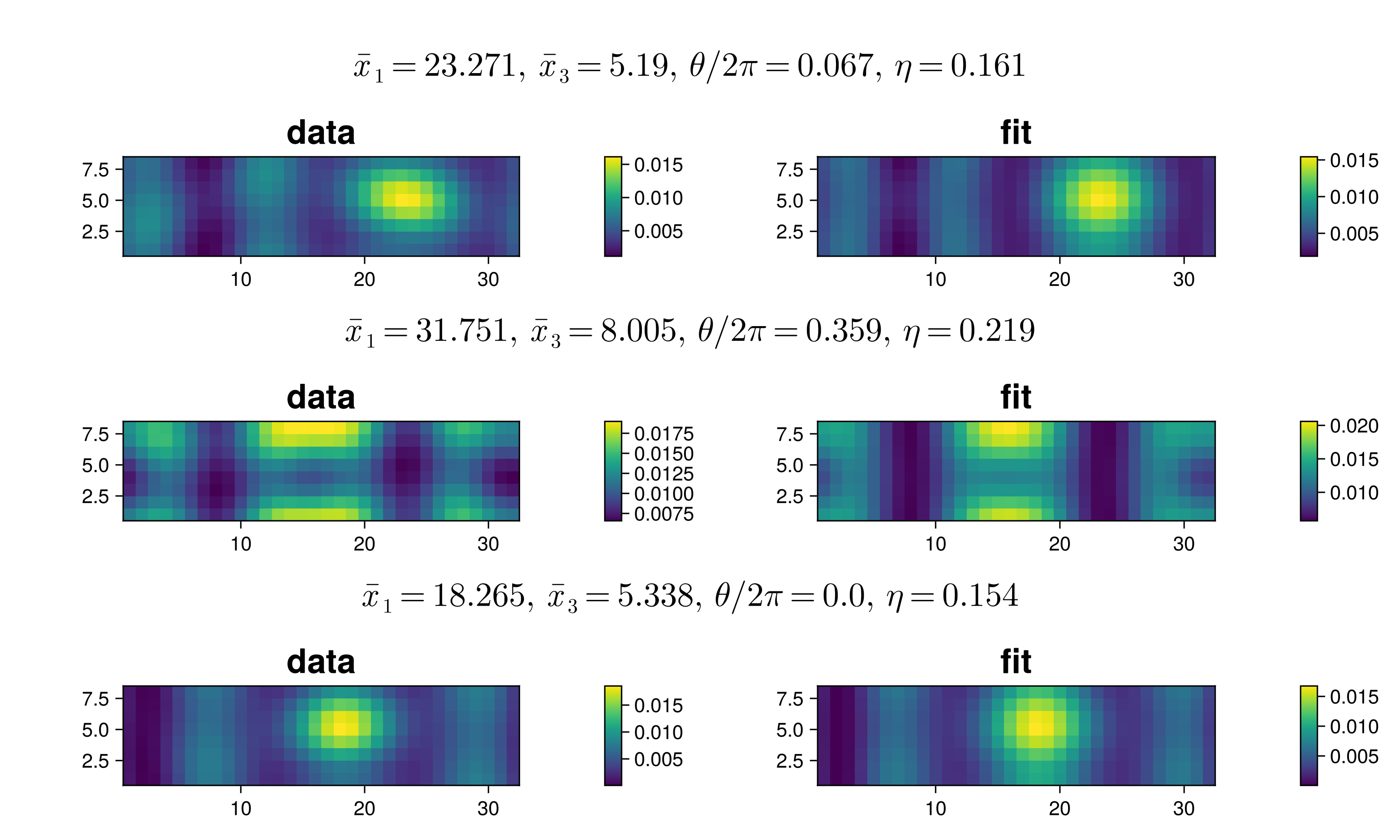} 
   \caption{The results of fitting $\Tr\(F_{13}F_{13}\)$  to the analytical expression (\ref{integrated F13 eta}) for three configurations (one in each row), obtained by minimizing the lattice action on a lattice of size $(32, 8, 8,8)$, corresponding to tuned $\mathbb T^4$ with $\Delta(2,1,2)=0$. The fits produce a mean squared deviation of $\sim10^{-6}$, showing excellent agreement with the analytical expression. \\
   {\it Left column:} the values of $\Tr\(F_{13}F_{13}\)$ obtained from lattice configurations, summed over the $2^{nd}$ and $4^{th}$ directions, as a function of $x_1$ and $x_3$.\\   
   {\it Right column:} the analytical expression (\ref{integrated F13 eta}), for values of the four moduli $\bar{x}_1,\bar x_3,\eta,\theta$ fitted to the lattice data using the least-square method.}
   \label{fig:eta_fit}
\end{figure}
\begin{figure}[h]
   \centering
   \includegraphics[width=3.5in]{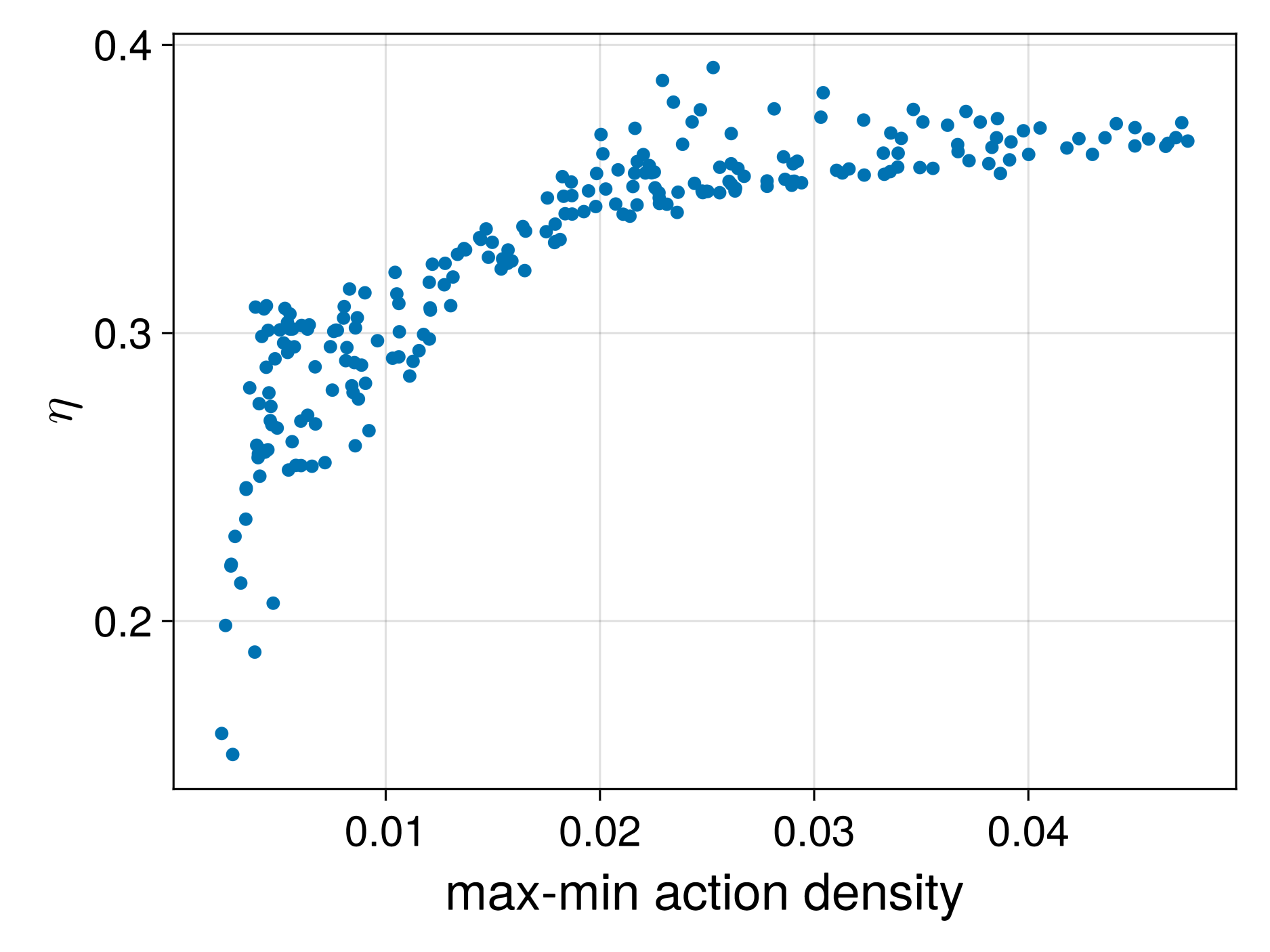} 
      \includegraphics[width=3.5in]{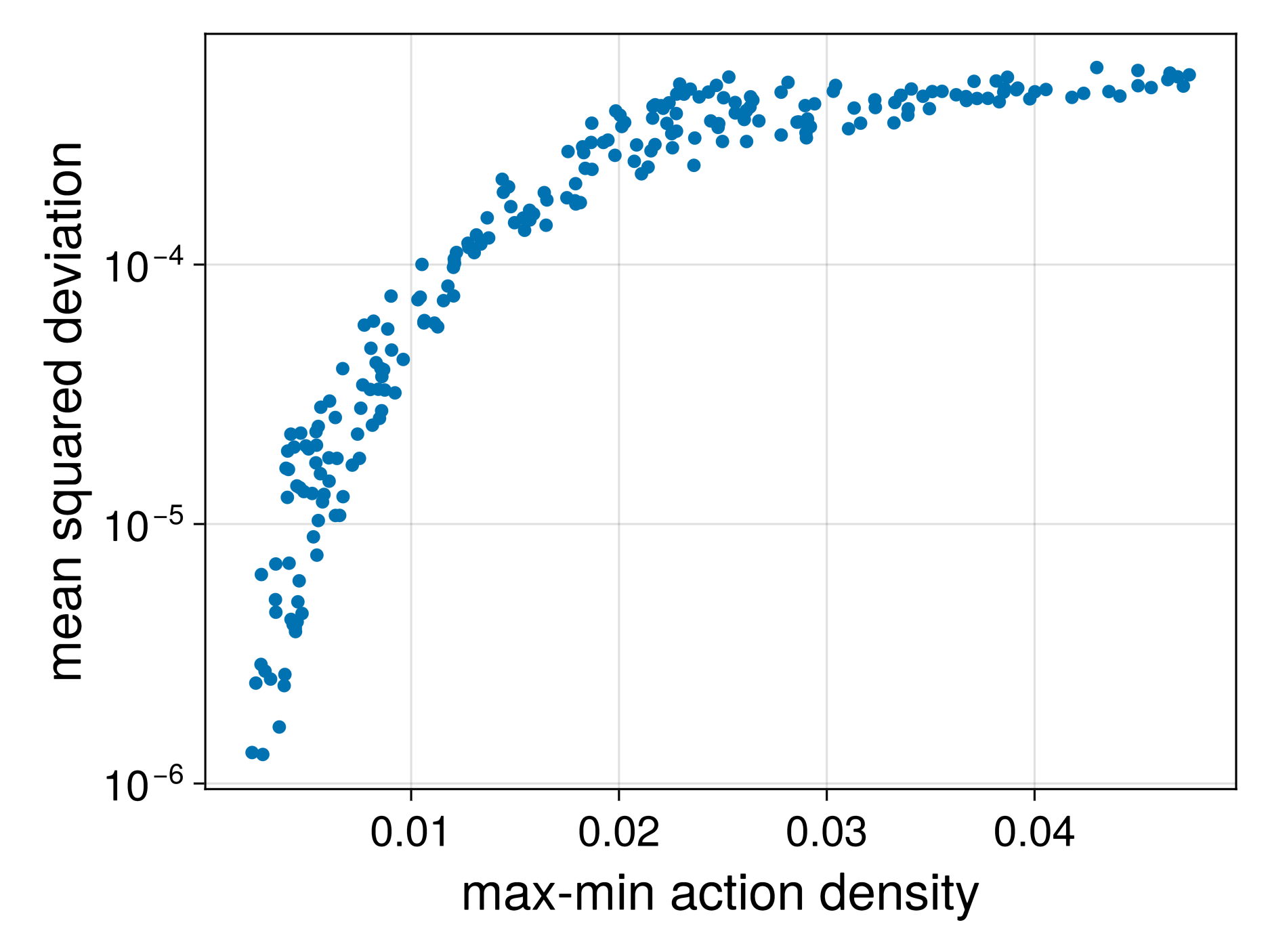} 
   \caption{To assess the quality of the lattice data fit to the analytical expression (\ref{integrated F13 eta}), we first define the uniformity of a given lattice configuration. This is measured as the absolute difference between the global maximum and minimum of the action density, with smaller differences indicating greater uniformity. We present the data for all the $233$ configurations for $r=2,k=1$, with lattice size $32,8,8,8$. \\
   {\it The top pannel:} the fit of the lattice configuration to the modulus $\eta$. The more uniform the configuration, and hence, closer to the constant-$F$ solution, the smaller the value of $\eta$.\\   
   {\it The bottom pannel:} the mean square deviation of the fit as a function of the uniformity of the solution. The more uniform the solution, the smaller the error.}
   \label{fig:eta goodness}
\end{figure}
 
We contrast these analytical expressions with the numerical lattice data of various configurations obtained for $SU(3)$ with $r=2, k=1$, on the tuned torus with $\Delta(2,1,2)=0$ and sides $(32, 8, 8,8)$. With the Wilson action density, we have
\begin{equation}
	f_\text{data}(n_1,n_3)\equiv \sum_{n_2=1}^{L_2}\sum_{n_4=1}^{L_4}\Tr\(F_{13}F_{13}\)(n)=\sum_{n_2=1}^{L_2}\sum_{n_4=1}^{L_4}2\Re\Tr\(\identity-\plaq{1}{3}\).
\end{equation}
The fitting procedure aims to minimize  
$
\sum_{n_1=1}^{L_1}\sum_{n_3=1}^{L_3}\left\lvert f_{\scriptsize\mbox{theory}}(x_1,x_3;\eta,\theta,\bar x_1,\bar x_3)-f_\text{data}(n_1,n_3)\right\rvert^2
$
with respect to the four fitting parameters, $\bar{x}_1$ and $\bar{x}_3$,  $\eta$, and $\theta$. We used the first few integers $m'$ and $n'$ in equations (\ref{the G1 function}, \ref{the G3 function}), since contributions from larger integers (in absolute values)  diminish exponentially. The data is then fitted using the standard least squares method, which optimizes the fit by minimizing the sum of squared differences between theoretical predictions and observed data across all points in the $x_1$-$x_3$ plane. Additionally, the improved action described in \eqref{eqn:improved_action} of Appendix \ref{appx:lattice} is employed to eliminate next-to-leading-order corrections.  Figure \ref{fig:eta_fit} illustrates these results for three distinct lattice configurations, each selected to ensure nearly uniform integrated action densities across all four directions (similar to the configurations with almost uniform integrated action densities shown in Figure \ref{fig:SU3lattice2}). The small errors (mean squared deviation $\ll 1$) indicate a remarkable agreement between numerical simulations and analytical predictions.

To precisely define uniformity, we quantify it as the difference between the global maximum and minimum values of the action density within a given lattice configuration. A smaller difference indicates a more uniform configuration, and thus, closer to the constant-$F$ solution. Our fits of all the $233$ configurations for $r=2,k=1$ (lattice of size $32,8,8,8$) gives $\eta<0.4$ with mean squared deviation $< 10^{-3}$. We find that the more uniform the configuration, the smaller the value of $\eta$, and the smaller the value of the mean squared deviation; see Figure \ref{fig:eta goodness}. Our findings are consistent with two main points in our analytical treatment. First, in the analytical treatment, we carry out the analysis to the leading order in $\lambda$, i.e., to ${\cal O}(\lambda^0)$, meaning that we ignore the contribution of the non-linear terms in the $\lambda$-expanded perturbations about the constant-$F$ solution. Second, the leading-order $\lambda$-expansion gives rise to the noncompact modulus $\eta$, which indicates that such modulus can be trusted only for small enough values. Interestingly, the successful fit of all lattice data with $\eta < 0.4$, along with the plateau observed in both the $\eta$-fit and mean squared deviation, suggests that nonlinear effects may ultimately resolve the noncompactness of $\eta$.

\subsection{Remarks on the moduli space of  $SU(2)$ instantons for $r \ge 2$ on the twisted-$\T^4$}
\label{appx:su2}

Here, we discuss an issue similar to the one discussed above for multi-fractional instantons in $SU(N)$ with gcd$(k,r)\ne r$. There, we provided analytical, in section~\ref{sec:analyticgcdnotr}, and numerical, in section~\ref{sec:numericgcdnotr},  evidence, that when the number of constant holonomies of the tuned-$\T^4$ constant-$F$ solution, $4$gcd$(k,r)$, is smaller than   the number of moduli expected from the index theorem, $4 r$, the constant-$F$ solutions represent only a measure-zero set of the moduli space---in contrast to the gcd$(k,r)=r$ case where the only self-dual solutions of charge $r/N$ are the constant-$F$ ones, as we argued in section \ref{sec:tunedgcd=r}. In other words, there are $4 r$$-4$gcd$(k,r)$ extra moduli whose nonzero values correspond to self-dual deformations of the solution which make its field strength
$x_\mu$-dependent. 

Here, we note that a similar problem also arises in $SU(2)$ theories with $Q={r \over 2}$, for any integer $r\ge 2$, as we describe in detail below. Little is known about these solutions from the numerical side:   with $r \ge 2$ the minimum action lattice configuration is either the well-studied one with $Q = {1 \over 2}$, if $r$ is odd, or the trivial vacuum with $Q=0$, if $r$ is even. Thus, these instantons ``fall through'' the lattice, complicating their numerical study. 

On the other hand, on the analytic side,\footnote{The only other analytic result we know of refers to (in our language)  $Q=1$ instantons of $SU(2)$ on the twisted torus, arguing that, with the translational moduli factored out, the moduli space  is a $K3$ surface    \cite{BraamTodorov}.} explicit constant-$F$ solutions, for any $r \ge 2$, exist on tuned tori, such that $\Delta(r,1,1)=0$, or from (\ref{deltadef}), $r L_3 L_4 = L_1 L_2$. These are, of course, given by our backgrounds of eqn.~(\ref{Ukbackground}, \ref{Ulbackground}) with $k=\ell=1$, $N=2$, and the desired value of $r$. Now, for any $r \ge 2$, there are only $4$ constant holonomies, the translational moduli of the solution. It stands to reason to expect that, as happened for the multi-fractional instantons studied   in section \ref{sec:tunedgcdnotr}, there will be extra $4 r- 4$ moduli, which, when turned on, will correspond to non-constant self-dual solutions.

 In what follows, we give an argument in favour of this, which goes beyond our leading-order study of the self-duality condition of  \ref{sec:analyticgcdnotr}. At the outset, however, let us state that all we shall be able to argue is that the right number of extra moduli appears and that the solution becomes $x_\mu$-dependent. Any claims about the global structure of the moduli space (like for the $r=2$  example studied in \cite{BraamTodorov}) remain beyond our current scope. 

We begin by parameterizing the fluctuations around the constant-$F$ solution (\ref{Ukbackground}, \ref{Ulbackground}) with $k=\ell=1$, $N=2$, any $r>1$, which is labelled $\bar A_\mu$ below. The fluctuations are similar to (\ref{solution21}), but simplify in the $SU(2)$ case:
\begin{equation}\label{solution31}
A_\mu = \bar A_\mu  +  \left(\begin{array}{cc} 2 \pi \cS_\mu & \cW_\mu\cr \cW_\mu^\dagger  & - 2 \pi \cS_\mu \end{array} \right)~, ~~ \partial_\mu S_\mu = 0, ~ D_\mu\cW_\mu = 0, ~D_\mu^* \cW_\mu^\dagger = 0,
\end{equation}
where we imposed the gauge condition (\ref{gaugecondition}) with the background covariant derivative  $D_\mu\cW_\nu$ defined in (\ref{wderivative}) (and with $D_\mu^* \cW_\nu^\dagger \equiv (\partial_\mu - i 4 \pi   \bar A_\mu^\omega) \cW_\nu^\dagger$). We stress that we include the constant holonomies $\phi_\mu$ in the constant-$F$ background $\bar A_\mu \equiv \omega \bar A_\mu^\omega$  and use $\cS_\mu$ to denote only $x$-dependent fluctuations in the Cartan direction. The field strength then has the form, as in \cite{Anber:2023sjn}, 
\begin{eqnarray}\label{field33}
F_{\mu\nu} = \bar F_{\mu\nu}  + \left(\begin{array}{cc}2 \pi \partial_{[\mu} \cS_{\nu]}  + i \cW_{[\mu} \cW^\dagger_{\nu]}&  
D_{[\mu} \cW_{\nu]} + i 4 \pi   \cS_{[\mu} \cW_{\nu]} \cr D_{[\mu} \cW^\dagger_{\nu]} - i 4 \pi   \cS_{[\mu} \cW^\dagger_{\nu]} &- 2 \pi \partial_{[\mu} \cS_{\nu]}+ i \cW^\dagger_{[\mu} \cW_{\nu]} \end{array} \right),
\end{eqnarray}
where $\bar F$ is the field strength of the constant-$F$ solution and $[\mu...\nu]$ denotes antisymmetrization with respect to these indices.
The boundary conditions on $\cW_\mu$ are as stated in (\ref{BCS W}) and $\cS_\mu$ is periodic in all $\T^4$  directions. Now, we impose self-duality of (\ref{field33}) and the background gauge condition (\ref{gaugecondition}). 
Before we continue with details, let us explain our strategy.

{\flushleft \bf{Strategy:}} We first note that the self-duality condition for the off-diagonal component in (\ref{field33}), for any periodic $\cS_\mu$, involves a linear differential operator acting on the complex vector field $\cW_\mu$. This operator has a background $U(1)$ gauge field, which has a component, $\bar A_\mu^\omega$, giving rise to  nontrivial fluxes (or Chern classes) through the various planes in the $\T^4$, and an unknown  periodic component, $\cS_\mu$, which does not give rise to nonzero Chern classes. We shall argue below that there is an index theorem that determines the number of normalizable solutions of the  $\cW_\mu$ self-duality condition in terms of the $U(1)$ fluxes, for any periodic $\cS_\mu$. Each solution of the self-duality equation for $\cW_\mu$ thus comes with a number of (real) moduli, which we argue equals $4 r$. Plugging this solution (which depends on the unknown but periodic $\cS_\mu$ in a nontrivial way) into the  self-duality condition for the diagonal elements in (\ref{field33}) then leads to a nonlinear first order equation for $\cS_\mu$. Solving this equation is beyond our means. However,  consistency of this equation  requires   the vanishing of the integral of the r.h.s. over the torus and restricts the number of moduli from $4r$ to $4 r - 3$. Adding the constant gauge transformations in the Cartan subalgebra of $SU(2)$, which is a symmetry of both the fluctuation equation and the gauge condition, leaves only $4 r -4$ extra moduli. In addition to the $4$ constant holonomies, this makes up for precisely $4r$ moduli. These $4r -4$ ones correspond to self-dual deformations of the constant-$F$ solution which make the field strengths of the instanton $x$-dependent.

{\flushleft \bf{The details:}} We begin by stating the self-duality condition for the ``field strength of $\cW_\mu$,'' namely the one entering the off-diagonal term in (\ref{field33}). We introduce, in addition to the operator $D_\mu$ entering the gauge condition, the operator ${\cal{D}}_\mu$ which has an additional $U(1)$ background field, the periodic $\cS_\mu$: 
\begin{eqnarray}\label{curlyD}
{\cal{D}}_\mu = \partial_\mu + i 4 \pi  (\bar A_\mu^\omega + \cS_\mu), ~~D_\mu  =  \partial_\mu + i 4 \pi   \bar A_\mu^\omega, 
\end{eqnarray}
in terms of these operators, the self-duality of the off diagonal component of (\ref{field33}) reads:
\begin{eqnarray}\label{Wselfduality1}
{\cal{D}}_{[\mu} \cW_{\nu]} - {1 \over 2} \epsilon_{\mu\nu\lambda\sigma}{\cal{D}}_{[\lambda} \cW_{\sigma]} &=&0\nonumber \\
D_\mu \cW_\mu &=&0.
\end{eqnarray}
Following \cite{Schwarz:1977az} (see also \cite{vanBaal:1984ar}), this can be recast as a problem for finding the zero modes of a differential operator $T$$: \cW_\mu \rightarrow (f_{\mu\nu}, g)$. $T$ is a map from the space of complex vector fields ($\cW_\mu$) on $\T^4$, with boundary conditions (\ref{BCS W}), to the space of complex self-dual antisymmetric $2$-tensors ($f_{\mu\nu}$) and scalars ($g$).\footnote{On $\T^4$, obeying boundary conditions determined by (\ref{BCS W}), the definition of $T$ and eqn.~(\ref{Wselfduality1}).} Explicitly:
\begin{eqnarray}
\label{defofT}
T: \cW_\mu \rightarrow (f_{\mu\nu} = {\cal{D}}_{[\mu} \cW_{\nu]} - {1 \over 2} \epsilon_{\mu\nu\lambda\sigma}{\cal{D}}_{[\lambda} \cW_{\sigma]}, ~g = D_\mu \cW_\mu)~.
\end{eqnarray}
From eqn.~(\ref{defofT}), it is clear that the solutions of the self-duality and background gauge conditions (\ref{Wselfduality1}) are the zero modes of $T$. Defining the inner products in the two spaces as 
$(\cW', \cW) \equiv \int\limits_{\T^4} \cW^{'  *}_\mu \cW_\mu$ and $((f', g'), (f,g)) \equiv \int\limits_{\T^4}({1 \over 4} f_{\mu\nu}^{' *} f_{\mu\nu} + g^{' *} g)$, the adjoint operator $T^\dagger$ is:
\begin{eqnarray}
\label{defofTdagger}
T^\dagger: (f_{\mu\nu}, g) \rightarrow \cW_\mu = - {\cal{D}}_\nu f_{\nu\mu} - D_\mu g~.
\end{eqnarray}
The operators $T^\dagger T$ and $T T^\dagger$  can also be worked out, as in \cite{Schwarz:1977az,vanBaal:1984ar}.
For use below, we note that $T T^\dagger$ is the second order differential operator acting on $(f_{\mu\nu}, g)$, which depends on $\bar A_\mu^\omega$ as well as $\cS_\mu$:
\begin{eqnarray}\label{TTdagger}
TT^\dagger: (f_{\mu\nu}, g) \rightarrow (f_{\mu\nu}' &=& - {\cal{D}}_\mu ({\cal{D}}_\lambda f_{\lambda\nu} - D_\nu g) +  {\cal{D}}_\nu ({\cal{D}}_\lambda f_{\lambda\mu} - D_\mu g) + \epsilon_{\mu\nu\eta\sigma} {\cal{D}}_\eta ({\cal{D}}_\lambda f_{\lambda\sigma} - D_\sigma g),  \nonumber \\ g' &=& - D_\mu {\cal{D}}_\nu f_{\nu\mu} - D_\mu D_\mu g)
\end{eqnarray}
As usual the index is defined as  ${\rm dimker} (T^\dagger T)- {\rm dimker}(T T^\dagger)$ and, as per the index theorem, depends only on the topological properties of the manifold on which the operators are  defined as well as on the topological properties (Chern classes) of the gauge background. For the abelian gauge background in (\ref{Wselfduality1})  only the second Chern classes (quantized fluxes through various two-planes) are nonzero.
 We now will argue, first, that generally
\begin{eqnarray}
\label{indexT}
{\rm dimker} (T^\dagger T) - {\rm dimker}(T T^\dagger) = 2 r
\end{eqnarray}
and, second, that (at least for sufficiently small $\cS_\mu$)
\begin{eqnarray}
{\rm dimker}(T T^\dagger)=0. 
\end{eqnarray}

To show this, as we now explain, the  operator $T T^\dagger$, with $\cS_\mu=0$, was studied in \cite{Anber:2023sjn}, using a different language, and was shown to have no normalizable zero modes.
The important point is that, for $\cS_\mu=0$ (recall that $\cS_\mu$ is, by definition, an $x$-dependent periodic background) the operator $T$  maps to the abelian Dirac operator, which we call $\bar D$, with a  background $U(1)$ gauge field $4 \pi   A_\mu^\omega$, acting on {\it two sets} of undotted two-component Weyl spinors. These Weyl spinors are equivalent to the four complex-dimensional space of $\cW_\mu$ through the quaternion map $\cW_\mu \rightarrow \cW =  \sigma_\mu \cW_\mu$ already introduced in section~\ref{sec:tunedgeneral}. Further, as eqn.~(\ref{main equations we need to solve k eq r}) there shows,  for $\cS=0$, the self-duality equations for $\cW$ are written as $\bar D \cW = 0$. This quaternionic equation is equivalent to {\it two} Dirac equations, with the Dirac operator acting on undotted\footnote{Recalling that the Euclidean action density, see \cite{Anber:2023sjn} for spinor notation, for a $U(1)$-charged Weyl fermion is proportional to $\bar\lambda_{\dot\alpha} \bar\sigma_\mu^{\dot\alpha \alpha} D_\mu \lambda^\alpha$.} two-component spinors $\lambda_\alpha$: 
\begin{equation}\label{barDU1}
T \rightarrow \text{two copies of} \; \bar D: \bar D \lambda = 0, ~ \text{or} ~ \bar D^{\dot\alpha \alpha}  \lambda_\alpha = 0, ~\text{with} \; \bar D  \equiv \bar\sigma_{\nu}^{\dot\alpha \alpha}  D_\mu  
\end{equation}
with $D_\mu$ from (\ref{curlyD}). The conjugate operator $T^\dagger$ maps then to the conjugate Dirac operator $D$, acting on  dotted spinors   $\bar\lambda^{\dot\alpha}$,  as
$
 \sigma^\mu_{\alpha \dot\alpha} D_\mu^* \bar\lambda^{\dot\alpha},
$
where $D_\mu^* = \partial_\mu - i 2 \pi N A_\mu^\omega$, as already defined after (\ref{solution31}). The equation for zero modes of $D$ then has the form
\begin{equation}\label{DU1}
T^\dagger \rightarrow \text{two copies of} \;  D: D \bar\lambda = 0, ~ \text{or} ~  D_{\alpha \dot\alpha} \bar\lambda^{\dot\alpha} = 0, ~\text{with} \;  D  \equiv  \sigma_{\nu \; \alpha \dot\alpha}  D^*_\mu~.
\end{equation}
These equations were studied in \cite{Anber:2023sjn}: in particular, we refer to eqn.~(A.3) there for (\ref{barDU1}) and the last two lines in eqn.~(3.13) for (\ref{DU1}) (in both cases, the indices $B', C$ in \cite{Anber:2023sjn} have to be put to unity). The connection to the $U(1)$ index theorem was not emphasized there, and instead these equations were treated as equations for the zero modes of the  ``off-diagonal components'' of the adjoint fermion.
To summarize, 
it was shown there that $\bar D$ has $r$ zero modes while $D$ has no zero modes. The conclusion from these computations, phrased in terms of the abelian operators defined above, is that
${\rm dimker} (D \bar D) - {\rm dimker}(\bar D D) = r$. The index is determined by the second Chern classes (quantized nonzero fluxes) of the abelian background $4 \pi   A_\mu^\omega$ through the various two-planes of $\T^4$.
Finally, going back to $T$, $T^\dagger$ via the maps (\ref{barDU1}, \ref{DU1}), we arrive at eqn.~(\ref{indexT}), for $\cS=0$.

Thus, the index (\ref{indexT})  is  $2 r$, determined by the fluxes of the $U(1)$ gauge field through the various $\T^4$ planes, and, further, the operator $TT^\dagger$ has no zero modes for $\cS_\mu=0$. Thus, for $\cS_\mu =0$, the index determines the number of normalizable zero modes of $\cW_\mu$, which is thus $2r$. This introduces $4 r$ real coefficients determining the solution of the self-duality and gauge conditions (\ref{Wselfduality1}), as in eqn.~(\ref{wsolution}) below. Now, turning on an arbitrary periodic $x$-dependent background $\cS_\mu \ne 0$ should not change the index, as the periodic abelian field does not introduce any nonzero fluxes through the $\T^4$ planes.
We have not studied $T T^\dagger$  for arbitrary $\cS_\mu$ and have not shown that there are no zero modes of $T T^\dagger$ from  (\ref{TTdagger}).\footnote{Admittedly, we have not shown that  no zero modes of $T T^\dagger$ appear for {\it any} normalizable periodic $\cS_\mu$; hence, this is the weakest point of our argument.  To study this, in future work,  the operator $T T^\dagger$ for general $\cS_\mu \ne 0$ should be studied, in the form of eqn.~(\ref{TTdagger}),  where it does not map to the Dirac operators discussed above. }  However, perturbation theory suffices to argue that a small $x$-dependent periodic background $\cS_\mu$ does not lead to the appearance of zero modes of the Hermitean operator $T T^\dagger\vert_{\cS=0}$, which has a nonzero gap in the spectrum.

Thus, we argue that the solutions of the self-duality conditions (\ref{Wselfduality1}) for $\cW_\mu$ take the form
\begin{equation}\label{wsolution}
\cW_\mu[C, \cS] = \sum\limits_{A = 1}^{2 r} C_A \Phi_\mu^A[\cS]\,,
\end{equation}
where $C=\{C_A\}$ are $2 r$ complex coefficients and the functions $\Phi_\mu^A$ are orthogonal in the metric introduced after eqn.~(\ref{defofT}); these functions have been worked out for $\cS_\mu=0$ (they can be read off Appendix~\ref{appx:phip}), but for general periodic $\cS_\mu$ are expected to have   complicated dependence on $\cS$, as indicated above.

Now, given (\ref{wsolution}), we come to the self-duality condition for the  diagonal components of (\ref{field33}). They take a particularly simple form when written using the quaternionic notation introduced after (\ref{selfduality1}). Reducing (\ref{main equations we need to solve k eq r}) to $SU(2)$ and explicitly including the unit quaternion $\sigma_4$, we find
\begin{equation}\label{sequation}
2 \pi \bar\partial \cS = i \cW_\mu \cW_\mu^* \sigma_4 - i \bar\cW \cW^\dagger~.
\end{equation}
Since $\cS$ is periodic on the $\T^4$ this quaternionic equation only has a solution if the integral of the r.h.s. over $\T^4$ vanishes (or equivalently, the r.h.s. is orthogonal to the zero modes of  the adjoint of $\bar\partial$, as per the remark in footnote \ref{foot1}). These conditions, upon substituting (\ref{wsolution}) for $\cW_\mu$,
can be written  in quaternionic form as
\begin{eqnarray}\label{sconstr1}
\int\limits_{\T^4} (\cW_\mu[C, \cS] (\cW_\mu[C, \cS])^* \sigma_4 -  \bar\cW[C, \cS] \cW[C, \cS]^\dagger) = 0~.
 \end{eqnarray}
The above equations impose one real and one complex constraint on the $4 r$ real components of $C_A$ in (\ref{wsolution}), reducing the number of extra moduli to $4 r -3$. Modding out the constant gauge transformations in the Cartan, this implies that, including the $4$ holonomies $\phi_\mu$ (the translational moduli), there are a total of $4 r - 4 + 4 = 4r$ moduli, as appropriate for the self-dual $SU(2)$ instanton   of topological charge $r/2$. The $4 r-4$ extra moduli are supposed to make the constant-$F$ solution $x$-dependent, a behaviour similar to the one of the gcd$(k,r)\ne r$ constant-$F$ solutions discussed earlier in this section.

This reduction of the number of parameters describing the self-dual manifold passing through the constant-$F$ solution is all we can argue here. To find the global structure of the moduli space requires understanding the solutions of the nonlinear equation (\ref{sequation}, \ref{wsolution}), subject to (\ref{sconstr1}), a challenging task beyond the scope of this paper.

\section{Testing the $\Delta$-expansion for multi-fractional instantons with $Q={2\over 3}$}
\label{sec:testdelta}

In this section, we use numerics to study solutions on the slightly detuned $\T^4$ and test the analytical approach of   \cite{GarciaPerez:2000aiw}, the $\Delta$-expansion, for the multi-fractional instantons found in \cite{Anber:2023sjn}. This approach constructs analytic solutions on the detuned $\T^4$, i.e. with $\Delta(r,k,\ell)\ll 1$, in a small-$\Delta$ expansion around the constant-$F$ solution. The use of the analytical solutions for multi-fractional instantons obtained via the $\Delta$ expansion was crucial for the calculation of the higher-order gaugino condensate on a small-$\T^4$.

\begin{figure}[h]
   \centering
   \includegraphics[width=6.1in]{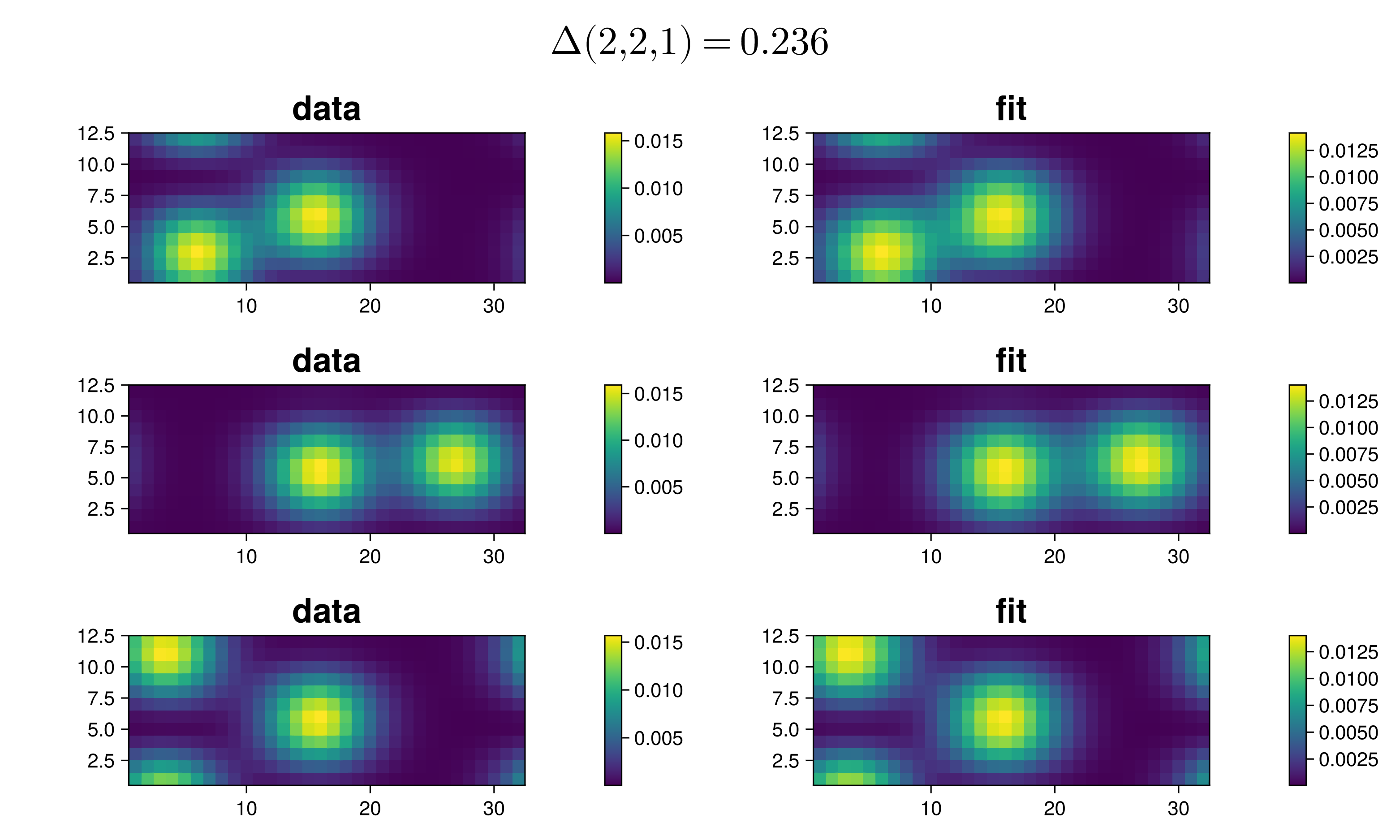} 
   \caption{The results of fitting $\Tr\(F_{13}F_{13}\)$  to the $\Delta$ expansion for three configurations (one in each row), obtained by minimizing the lattice action on a lattice of size $(32,4,12,12)$, corresponding to $\Delta(2,2,1)=0.236$. The fits produce a mean squared deviation of $\sim 10^{-7}$, showing excellent agreement with the $\Delta$ expansion. \\
   {\it Left column:} the values of $\Tr\(F_{13}F_{13}\)$ obtained from lattice configurations, summed over the $2^{nd}$ and $4^{th}$ directions, as a function of $x_1$ and $x_3$.\\   
   {\it Right column:} the $\Delta$-expansion analytical solution \eqref{eqn:Delta_fit_13},  for values of the  four parameters, the moduli $\bar{x}_1^{C'=1,2}$ and $\bar{x}_3^{C'=1,2}$,  fitted to the lattice data using a procedure described in the text. The values of the moduli $(\bar{x}_1^{C'=1},\bar{x}_3^{C'=1},\bar{x}_1^{C'=2},\bar{x}_3^{C'=2})$ for the three configurations (from top to bottom) are (5.97,8.87, 15.71,11.81), (15.87,11.57, 26.90,12.33), and (3.41,5.22, 15.77,11.79) respectively.}
   \label{fig:Delta_fit}
\end{figure}

The $\Delta$ expansion was tested numerically already in the original papers, for $SU(2)$ instantons of charge $1/2$ and was found to work well for $\Delta$ as large as $0.09$. Here we want to perform the first such tests for multi-fractional instantons. 
Due to numerical constraints we focus on the simplest theory that permits multi-fractional instantons, i.e. $SU(3)$. We consider $N=3$, $r=2$, $k=2$, i.e. instantons of charge $2/3$. It was analytically shown in \cite{Anber:2023sjn} that within the leading order in $\Delta$, these solutions have an ``instanton-liquid''-like structure, consisting of $2$ closely packed lumps.

\begin{figure}[h]
   \centering
   \includegraphics[width=4.6in]{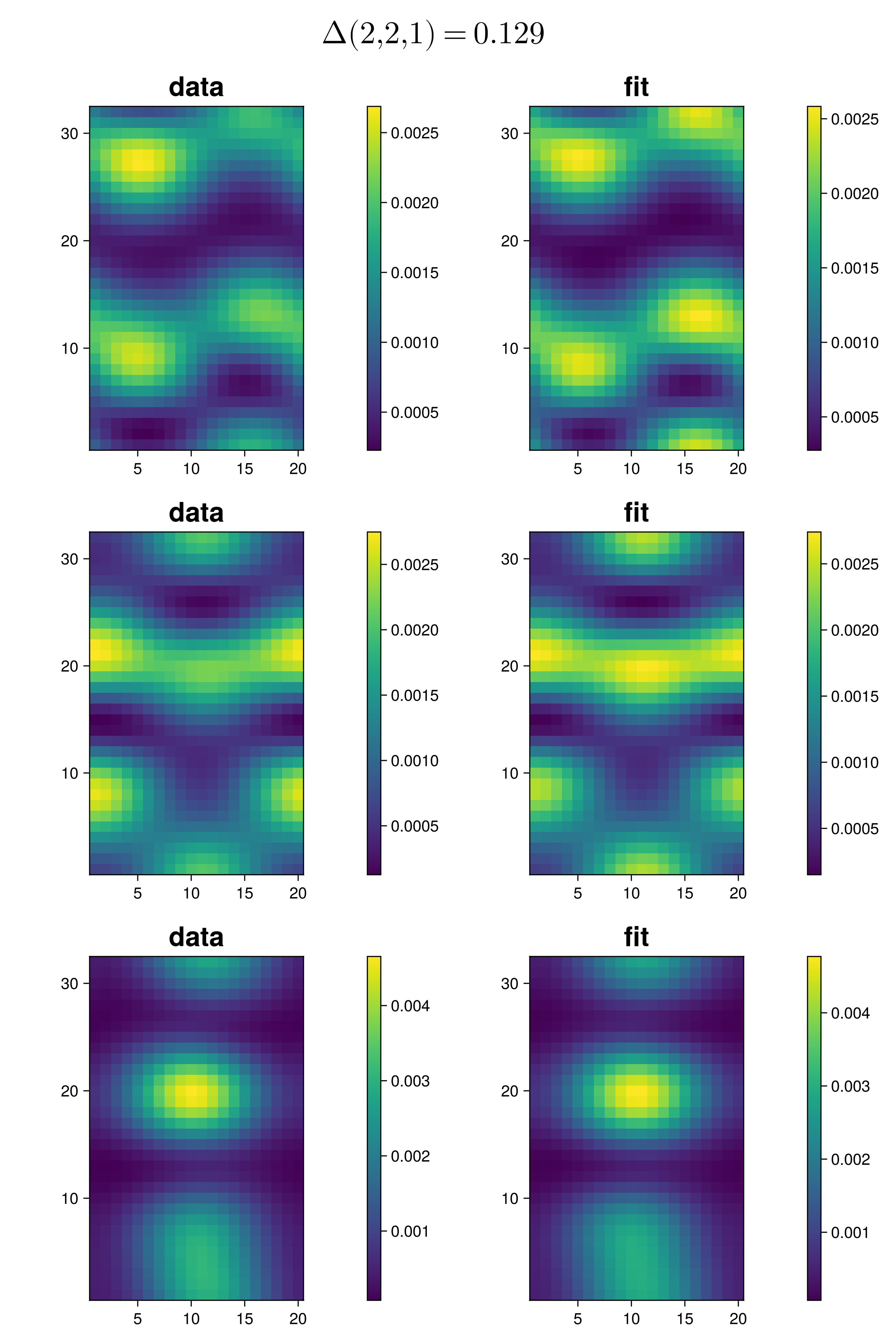} 
   \caption{ 
   The results of fitting $\Tr\(F_{13}F_{13}\)$ to the $\Delta$ expansion for three configurations on a lattice of size $(20,32,8,12)$, corresponding to $\Delta(2,2,1)=0.129$.  The fits produce a mean squared deviation of $\sim 10^{-8}$, showing excellent agreement with the $\Delta$ expansion. The  left and right columns plot the lattice data and the best fit to the theoretical expression, respectively (as in Figure~\ref{fig:Delta_fit}), for three different lattice configurations. The apparent more than two-lump structure on two of the plots for  $\Delta =0.129$   is a result of the overlap of the terms in eqn.~(\ref{eqn:Delta_fit_13}) for the best-fit values of the moduli. The values of the moduli $(\bar{x}_1^{C'=1},\bar{x}_3^{C'=1},\bar{x}_1^{C'=2},\bar{x}_3^{C'=2})$ for the three configurations (from top to bottom) are (5.36,1.83, 16.14,6.75), (0.96,14.99, 11.14,25.87), and (9.84,13.40, 11.27,26.00) respectively.}  
   \label{fig:Delta_fitSmall}
\end{figure}

Now, while the $\Delta$-expansion for gcd$(r,k)=r$ is a well defined expansion, putting it in practice is challenging and only a few quantities can be explicitly calculated to leading order in $\Delta$, as explained in \cite{Anber:2023sjn}. Here we focus on the local gauge invariants of the form $\tr F_{\mu\nu} F_{\mu\nu}$, with no sum over $\mu,\nu$, and for $\mu,\nu$ not both belonging to a two plane where a nontrivial twist has been applied (i.e., in our convention, the $1$-$2$ or $3$-$4$ plane). The reason is that in these planes, there is an order $\Delta^0$ field strength from the abelian solution and there are also contributions from the leading-order field strength of $\cS_\mu$, which is challenging to obtain in closed form.

Here, we use the results of \cite{Anber:2023sjn} to calculate the quantity $\Tr\(F_{13}F_{13}\)$. This quantity vanishes for $\Delta=0$, and to leading order in $\Delta$ is given by 
\begin{eqnarray}\label{invariant1}
	\Tr\(F_{13}F_{13}\)&=&\frac{1}{V}\frac{L_1L_3}{L_4^2}\frac{\(2\pi\)^3}{3}\Delta     \\
	&&   \times \sum_{C'=1}^2  \bigg\vert \sum_{m \in\Z}e^{i\(2\pi\frac{x_2}{L_2}+\phi_1^{C'}L_1+\pi\) m} e^{-\frac{\pi L_1}{L_2}\(\frac{x_1}{L_1}-\frac{L_2}{2\pi}\phi_2^{C'}-m-\frac{2C'-3}{4}\)^2}\bigg\vert^2 \nonumber \\
 	&& \qquad \times \;  \bigg\vert \sum_{n \in \Z} e^{i\(2\pi\frac{x_4}{L_4}+\phi_3^{C'}L_3\) n}  e^{-\frac{\pi L_3}{L_4}\(\frac{x_3}{L_3}-\frac{L_4}{2\pi}\phi_4^{C'}-n\)^2}  \(\frac{x_3}{L_3}-\frac{L_4}{2\pi}\phi_4^{C'}-n\) \bigg\vert^2\,, \nonumber
	\end{eqnarray}
where $V=L_1L_2L_3L_4$ is the volume. As appropriate for a solution of charge $2/3$, there are $8$ moduli, denoted by $\phi_\mu^{C'=1,2}$. 

The formula (\ref{invariant1}) can be interpreted as due to the contributions of two identical strongly overlapping ``lumps," the two  non-negative functions appearing in the sum over $C'=1, 2$, whose only difference is that their locations on  the $\T^4$ are different, determined by the moduli $\phi_\mu^1$ and $\phi_\mu^2$, respectively. 
Four linear combinations of the eight moduli correspond to translations of the center-of-mass of the two-lump solution, while the other four can be pictorially described as a relative separation between the lumps (the global structure of the $\phi_\mu^{C'}$ moduli space was elucidated in detail in \cite{Anber:2024mco}).

To facilitate comparison to the lattice data, we 
integrate (\ref{invariant1}) over $x_2$ and $x_4$. This eliminates the oscillating exponents in (\ref{invariant1}) and renders the integral independent of the moduli $\phi_1^{C'}$ and $\phi_3^{C'}$, corresponding to (overall and relative) translations in $x_2$ and $x_4$. Let us define $\bar{x}_1^{C'}=\frac{L_1L_2}{2\pi}\phi_2^{C'}+L_1\frac{2C'-3}{4}$ and $\bar{x}_3^{C'}=\frac{L_3L_4}{2\pi}\phi_4^{C'}$, giving us, from eqn.~(\ref{invariant1})
\begin{align}
	&f_\text{theory}(x_1,x_3; \bar{x}_1^{i},\bar{x}_3^{i})\equiv \int dx_2 dx_4 \Tr\(F_{13}F_{13}\)\label{eqn:Delta_fit_13}\\
	&\qquad = \frac{1}{L_4^2}\frac{\(2\pi\)^3}{3}\Delta\sum_{C'=1}^2\sum_{m,n\in\Z}e^{-\frac{2\pi L_1}{L_2}\(\frac{x_1-\bar{x}_1^{C'}}{L_1}-m\)^2}e^{-\frac{2\pi L_3}{L_4}\(\frac{x_3-\bar{x}_3^{C'}}{L_3}-n\)^2}\(\frac{x_3-\bar{x}_3^{C'}}{L_3}-n\)^2\nonumber.
\end{align}
We fit the four parameters $\bar{x}_1^{C'}$ and $\bar{x}_3^{C'}$ for $C'=1,2$ to lattice data.  With the Wilson action density we have
\begin{equation}
	f_\text{data}(n_1,n_3)\equiv \sum_{n_2=1}^{L_2}\sum_{n_4=1}^{L_4}\Tr\(F_{13}F_{13}\)(n)=\sum_{n_2=1}^{L_2}\sum_{n_4=1}^{L_4}2\Re\Tr\(\identity-\plaq{1}{3}\).
\end{equation}
The fitting procedure aims to minimize  
$
\sum_{n_1=1}^{L_1}\sum_{n_3=1}^{L_3}\left\lvert f_\text{theory}(n_1,n_3;\bar{x}_1^{C'},\bar{x}_3^{C'})-f_\text{data}(n_1,n_3)\right\rvert^2
$
with respect to the four fitting parameters, $\bar{x}_1^{C'}$ and $\bar{x}_3^{C'}$, for $C' = 1,2$. Only the first few values of $m$ and $n$ in equation \eqref{eqn:Delta_fit_13} are considered, as contributions from higher values are exponentially suppressed. Additionally, the improved action described in \eqref{eqn:improved_action} of Appendix \ref{appx:lattice} is employed to eliminate next-to-leading-order corrections. The data is fitted using the standard least squares method, minimizing the squared difference between theory and data across all $x_1$-$x_3$ plane points.

We show results for two different values of $\Delta$. The result for $\Delta(2,2,1)=0.236$, on a $(32,4,12,12)$ lattice is shown on Figure~\ref{fig:Delta_fit}. A different fit, this time for a smaller $\Delta(2,2,1)=0.129$ on a lattice $(20,12,8,32)$ is shown on Figure~\ref{fig:Delta_fitSmall}. The fits provide a remarkable agreement between the lattice data and the leading-order $\Delta$-expansion analytic result of \cite{Anber:2023sjn}.

{\bf {\flushleft{Acknowledgments:}}}  We would like to thank Margarita Garc\' ia P\' erez, Antonio Gonz\' alez-Arroyo, and Rajamani Narayanan for the many insightful discussions. M.A. and E.P. express their gratitude for the warm hospitality of the Instituto de Física Teórica, UAM-CSIC, Madrid. Additionally, M.A. thanks the University of Toronto for the warm hospitality during the completion of this work.  M.A. is supported by STFC through grant ST/X000591/1.   A.C. and E.P. are supported by a Discovery Grant from NSERC. 
 
 \appendix

 \section{The functions $\Phi^{(p)}$ and ${\cal G}^{(p)}_{1}$, ${\cal G}^{(p)}_{3}$}
 \label{appx:phip}
 
 We now present the normalizable solution \cite{Anber:2023sjn} of the quaternion-form eqn.~(\ref{orderlambdazero}) for general $r,k,\ell$:
 \begin{eqnarray}
  {\bar D} {\cal W}^{(0)k\times \ell}&=&0~.
\label{Wequation1}
\end{eqnarray}
 First, the four components of four-vector $\cW_\mu^{(0) \; k \times \ell}$ solving (\ref{Wequation1}) are related as follows:
  \begin{eqnarray}\label{assertion1}
{\cal W}_4^{(0)k\times \ell}=i{\cal W}_3^{(0)k\times \ell}\,, \quad {\cal W}_2^{(0)k\times \ell}=i{\cal W}_1^{(0)k\times \ell}\,
\end{eqnarray}
The solutions for ${\cal W}_4^{(0)k\times \ell}, {\cal W}_2^{(0)k\times \ell}$ are then given in terms of $r\over {\rm gcd}(k,r)$ functions $\Phi^{(p)}_{C'C}$ (shown further below in (\ref{form of Phi})): %
\begin{eqnarray}
\nonumber
\left({\cal W}_2^{(0)k\times \ell}\right)_{C'C}=V^{-1/4} \sum_{p=0}^{\scriptsize \frac{r}{\mbox{gcd}(k,r)}-1}{\cal C}^{[C'+pk]_r}_2\Phi^{(p)}_{C'C}(x,\hat\phi) =: {W}_{2 \; C'C}\,,\\
\left({\cal W}_4^{(0) k\times \ell}\right)_{C'C}=V^{-1/4} \sum_{p=0}^{\scriptsize \frac{r}{\mbox{gcd}(k,r)}-1}{\cal C}^{[C'+pk]_r}_4\Phi^{(p)}_{C'C}(x,\hat\phi)  =:{ W}_{4 \; C'C}\,,
\label{expressions of W2 and W4 with holonomies}
\end{eqnarray}
where   the volume factor $V=L_1 L_2 L_3 L_4$ is included for  convenience. 
 The solutions are parameterized by $2r$ arbitrary complex coefficients ${\cal C}^{[C'+pk]_r}_2$ and ${\cal C}^{[C'+pk]_r}_4$. They are not determined at the linear order of the $\lambda$- or $\Delta$-expansion, as discussed in the main text (and in \cite{Anber:2023sjn}). 

We also note that the solution we present below includes the dependence on the moduli $\phi_\mu^{C'}$ of (\ref{Ukbackground}, \ref{Ulbackground}, \ref{SUNmoduli}) in the background covariant derivative in (\ref{Wequation1}). The precise relation between $\phi_\mu^{C'}$ and the $\hat\phi_\mu^{C'}$ appearing in (\ref{form of Phi}) below is as follows\footnote{Briefly, it is determined by the adjoint action of the background covariant derivative and the fact that $\cW$ is the $k \times \ell$ component inside $SU(N)$, as per (\ref{solution21}).}
\begin{equation}
\label{phis}
\hat\phi_\mu^{C'} = - 2\pi (\ell \phi^{C'}_\mu + k \tilde\phi_\mu)
= - 2\pi (\ell \phi^{C'}_\mu + \sum\limits_{A'=1}^k \phi^{A'}_\mu)~,
\end{equation}
and also obeys  (\ref{SUNmoduli}).
To complete the solution, the ${r \over {\rm{gcd}}(k,r)}$ functions $\Phi^{(p)}$ are given by (a derivation is in Appendix A of \cite{Anber:2023sjn}):\begin{eqnarray}
\nonumber
\Phi^{(p)}_{C' B}(x,\hat\phi)&=& \sum_{\scriptsize m=p+\frac{rm'}{\mbox{gcd}(k,r)},\, m'\in \mathbb Z}~~\sum_{n'\in \mathbb Z}e^{\frac{i2\pi x_2 }{L_2}(m+\frac{2	C'-1-k}{2k})}e^{\frac{i2\pi x_4 }{L_4}(n'-\frac{2	B-1-\ell}{2\ell})}\\
\nonumber
&&~~\times e^{-i\frac{\pi (1-k)}{k}\left(C'-\frac{1+k(1-2m)}{2}\right)} e^{i\frac{\pi(1-\ell)}{\ell}\left(B-\frac{1+\ell(2n'+1)}{2}\right)}\\
\nonumber
&&~~\times \; e^{-\frac{\pi r}{k L_1 L_2}\left[x_1-\frac{k L_1 L_2}{2\pi r}(\hat \phi_2^{[C']_r}-i\hat \phi_1^{[C']_r})-\frac{L_1}{r}\left(km +\frac{2C'-1-k}{2}\right)\right]^2}\\
&&~~\times \; e^{-\frac{\pi }{\ell L_3 L_4}\left[x_3-\frac{\ell L_3 L_4}{2\pi }(\hat \phi_4^{[C']_r}-i\hat \phi_3^{[C']_r})-L_3\left(\ell n' -\frac{2B-1-\ell}{2}\right)\right]^2}\,. \label{form of Phi}
\end{eqnarray}
The explicit form of the functions $\Phi^{(p)}$ will be useful  in our study of the properties  of the self-dual fractional instantons moduli spaces. We shall need the property \cite{Anber:2023sjn} that:
\begin{equation}
\label{property1}
\int_{\mathbb T^4} \sum\limits_{C=1}^\ell \Phi^{(p)}_{C'C}(x) \Phi^{(p') *}_{C'C}(x) = const.\;  \delta^{p,p'}
\end{equation}
where the constant is nonzero and depends on $\hat\phi_\mu^{C'}$ only.

In section~\ref{sec:lambdafit}, we also use
the functions ${\cal G}^{(p)}_{1,C',C}(x,\hat\phi)$ and ${\cal G}^{(p)}_{3,C',C}(x,\hat\phi)$
in our calculation of the gauge-invariant density in the framework of the $\lambda$-expansion on the tuned-$\T^4$ for gcd$(k,r)\ne r$. 
These functions are related to  derivatives of $\Phi^{(p)}$, as described in \cite{Anber:2023sjn}. Their explicit form, taking $k=1, \ell = 2$ ($N=3$) is:
\begin{eqnarray}
\label{the G1 function}
{\cal G}^{(p)}_{1,C'=1,C}(x,\hat\phi)&=&-\frac{4\pi  }{L_1 L_2}{\cal J}  e^{-i\hat \phi_1^{} x_1}  e^{-i\hat \phi_3^{} x_3}\\
\nonumber
&&\times  \sum_{\scriptsize m=p+2m',\, m'\in \mathbb Z}\sum_{n'\in \mathbb Z} e^{i \left(\frac{ 2\pi x_2 }{L_2}+\frac{L_1 }{2}\hat \phi_1^{}\right)(m)}e^{i\left(\frac{2\pi x_4 }{L_4}+2 L_3\hat \phi_3^{}\right)(n'-\frac{2	C-3}{4})}\\
\nonumber
&&\times e^{-i\frac{\pi}{2}\left(C-\frac{1+2(2n'+1)}{2}\right)} \left( x_1-\frac{ L_1 L_2\hat \phi_2^{}}{4\pi } -\frac{L_1 m}{2}\right) ~ e^{-\frac{2\pi }{ L_1 L_2}\left[x_1-\frac{ L_1 L_2}{4\pi }\hat \phi_2^{}-\frac{L_1}{2}\left(m \right)\right]^2} \nonumber \\
&&\times e^{-\frac{\pi }{2 L_3 L_4}\left[x_3-\frac{2 L_3 L_4}{2\pi }\hat \phi_4^{}-L_3\left(2 n' -\frac{2C-3}{2}\right)\right]^2}\,,\quad C=1,2\,, \nonumber 
\end{eqnarray}
where 
$$
{\cal{J}}^2 = {2 \sqrt{V} \over L_2 L_4}~,
$$
and 
\begin{eqnarray}\label{the G3 function}
{\cal G}^{(p)}_{3,C'=1,C}(x,\hat\phi)&=&-\frac{\pi }{L_3 L_4}{\cal J}  e^{-i\hat \phi_1^{} x_1}  e^{-i\hat \phi_3^{} x_3}\\
\nonumber
&&\times  \sum_{\scriptsize m=p+2m',\, m'\in \mathbb Z}\sum_{n'\in \mathbb Z} e^{i \left(\frac{ 2\pi x_2 }{L_2}+\frac{L_1 }{2}\hat \phi_1^{}\right)(m)}e^{i\left(\frac{2\pi x_4 }{L_4}+2 L_3\hat \phi_3^{}\right)(n'-\frac{2	C-3}{4})}\\
\nonumber
&&\times e^{-i\frac{\pi}{2}\left(C-\frac{1+2(2n'+1)}{2}\right)} \left( x_3-\frac{2 L_3 L_4\hat \phi_4^{}}{2\pi } -L_3\left(\ell n'-\frac{2C-3}{2}\right)\right)\\
\nonumber
&&\times e^{-\frac{2\pi }{ L_1 L_2}\left[x_1-\frac{ L_1 L_2}{4\pi }\hat \phi_2^{}-\frac{L_1}{2}\left(m \right)\right]^2} ~e^{-\frac{\pi }{2 L_3 L_4}\left[x_3-\frac{2 L_3 L_4}{2\pi }\hat \phi_4^{}-L_3\left(2 n' -\frac{2C-3}{2}\right)\right]^2}\,,\quad C=1,2\,. \nonumber 
\end{eqnarray}
The functions ${\cal G}^{(p)}_{1,C',C}$ and ${\cal G}^{(p)}_{2,C',C}$ satisfy the normalization conditions (with the r.h.s. independent of the moduli):
\begin{eqnarray}\nonumber
&&\int_{\mathbb T^4}\sum_{C=1}^{\ell} {\cal G}^{*(p)}_{1,C',C} {\cal G}^{(p')}_{1,C',C}\propto\delta_{pp'}\,,\quad \int_{\mathbb T^4}\sum_{C=1}^{\ell} {\cal G}^{*(p)}_{3,C',C} {\cal G}^{(p')}_{3,C',C}\propto\delta_{pp'}\\
&&\int_{\mathbb T^4}\sum_{C=1}^{\ell} {\cal G}^{*(p)}_{1,C',C} {\cal G}^{(p')}_{3,C',C}=0\,.
\end{eqnarray}

 \section{Moduli dependence of the $SU(3)$ Wilson lines for the constant-$F$ solutions with $r=k$ on the tuned-$\T^4$ }
 \label{appx:wilson}
 
 For completeness, here we give the Wilson  lines for $SU(3)$ in the form $(\Re W_\mu, \Im W_\mu)(x=0,z_\mu)$ in the case $N=3$ and $k=r=1$,  plotted on Figure~\ref{fig:r=k=1wilson}:
\begin{eqnarray}\nonumber
(\Re W_1, \Im W_1)&=&\left(\cos(4\pi z_1)+2\cos(2\pi z_1), -\sin(4\pi z_1)+2\sin(2\pi z_1)\right)\,,\\\nonumber
(\Re W_2, \Im W_2)&=&\left(\cos(4\pi z_2)+2\cos(2\pi z_2), -\sin(4\pi z_2)+2\sin(2\pi z_2)\right)\,,\\\nonumber
(\Re W_3, \Im W_3)&=&\left(\cos(4\pi z_3), -\sin(4\pi z_3)\right)\,,\\
(\Re W_4, \Im W_4)&=&\left(\cos(4\pi z_4), -\sin(4\pi z_4)\right)\,. \label{wilsoneqns1}
\end{eqnarray}
Here, we have relabeled the four moduli of eqn.~(\ref{SUNmoduli}) as  $\phi_\mu^{C'=1}=4 \pi z_\mu$ where the range of  $z_\mu$ is $z_{1,2}\in [0,1]$ and $z_{3,4}\in [0,\frac{1}{2}]$, as per \cite{Anber:2024mco}. The  curves in the four $(\Re W_\mu, \Im W_\mu)(x=0,z_\mu)$ planes, traced as $z_\mu$ is varied are shown by the continuous lines on Figure~\ref{fig:r=k=1wilson}.

For the case of $N=3$ and $r=k=2$, the Wilson lines, also given in the form $(\Re W_\mu, \Im W_\mu)(x=0,\phi_\mu^{C'})$, have a more complicated expression \cite{Anber:2024mco}. This is because there are now $8$ moduli $\phi_\mu^{C'=1}$ and $\phi_\mu^{C'=2}$, as per (\ref{SUNmoduli}). However, their ranges and meaning\footnote{As explained in \cite{Anber:2024mco}, when the $r=k=2$ solution is considered on the $\Delta \ne 0$ deformed-$\T^4$, $z_\mu$ becomes the center of mass coordinate of the two-lump instanton and $a_\mu$ is the relative separation coordinate.} are more transparent upon changing variables as $
\phi_\mu^1 = 2\pi (z_\mu - {a_\mu})$, $\phi_\mu^2 =  2\pi (z_\mu + {a_\mu})$, where the ranges of $z_\mu$  and $a_\mu$  are $z_\mu\in [0,1]$ and $a_\mu \in [0,\frac{1}{2}]$.
Now, we find, from Appendix D of \cite{Anber:2024mco}:
\begin{eqnarray}\nonumber
(\Re W_1, \Im W_1)&=&\left(-\cos\left(2\pi(-z_1+ { a_1} )\right)-\cos\left(2\pi(z_1+a_1)\right)+\cos(4\pi z_1),\right.\\\nonumber
&&\left.-\sin\left(2\pi(-z_1+a_1)\right)+\sin \left(2\pi(z_1+a_1)\right)+\sin(4\pi z_1) \right)\,,\\\nonumber
(\Re W_2, \Im W_2)&=&\left(\cos\left(2\pi(-z_2+a_2-\frac{1}{2})\right)+\cos\left(2\pi(z_2+a_2-\frac{1}{2})\right)+\cos(4\pi z_2),\right.\\\nonumber
&&\left.\sin\left(2\pi(-z_2+a_2-\frac{1}{2})\right)-\sin \left(2\pi(z_2+a_2-\frac{1}{2})\right)+\sin(4\pi z_2) \right)\,,\\\nonumber
(\Re W_3, \Im W_3)&=&\left(\cos\left(2\pi(-z_3+a_3)\right)\cos\left(2\pi(z_3+a_3)\right)+\cos(4\pi z_3),\right.\\\nonumber
&&\left.\sin\left(2\pi(-z_3+a_3)\right)-\sin \left(2\pi(z_3+a_3)\right)+\sin(4\pi z_3) \right)\,,\\\nonumber
(\Re W_4, \Im W_4)&=&\left(\cos\left(2\pi(-z_4+a_4)\right)+\cos\left(2\pi(z_4+a_4)\right)+\cos(4\pi z_4),\right.\\
&&\left.\sin\left(2\pi(-z_4+a_4)\right)-\sin \left(2\pi(z_4+a_4)\right)+\sin(4\pi z_4) \right)\,. \label{wilsonr=k=2}
\end{eqnarray}
When $a_\mu=0$, as a function of $z_\mu$, the above equations trace the solid curves shown on each   $(\Re W_\mu, \Im W_\mu)(x=0,z_\mu,0)$ plane on Figure~\ref{fig:r=k=2wilson}. As $a_\mu$ is varied away from $0$, they fill the inside of the envelope of the solid curves on the figure.

\section{Numerical studies on the lattice}\label{appx:lattice}
To study fractional instantons numerically, e.g.~as in \cite{Montero:2000mv}, we consider a lattice of size $(L_1,L_2,L_3,L_4)$. Link variables are related to gauge fields in the usual way via $\tilde{U}_\mu(n)=e^{ia A_\mu(n)}$, where lattice points are labeled by integer coordinates $n_\mu = \{1, ..., L_\mu \}$. The plaquettes summed over in the lattice action emanate   from all lattice points with coordinates  $n_\lambda = \{1,...,L_\lambda\}$ in all positive directions in all $\mu$-$\nu$ planes. We denote by $\tilde{U}_{\mu\nu}(n)$  the plaquette in the $\mu$-$\nu$ plane at point $n$,
\begin{equation}\label{plaquette}
	\tilde{U}_{\mu\nu}(n)=\plaq{\mu}{\nu}=\tilde{U}_\mu\(n\)\tilde{U}_\nu\(n+\hat{\mu}\)\tilde{U}_\mu^\dagger\(n+\hat{\nu}\)\tilde{U}_\nu^\dagger\(n\)~.
\end{equation}
To define a plaquette  whose origin is at the edge of the lattice, i.e. has any coordinate with $n_\mu = L_\mu$, we need to impose boundary conditions on the link variables, relating their values   at $n_\mu = L_\mu+1$ to those at  $n_\mu=1$. Instead of the usual periodic boundary conditions, we 
  subject  the  link variables to  the twisted boundary conditions, relating their values at $n_\mu=1$ and $n_\mu = L_\mu+1$, as in the continuum eqn.~(\ref{conditions on gauge field}),
\begin{equation} \label{bc1}
	\tilde{U}_\mu\(n+L_\nu\hat{\nu}\)=\Omega_\nu\(n\)\tilde{U}_\mu\(n\)\Omega^\dagger_\nu\(n+\hat{\mu}\),
\end{equation}
where the transition functions $\Omega_\nu$ obey  (\ref{cocycle}) (recalling that $\Omega_\nu$  is independent on $n_\nu$).\footnote{Formally, the definition (\ref{bc1}) defines fields over an infinite cover of the finite lattice, but we shall only need the subset of link variables obeying (\ref{bc1}) that enter the lattice action (\ref{action1}).}  However, when written using the independent $\tilde U_\mu$ variables, after imposing (\ref{bc1}), the plaquette action acquires explicit dependence on the transition functions\footnote{As is easiest to verify in e.g.~a single-plaquette two-dimensional world with $L_1=L_2=1$.} and the measure has to include integration over these with the right cocycle condition.
One can, however, make a change of variables:
\begin{equation}\label{bc2}
	\tilde U_\mu(n)=\begin{cases}
		 {U}_\mu(n) & n_\mu\neq L_\mu\\
		 {U}_\mu(n)  \Omega^\dagger_\mu(n) \;& n_\mu=L_\mu
	\end{cases}~.
\end{equation}
 A careful application of (\ref{bc1}) to the lattice action (given by the first term in (\ref{action1}) below), followed by (\ref{bc2}), shows that when expressed in terms of $U_\mu$, the  Wilson action does not have an explicit dependence on $\Omega_\mu$ and is, instead,  given by the second term below
\begin{equation} \label{action1}
	S_{\text{Wilson}}=\frac{2}{g^2}\sum_{n_\lambda =1}^{L_\lambda} \sum_{\mu<\nu}\Re\Tr\(\identity-\tilde{U}_{\mu\nu}(n)\)=\frac{2}{g^2}\sum_{n_\lambda=1}^{L_\lambda} \sum_{\mu<\nu}\Re\Tr\(\identity-B_{\mu\nu}(n)U_{\mu\nu}(n)\). 
\end{equation}
Here, 
the plaquette $U_{\mu\nu}(n)$ is defined in the same way as (\ref{plaquette}), with the  boundary plaquettes evaluated using $U_\mu(n + L_\nu \hat \nu ) = U_\mu(n)$ instead of (\ref{bc1}). Most importantly, the dependence on the twist (\ref{cocycle}) is encoded in $B_{\mu\nu}(n)$, the center symmetry $\Z_N$-background,
\begin{equation} \label{twoform}
	B_{\mu\nu}(n)=\begin{cases} e^{-2\pi in_{\mu\nu}/N} & n_\mu=L_\mu \text{ and }n_\nu=L_\nu\\
	1 & \text{otherwise}
	\end{cases}.
\end{equation}
As made clear from (\ref{action1}), a phase $e^{-2\pi i n_{\mu\nu}/N}$ 
is now included in the action, at the point $(n_\mu,n_\nu)=(L_\mu,L_\nu)$ in the $\mu$-$\nu$ plane, for all values of the other $n_\lambda$, $\lambda\ne \mu,\nu$ coordinates. With the path integral over periodic $U_\mu(n)$, the modern interpretation \cite{Kapustin:2014gua} of the 't Hooft twist as arising from a nontrivial topological background\footnote{``Topological'' means that there is no $\Z_N$ flux of the two-form $B_{\mu\nu}$ through any cubes. The background is instead characterized by the nontrivial $\Z_N$ holonomies (determined by $n_{\mu\nu}$(mod $N$)) of  the two-form field over the noncontractible two-planes of the torus.} for the two-form  $\Z_N$ gauge field  gauging the $1$-form center symmetry, i.e.~the plaquette-based (\ref{twoform}), is quite explicit. An equivalent interpretation is that a nondynamical center vortex wrapping the  $\lambda\ne \mu,\nu$ directions of the torus and located at $(n_\mu,n_\nu)=(L_\mu,L_\nu)$ is imposed by the twist \cite{Greensite:2011zz}.

To find fractional instantons, we minimize the $SU(N)$ Wilson action starting from random configurations, distributed according to the $SU(N)$ Haar measure. The action is minimized using cooling via $SU(2)$ subgroups \cite{Cabibbo:1982zn,Okawa:1982ic}. The cooling procedure iteratively minimizes the action with respect to each link variable in the following way. The part of the action which depends on $U_\mu(n)$ is given by
\begin{align}\nonumber
    &-2\Re\sum_{\nu\neq \mu}\Tr\left(U_\mu(n)\left(B_{\mu\nu}(n)U_\nu(n+\hat{\mu})U^\dagger_\mu(n+\hat{\nu})U^\dagger_\nu(n) \right.\right.\\\nonumber
    &\left.\left.     \quad\quad \quad\quad+ B_{\mu\nu}^*(n-\hat{\nu})U^\dagger_\nu(n+\hat{\mu}-\hat{\nu})U^\dagger_\mu(n-\hat{\nu})U_\nu(n-\hat{\nu})\right)\right)\\
    &\quad=-2\Re\Tr(U_\mu(n)M_\mu(n)),
\end{align}
where $M_\mu(n)$ is the staple defined as
\begin{equation}
    M^\dagger_\mu(n)=\sum_{\nu\neq \mu}\left(\ \stapup + \stapdw\right).
\end{equation}
The case of $N=2$ is special since the sum of $SU(2)$ matrices is given by an $SU(2)$ matrix multiplied by a non-negative real number, meaning that $M_\mu^\dagger(n)=\alpha V$ for $\alpha\geq 0$ and $V\in SU(2)$. Since the trace of an $SU(2)$ element is real and bounded by $\pm2$, the optimal update is $U_\mu(n)\rightarrow V$. For $N>2$ this procedure is generalized, by decomposing the $SU(N)$ elements into $N(N-1)/2$ $SU(2)$ subgroups and performing the $SU(2)$ procedure there. One may generalize this procedure to any plaquette based action, i.e.~changing the shape of the plaquettes used, by appropriately modifying the staples.

Once the action is sufficiently close to the BPS limit, we switch to cooling, via the same method by appropriately modifying the staples $M_\mu(n)$, an improved action by introducing $2\times2$ plaquettes \cite{GarciaPerez:1993lic,deForcrand:1995qq} to eliminate the next-to-leading-order, $\mathcal{O}(a^6)$, terms in the Wilson action (i.e.~we take $\epsilon=0$ in (\ref{eqn:improved_action})).  The improved action takes the form 
\begin{equation}
	\frac{S_{\text{improved}}}{2/g^2} = \sum_{n_\lambda=1}^{L_\lambda}\sum_{\mu,\nu}\[\frac{4-\epsilon}{3}\Tr\(\identity - \plaq{\mu}{\nu}\) + \frac{\epsilon-1}{48}\Tr\(\identity - \twoplaq \)\].\label{eqn:improved_action}
\end{equation}
After minimizing the action, the topological charge of the resulting configuration is calculated using a n\"{a}ive, but sufficient, discretization of the continuum topological charge. Configurations are confirmed to be BPS by checking that the BPS limit, $S=\frac{8\pi^2\left\lvert Q_{\text{top}}\right\rvert}{g^2}$, is satisfied (the configurations here are all within 1\% of the BPS action), using the simplest (naive) lattice definition of the topological charge,
\begin{equation}
	Q_\text{top}=-\frac{1}{32\pi^2}\sum_{n_\lambda=1}^{L_\lambda}\sum_{\mu,\nu,\alpha,\beta}\epsilon_{\mu\nu\alpha\beta}\Tr\(\plaq{\mu}{\nu}\cdot\plaq{\alpha}{\beta}\).
\end{equation}

In our numerical study for $SU(3)$ gauge group, we only take the twists $n_{12}=n_{34}=1$ to be nonzero (mod $3$). This corresponds to a topological charge $Q= -{1 \over 3}\; (\rm{mod}\; 1)$. Thus, the numerical minimization procedure described above, starting from a random configuration, gives rise to minimum-action self-dual configurations whose topological charges equal either $Q=-1/3$ or  $Q=2/3$. We note that with our cooling procedure, $Q=2/3$ configurations are stable (even allowing for $10^5$ cooling sweeps of the lattice they do not decay to $Q=-1/3$).

\bibliography{ModuliT4.bib}

  \bibliographystyle{JHEP}

\end{document}